\documentclass[lettersize,journal]{IEEEtran}
\usepackage{amsmath,amsfonts}

\usepackage{array}
\usepackage[caption=false,font=normalsize,labelfont=sf,textfont=sf]{subfig}
\usepackage{textcomp}
\usepackage{stfloats}
\usepackage{url}
\usepackage{verbatim}
\usepackage{graphicx}
\usepackage{xcolor}
\usepackage{cite}
\usepackage{multirow}
\usepackage{makecell}
\usepackage{amsmath,amssymb,amsfonts}
\usepackage{graphicx}
\usepackage{hyperref}
\usepackage{tabularray}
\usepackage{threeparttable}

\usepackage{algorithm}
\usepackage{algpseudocode}
\floatname{algorithm}{Procedure}

\usepackage{enumitem}
\setlist[enumerate,1]{          %   1 = outermost level
  label=(\textit{\roman*}),              %   (i), (ii), …
  leftmargin=11pt,               %   no extra indent
  labelsep=0.2em,               %   gap between label and text
}

\usepackage{booktabs}

\usepackage{fancyhdr} % For customizing headers and footers

\fancypagestyle{firstpagestyle}{
  \fancyhf{} % Clear all header and footer fields for this style
  \fancyhead[C]{%
    \fontsize{9}{11}\selectfont% Set font size
    © 2026 IEEE. Personal use of this material is permitted. Permission from IEEE must be obtained for all other uses, in any current or future media, including reprinting/republishing this material for advertising or promotional purposes, creating new collective works, for resale or redistribution to servers or lists, or reuse of any copyrighted component of this work in other works.
  }
}

\fancypagestyle{subsequentpages}{
  \fancyhf{} % Clear all header and footer fields for this style
  \fancyhead[L]{%
    \fontsize{9}{11}\selectfont% Set font size
    \textsc{Aftabi \MakeLowercase{\textit{et al.}}: DynaMark: An RL Framework for Dynamic Watermarking in Industrial MTC's, \today}%
  }
  \fancyhead[R]{\thepage} % Add page number on the right
}

\usepackage{orcidlink}

\usepackage{booktabs}

\usepackage{amsthm}
\theoremstyle{plain} 
\newtheorem{theorem}{Theorem}
\newtheorem{remark}{Remark}
\newtheorem{lemma}{Lemma}

\newcommand{\vx}{\mathbf{x}}
\newcommand{\vy}{\mathbf{y}}
\newcommand{\vu}{\mathbf{u}}
\newcommand{\vv}{\mathbf{v}}
\newcommand{\vw}{\mathbf{w}}
\newcommand{\ve}{\mathbf{e}}
\newcommand{\vh}{\mathbf{h}}
\newcommand{\N}{\mathcal{N}}
\newcommand{\vr}{\mathbf{r}}
\newcommand{\R}{\mathbb{R}}

\DeclareMathOperator{\E}{\mathbb{E}}

\DeclareMathOperator{\tr}{tr}

\begin{document}

\title{\vspace{2ex}DynaMark: A Reinforcement Learning Framework for Dynamic Watermarking in Industrial Machine Tool Controllers}

\author{
Navid Aftabi\orcidlink{0000-0002-2063-1622}, Abhishek Hanchate\orcidlink{0000-0001-8438-3325}, Satish Bukkapatnam\orcidlink{0000-0003-3312-8222}, and Dan Li\orcidlink{0000-0001-6016-454X}
\vspace{-4ex}
        % <-this % stops a space
% \thanks{}% <-this % stops a space
\thanks{Navid Aftabi and Dan Li are with Industrial \& Systems Engineering Department, University of Washington, Seattle, WA 98195 USA (e-mail: \href{mailto:aftabi@uw.edu}{aftabi@uw.edu}, \href{mailto:dli27@uw.edu}{dli27@uw.edu}). }%
\thanks{Abhishek Hanchate and Satish Bukkapatnam are with Wm Michael Barnes ’64 Department of Industrial and Systems Engineering, Texas A\&M University, College Station, TX 77843 USA. (e-mail: \href{mailto:abhishek.hanchate@tamu.edu}{abhishek.hanchate@tamu.edu}, \href{mailto:satish@tamu.edu}{satish@tamu.edu}).}
}

% \markboth{Journal of \LaTeX\ Class Files,~Vol.~14, No.~8, August~2021}{Aftabi \MakeLowercase{\textit{et al.}}: DynaMark: An RL Framework for Dynamic Watermarking in Industrial MTC's}

% \IEEEpubid{}

\maketitle

\thispagestyle{firstpagestyle}

\begin{abstract}
Machine tool controllers are vulnerable to replay attacks that conceal abnormal behavior by streaming previously recorded measurements while maliciously manipulating control commands. Watermarking can expose replay by injecting authentication signals, yet static watermark statistics impose a fragile trade-off between detection sensitivity, actuation energy, and control performance, especially under nonlinear and time-varying dynamics. To address this challenge, we propose DynaMark, a reinforcement learning framework that casts adaptive watermarking design—Dynamic Watermarking—as a Markov decision process, and learns an adaptive watermarking policy directly from measured data and detector feedback. The policy selects time-varying watermark statistics to remain performance-aware during nominal operation and to intensify excitation when replay-induced inconsistencies emerge. We provide a local analytical characterization of how replay alters the detector statistic under nonlinear and time-varying measurement-space dynamics, enabling an online belief update that summarizes detection confidence and serves as a compact state input to the learning algorithm. We further establish a closed-loop mean-square boundedness guarantee for the resulting watermarked operation under mild local regularity conditions and bounded watermark budgets. Experiments on a Siemens Sinumerik 828D controller digital emulator, a nonlinear mass–spring–damper benchmark, and a physical smart stepper-motor testbed demonstrate that DynaMark achieves rapid replay detection while maintaining favorable energy and control-performance trade-offs compared with static-variance, non-RL belief-adaptive, and linear-surrogate baselines.
\end{abstract}

\begin{ntp}
Networked sensors and controllers in machine tools enable high performance but also expose systems to replay attacks, where recorded normal measurements are fed to monitoring software while the machine is driven incorrectly. This paper introduces an adaptive watermarking method that adds small random variations to command signals and adjusts their strength online: it stays minimal during normal operation to avoid unnecessary energy use and quality loss, and increases only when the monitor indicates elevated risk, making replay harder to hide. Deployment requires routine data logs, a lightweight detector, and a safe cap on watermark intensity. Limitations include model fitting from nominal data to build a simulator for RL optimization and process-specific tuning. Future work will add online model updates and automated recovery after alarms.
\end{ntp}


\pagestyle{subsequentpages}

\section{Introduction}
\label{sec:introduction}
\IEEEPARstart{T}{he} digital transformation, real-time analytics, and Artificial Intelligence (AI) are advancing manufacturing toward interconnected Industry~4.0 ecosystems, but cybersecurity remains deficient, leaving legacy plant floor assets exposed to sophisticated threats \cite{TUPTUK201893,Mullet9345803,9247392}. Notable incidents like WannaCry shutdowns at auto plants and the 2019 LockerGoga attack on Norsk Hydro highlight the vulnerability of Machine Tool Controllers (MTC) in managing Computer Numerical Control (CNC) machinery and other equipment on the plant floor~\cite{9247392}. Compounding this vulnerability, most MTCs have nonlinear, closed architectures, limiting insight into their mechanisms and restricting efforts to understand and mitigate their security risks~\cite{TIWARI2023695}. Among cyberattacks, replay attacks are especially dangerous as they need no model knowledge; attackers can just record and replay legitimate measurement streams, bypassing intrusion detection and risking part quality and catastrophic damage~\cite{9177270,Mo7011170}. A common method to detect replay attacks is using authentication signals like watermarking, which are unknown to attackers~\cite{9177270}. 

Physical watermarking verifies system integrity and authenticity, similar to how traditional watermarks prevent piracy and confirm ownership. Watermarking embeds unique authentication signals into the system to distinguish legitimate from replayed data~\cite{Zhou9765754}. However, the effectiveness of this method depends heavily on the careful design of these signals. High detection accuracy may degrade control performance, as overly sensitive detection mechanisms can disrupt normal controller operation~\cite{Mo7011170,Liu9061046}. Static or poorly tuned watermarks hinder performance or fail to address evolving attacks~\cite{Mo7011170,Liu9061046,Xudong10313050}. This tradeoff motivates an adaptive watermarking paradigm, an approach that offers greater flexibility but increases complexity, posing additional challenges for nonlinear systems~\cite{Porter9187955}. A promising approach would use adaptive watermarking to detect replay attacks, balancing detection accuracy and system responsiveness. In this paper, Dynamic Watermarking (DWM) refers to adaptive watermarking (online selection of watermark statistics using system feedback) whereas offline fixed-statistics designs are termed static or stationary watermarking.

\subsection{Related Works}\label{sec:litrev}

\subsubsection{MTC Cybersecurity}
The shift to IoT-enabled, data-driven Industry 4.0 workflows has expanded the cyberattack surface of MTCs~\cite{TUPTUK201893,9247392,Mullet9345803}. MTCs are vulnerable due to outdated systems without regular security updates, low operator awareness, and dense network connections~\cite{Mullet9345803,9247392}. Manufacturing facilities should integrate IoT-specific security, such as multi-layer authentication, tamper-resistant encryption, and real-time surveillance, into their operational technology infrastructure~\cite{Mullet9345803}. AI and machine learning algorithms can analyze controller data to detect nuanced anomalies while adjusting defense policies. Recent cybersecurity studies on CNC machines have further explored deep learning-based models for detecting cyber-physical attacks, but these mainly serve as passive anomaly detectors rather than active authentication or watermarking mechanisms~\cite{11050046}. Viewing cybersecurity as a core design element, rather than just an operational expense, is crucial for resilient smart manufacturing and MTC operations~\cite{Mullet9345803,9247392}.

\subsubsection{Attacks on MTCs}\label{sec:2.2}
Cyberattacks on MTCs can cause not only data breaches but also significant physical damage. Among these, \emph{deception attacks} are especially dangerous because they exploit the trust between cyber and physical components~\cite{9177270,Porter9187955}. In a deception attack, an adversary alters system data to cause harmful actions by the system or its users. Three major types of deception attacks on MTCs~\cite{Porter9187955} commonly discussed in the literature are:
\begin{itemize}[leftmargin=10pt]
    \item \emph{Flip attacks:} Flip attacks compromise data and control signal integrity in industrial control systems and MTCs by reversing their sign. In MTCs, this sign reversal in actuator commands leads to large errors and instability as opposing signals accumulate over time~\cite{9247392}.
    \item \emph{False-Data Injection (FDI) attacks:} FDI attacks undermine data integrity in Cyber-Physical Systems (CPS) by injecting deceptive data into data streams. This can mislead control mechanisms into harmful actions by exploiting the trust among sensors, controllers, and actuators, creating discrepancies between perceived and actual states~\cite{TUPTUK201893,9247392,Du9541308}.
    \item \emph{Replay attacks:} In a replay attack, adversaries capture and retransmit valid signals to trick the system without prior system knowledge. They record nominal operational data during normal operation and later replay it. This makes the detector and controller operate on a falsified but statistically consistent measurement stream. Meanwhile, attackers manipulate the actuator to drive the plant’s actual behavior away from correct operation. This masking of the measurement channel is precisely why replay attacks are difficult to detect with conventional integrity- and authenticity-based mechanisms~\cite{9177270}.
\end{itemize}
Replay attacks are particularly challenging because the detector and controller continue to observe a benign-looking measurement stream even while the physical plant evolves under manipulated actuation. By contrast, direct falsification mechanisms such as injection or flip attacks often create immediate inconsistencies in the observed measurements themselves.

\subsubsection{Replay Attack Detection}
A common replay-attack detection method uses watermarking, assumed unknown to adversaries~\cite{9177270,Shahabadi10976679}. It enables prompt detection by disrupting the watermarking during an attack. Physical watermarking is classified into input-added, which alters input signals, and output-added, which alters output signals~\cite{Liu10643207}. This ensures detection of any interference, regardless of whether inputs or outputs are compromised. Mo et al.~\cite{Mo5394956} introduced static watermarking for replay attack detection in controlled systems by embedding authentication signals into control inputs, enhancing detection while analyzing the resulting detection-performance trade-offs. Later research has divided into three complementary streams:
\begin{enumerate}[label=(\roman*)]
    \item \textit{Operator–specified stationary watermarking design:} These frameworks show that combining random Gaussian signals can expose tampering even in noisy or partially observed plants, including single- and multi-input-multi-output extensions with robustness to non-Gaussian disturbances~\cite{Satchidanandan7738534,Satchidanandan8746227}, Linear Time-Invariant (LTI) to linear time-variant extensions~\cite{Porter9187955}, lightweight key-based recursion for unsecured channels~\cite{Song8984338}, and multi-layer industrial deployments that reduce false alarms and localize faults~\cite{Yang9954617}. However, these methods mostly assume LTI–Gaussian models and keep the operator-chosen watermark distribution stationary, leading to a static detectability–performance trade-off.
    \item \textit{Optimization-based offline or online stationary watermark design:} Systematic watermarking strategies balance detection accuracy and control performance using control-theoretic optimization. For LTI systems with full model knowledge, they choose a static watermark covariance offline by minimizing a Linear-Quadratic-Gaussian (LQG) cost or maximizing Kullback–Leibler (KL) divergence, producing static but provably optimal signals for cost or detection metrics \cite{Mo6612700,Mo7011170,Naha9772401,11107908,Xia2025,XIA2026112988}. These approaches, however, assume LTI–Gaussian dynamics and a single offline watermark power, so they cannot adapt to changing plant dynamics and are less effective against replay attacks in time-varying or nonlinear MTCs. For unknown LTI parameters, Liu et al.~\cite{Liu9061046} propose an online identification-and-design loop that updates the watermark covariance during a learning phase and provably converges to the optimal stationary covariance. Yet the steady-state watermark remains stationary and still relies on LTI–Gaussian assumptions, so it cannot provide the dynamic adaptation required for nonlinear or time-varying MTCs.

    \item \textit{Non-stationary, scheduled, and learning-based watermarking design:} 
    Recent works relax the stationary watermarking by changing when or how watermarks are applied: periodic schedules reduce control cost for discontinuous replay attacks~\cite{FANG2020108698} and have been extended to application-level periodic watermarking with compensation for unmanned marine vehicles~\cite{app14188298}; switching multiplicative schemes watermark sensor outputs through hybrid generator-remover filters with time-varying parameters unknown to the adversary~\cite{9157952}; and event-based physical watermarks use innovation-dependent stochastic or deterministic triggering rules to improve the detection--performance trade-off~\cite{Xudong10313050,Xudong10772602}. These methods show the benefit of non-stationary watermarking, but they are mainly based on predesigned schedules, switching and triggering rules, or known linear CPS models, and generally do not provide a detector-driven adaptation mechanism. More recently, RL has been used to regulate watermark timing and strength for replay detection in linear CPS~\cite{ZHU2026132655}. However, this approach is developed under the LTI assumption and regulates watermark activation or scalar strength, rather than learning a general watermark-statistics policy for nonlinear or time-varying MTC operation.
\end{enumerate}

\subsection{Research Gaps \& Contributions}

Despite cybersecurity advancements, detecting replay attacks in MTCs remains challenging. The review of existing literature identifies three unresolved shortcomings that are particularly pronounced in proprietary, closed-architecture MTCs:

\begin{enumerate}[label=\textbf{G\arabic*:},align=left,leftmargin=0pt]
 \item \emph{LTI-Gaussian dependence}. Most watermarking frameworks assume stationary LTI dynamics and \emph{i.i.d.} Gaussian noise. Time-variant dynamics and unmodeled nonlinearities common in modern plant floors and MTCs could undermine detection and performance assurances.

\item \emph{Limited adaptive watermark-statistics design}. Static covariances cannot adapt to operating changes, while non-stationary designs are mainly based on predesigned schedules, switching and triggering rules, or scalar watermark activation and strength adaptation. This leaves a fragile security-performance trade-off when watermark sensitivity varies across nonlinear or time-varying MTC regimes.

\item \emph{Lack of detection uncertainty quantification}. Existing watermark-based replay detectors typically make threshold, schedule, or triggering decisions from residual statistics, but do not maintain a principled online measure of detection confidence for closed-loop watermark adaptation. This limits adaptation, especially under time-varying or nonlinear dynamics where the statistic distribution can drift.

\end{enumerate}

We address research gaps G1-G3 by presenting \emph{DynaMark}, a model-augmented Reinforcement Learning (RL) framework that \emph{learns} and \emph{adapts} watermarking statistics and intensity online without requiring an analytic plant
transition model in the RL policy optimization. Using readily available measurements (e.g., position) and online detection-confidence, DynaMark iteratively refines a watermarking policy that adjusts DWM covariance to enhance replay-attack detection with least energy and control-performance overhead. The key contributions of this paper are as follows.

\begin{enumerate}
\item We formulate the DWM as a Markov Decision Process (MDP) and design an RL-based DWM framework that generates watermarking signals with adaptive intensity for replay attack detection on MTCs. This allows the watermark to adapt to the detector's confidence about the system state and adjust watermark intensity dynamically to achieve a balance between control performance and detection power. 

\item The unique design of the reward function considers control performance, energy consumption, and detection power simultaneously. Together with a bounded action set, this design enables systematic adjustment of the security–performance trade-off while enforcing actuator-safe watermark budgets.

\item To enable DWM under potentially time-varying and nonlinear dynamics, we characterize how replay-induced watermark mismatch alters the detector statistic and develop a Bayesian belief update that quantifies online detection confidence and serves as a compact input to the RL policy. We further establish a closed-loop mean-square boundedness guarantee under mild local conditions and bounded watermark budgets.

\item The effectiveness of the proposed framework is demonstrated on a Digital Emulator (DE) of a closed-architecture MTC, a nonlinear Mass-Spring-Damper (MSD) benchmark, and a real-world physical stepper-motor testbed, where DynaMark achieves faster detection and lower control performance degradation than benchmark watermarking schemes.

\end{enumerate}

The remainder of this paper is organized as follows: Section~\ref{sec:2} presents the MTC system dynamics, watermarking scheme, replay attack model, and detection-power analysis. Section~\ref{sec:3} details the DynaMark methodology, the RL-based algorithm, and stability results. Section~\ref{sec:4} reports experiments on the DE, the MSD, and the physical testbed, including comparisons with static watermarking baselines and existing benchmarks. Finally, it concludes in Section~\ref{sec:5}.

\section{Problem Setup}\label{sec:2}
\begin{figure}[t!]
    \centering
    \includegraphics[width=.85\linewidth]{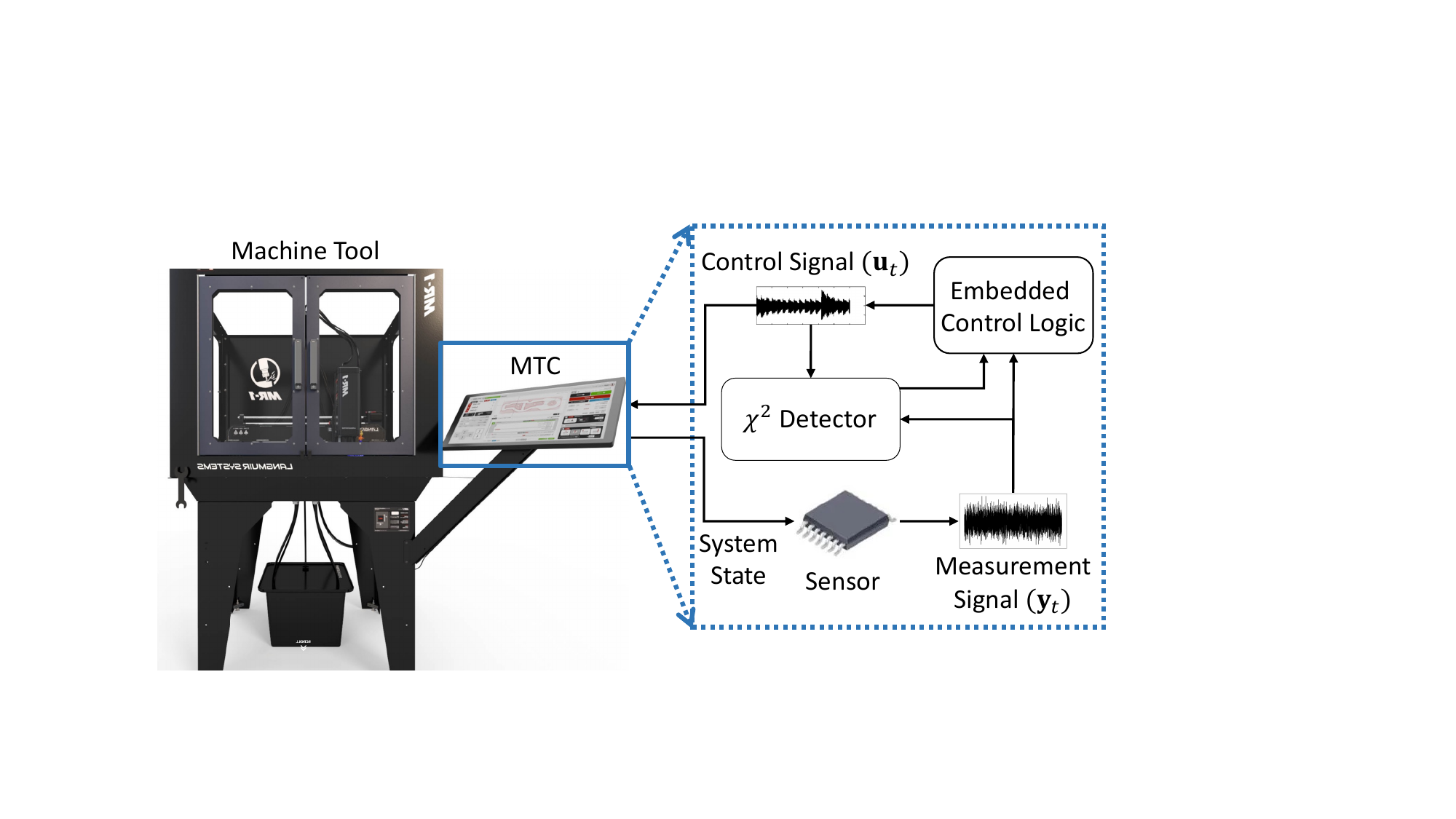}
    \caption{Flowchart of the interaction between machine tools, sensors, controllers, and the detector for real-time monitoring and control.}\label{fig:1}
\end{figure}

In this section, we introduce the MTC system dynamics, the DWM scheme, and the replay attack model. The modeling in this section is used to characterize the detector statistic and belief feedback provided to DynaMark. The RL policy optimization is not supplied with an analytic plant transition model. Fig.~\ref{fig:1} shows the flowchart of the machine tool monitoring and control process, and we explain each block in detail.

\subsection{System Model}\label{sec:3.1}
A sensor network monitors the machine tool, as shown in Fig.~\ref{fig:1}. Let $\vx_t\in\R^{m}$ be the (latent) physical state of the machine tool at time $t$, evolving according to the discrete-time stochastic nonlinear dynamics
\begin{equation}\label{eq:state}
    \vx_{t+1} = \mathcal{F}_t(\vx_t,\vu_t) + \vw_t,
\end{equation}
where $\mathcal{F}_t:\R^{m}\times\R^c\to\R^{m}$ is the state transition dynamics, $\vu_t\in\R^{c}$ is the control input from the control logic, and $\vw_t$ is a zero-mean disturbance capturing unmodeled dynamics and exogenous perturbations. The control logic, actuator, and detector do not directly access $\vx_t$; they only observe the measurement (reported output) $\vy_t\in\R^{n}$ given by
\begin{equation}\label{eq:meas}
    \vy_t = \mathcal{H}(\vx_t) + \vv_t,
\end{equation}
where $\mathcal{H}:\R^m\to\R^n$ is the sensing map and $\vv_t$ is measurement noise. The control logic and detector use only $\vy_t$, which forms the cyber interface vulnerable to adversarial manipulation (see \S~\ref{sec:3.3}). We therefore model the system dynamics in the measurement space using
\begin{equation}\label{eq:sys}
    \vy_{t+1} = \mathcal{G}_t(\vy_t,\vu_t) + \ve_t,
\end{equation}
where $\mathcal{G}_t:\R^{n}\times\R^{c}\to\R^{n}$ is a known (possibly time-varying or nonlinear) transition function, and $\ve_t\in\R^{n}$ aggregates the effects of $\vw_t$, $\vv_t$, and modeling mismatch in measurement coordinates. We assume $\ve_{t}$ is independent zero-mean Gaussian distributed, $\ve_{t}\sim\N(0,Q_t)$, with covariance $Q_t\in\mathbb{S}_{+}^{n}$ symmetric Positive Semidefinite (PSD). Under honest sensing with direct state readout $\vy_t\equiv \vx_t$, we have $\mathcal{H}(\vx_t)=\vx_t$ and $\vv_t\equiv 0$, so Eq.~\eqref{eq:state} implies $\mathcal{G}_t=\mathcal{F}_t$ and $\vw_t=\ve_t$.

A feedback controller computes control actions from observed measurements. We model the control logic as a (possibly nonlinear) function $f_c:\R^{n}\rightarrow\R^{c}$ parameterized by $\eta$, i.e., $\vu_t = f_c(\vy_{\{t\}};\eta)$, where $\vy_{\{t\}}=[\vy_0,\vy_1,\dots,\vy_t]$ is the output history. We assume the detector forms the one-step-ahead prediction
\begin{equation}\label{eq:pred}
    \widehat{\vy}_{t+1} = \mathcal{G}_t(\vy_t,\vu_t),
\end{equation}
and the corresponding residuals
\begin{equation}\label{eq:res}
    \vr_{t+1} = \vy_{t+1} - \widehat{\vy}_{t+1}.
\end{equation}
The detector computes $\{\vr_t\}$ and triggers an alarm using preset confidence levels and statistical metrics (see Fig.~\ref{fig:1} and \S~\ref{sec:3.2}). It then decides whether the control logic continues generating signals (normal operation) or stops (suspected attack).

\begin{remark}
Equation~\eqref{eq:sys} models machine tool dynamics in measurement space and underpins control-side estimation, detection, and belief updating in MTCs where $\vx_t$ is inaccessible and monitoring relies on $\vy_t$. The existence of $\mathcal{G}_t$ is assumed when measurements are informative enough to produce an accurate reduced-order model over the operating region. In practice, $\mathcal{G}_t$ and $Q_t$ may be obtained from nominal data via system identification, local linearization, piecewise modeling, or empirical calibration, and then used by the detector and belief-update modules. The RL policy optimization is not supplied with an analytic transition kernel rather it learns from sampled transitions.

\end{remark}

\subsection{Dynamic Watermarking Scheme}\label{sec:3.2}
The main idea of a physical watermark is to inject a random signal, $\phi_{t}$, which is called the watermark signal, into the system in Eq.~\eqref{eq:sys}. This signal is used to excite the system and check whether the system responds to the watermark signal in accordance with the dynamics of the system. We consider injecting watermarks into the control actions, i.e.,
\begin{equation}\label{eq:wm_scheme}
    \vu^{\phi}_{t} = \vu_{t} + \phi_{t}.
\end{equation}
In practice, watermarks are randomly drawn from a predefined distribution, typically Gaussian. Practitioners design this distribution to preserve controller performance and normal operations while improving detection. In DWM, however, the distribution or its parameters are adjusted based on the system state and detection performance. We reserve ``DWM'' for this adaptive setting. By contrast, offline covariance-design methods that produce a stationary covariance $U$ are referred to as ``static watermarking". In this study, we assume the watermark signals $\{\phi_{t}\}$ are independent zero-mean Gaussian random variables with covariance $U_{t}$, which is time-varying and (in the DynaMark setting) chosen adaptively.

With the DWM defined in Eq.~\eqref{eq:wm_scheme}, the system dynamics are governed by
$\vy_{t} = \mathcal{G}_{t-1}(\vy_{t-1},\vu^\phi_{t-1}) + \ve_{t-1}$, while the corresponding state estimate is obtained as
$\widehat{\vy}_{t} = \mathcal{G}_{t-1}(\vy_{t-1},\vu^\phi_{t-1})$. With residuals defined in Eq.~\eqref{eq:res}, their sequence inherits the same probabilistic attributes as $\ve_{t-1}$, namely $\mathbf{r}_{t}\sim \N(0,Q_{t-1})$. Therefore, the probability of obtaining the measurement $\vy_{t}$ is computed as $f_{\vy_{t}}(y) = (2\pi)^{-n/2} \det(Q_{t-1})^{-1/2} \exp\left(-1/2 g_{t}\right),$
where
\begin{equation}\label{eq:test}
    g_{t} = \vr_{t}^\top  Q_{t-1}^{-1} \vr_{t}.
\end{equation}
It is easy to show that $g_{t}\sim\chi^{2}_{n}$ with $n=|\vy_{t}|$ degrees of freedom~\cite{SAINI2022100617}. When this probability is low, it means the system is likely to be subject to a certain anomaly or attack. Therefore, the test for detecting an attack involves checking $g_t \lessgtr \Tilde{g}$ where $\Tilde{g}$ is an appropriate threshold. If $g_{t}$ exceeds the threshold, the $\chi^2$ detector will trigger an alarm. 

\subsection{Residuals Analysis under Replay Attack}\label{sec:3.3} 
Replay attacks aim to keep the detector statistically consistent with nominal operation by replacing the real-time measurement stream with previously recorded legitimate measurements while manipulating control input. We assume the adversary has access to all sensors and controllers and is capable of $(i)$ intercepting and modifying the control commands $\vu_t^a$ sent to the plant, and $(ii)$ observing and altering all sensor readings (let $\vy_t^{\prime}$ be the modified measurements). The attacker implements a two-stage strategy: $(i)$ record a sufficiently long benign trajectory $\{(\vy_t,\vu_t)\}$ under nominal closed-loop operation (i.e., without attacker intervention), then $(ii)$ replay the recorded measurements while applying a chosen actuator input sequence to the plant. Thus, during the attack, the detector and control logic receive the spoofed sequence $\vy_t^{\prime}$ while the plant receives $\vu_t^a$ via the actuators. It is conventional to interpret $\vy'_t$ as the output of the virtual (time-shifted) system dynamics in Eq.~\eqref{eq:replay_dyn}. This interpretation does not necessarily imply that the attacker runs a virtual system; rather, it serves as a conceptual construct for analysis.
\begin{equation}\label{eq:replay_dyn}
    \vy'_{t+1}=\mathcal{G}_t(\vy'_t,\vu'_t)+\ve'_t, \ \ \ \ \ve'_t\sim\N(0,Q_t),
\end{equation}
We denote the residual under replay by $\vr_{t}^A$, where $\vr_{t}^A:=\vy'_t-\widehat{\vy}'_t$. To characterize $\vr_{t}^A$, we locally linearize $\mathcal{G}_t$ around the operating point $(\widehat{\vy}_t,\Bar{\vu}_t)$, where $\Bar{\vu}_t$ is the control signal when $\widehat{\vy}_t$ is estimated. We assume $\mathcal{G}_t$ is continuously differentiable for all $t$ and define the Jacobians
$G_t:=\nabla_y \mathcal{G}_t(\widehat{\vy}_t,\Bar{\vu}_t)$ and $H_t:=\nabla_u \mathcal{G}_t(\widehat{\vy}_t,\Bar{\vu}_t)$.
For $(\vy,\vu)$ near $(\widehat{\vy}_t,\Bar{\vu}_t)$, a first-order Taylor expansion gives $\mathcal{G}_t(\vy,\vu)\approx \mathcal{G}_t(\widehat{\vy}_t,\Bar{\vu}_t)+G_t(\vy-\widehat{\vy}_t)+H_t(\vu-\Bar{\vu}_t),$ with higher-order terms negligible over the operating region of interest. Then, locally,
\begin{equation}\label{eq:lin_sys}
    \vy_{t+1} \approx G_t \vy_t + H_t \vu_t + \Tilde{\vu}_t + \ve_t,
\end{equation}
where $\Tilde{\vu}_t = \mathcal{G}_t(\widehat{\vy}_t,\Bar{\vu}_t) - G_t \widehat{\vy}_t - H_t \Bar{\vu}_t$.

Without DWM and under replay, $\vu_t$ is the nominal control input computed from $\vy'_t$. Because the attacker wants the fake readings $\vy'_t$ to resemble the normal $\vy_t$, it is natural to pair them with the nominal input sequence $\{\vu'_t\}$ that produced them. Additionally, the adversary chooses a replay segment from the same operating regime as the current trajectory so that, for all $t$ during replay, $(\vy'_t,\vu'_t)$ stays near $(\widehat{\vy}_t,\bar{\vu}_t)$, where the local linearization is accurate and the Jacobians vary negligibly. Thus, under consistent replay, $\vu'_t \approx \vu_t$ and the Jacobians along replay are approximately nominal. The detector then computes $\widehat{\vy}'_{t+1}=\mathcal{G}_t(\vy'_t,\vu_t)$, while the replayed measurement satisfies Eq.~\eqref{eq:replay_dyn}. Using Eq.~\eqref{eq:lin_sys}, the replay residual satisfies
$\vr_{t+1}^A\approx H_t(\vu'_t-\vu_t) + \ve'_t\approx \ve'_t$. Hence, they are statistically indistinguishable from nominal ones, and the replay attack cannot be detected by the test in~\eqref{eq:test} without DWM.

Under DWM in Eq.~\eqref{eq:wm_scheme}, $\vy'_t$ is generated by the nominal controller and includes its own watermark realization $\phi'_t\sim\N(0,U'_t)$. Since the attacker replays past measurements, they generally cannot match the current watermark realization $\phi_t\sim\N(0,U_t)$ in the replayed data. The lemma below describes the residual distribution during the replay attack.

\begin{lemma}\label{lem:replay_residual_nonlinear}
Suppose $\mathcal{G}_t$ is continuously differentiable for all $t$, the Jacobians along replay are approximately the same as nominal, and the first-order linearization remainder is negligible over the operating region. Let the detector form $\widehat{\vy}'_{t+1}=\mathcal{G}_t(\vy'_t,\vu_t^\phi)$, while the replayed measurement satisfies Eq.~\eqref{eq:replay_dyn}. Then, approximately, $\vr_{t+1}^A \approx \ve'_t + H_t(\phi'_t-\phi_t)$. Hence, (approximately) $\vr_{t+1}^A\sim\N(0, \mathcal{S}_{t+1})$ with $\mathcal{S}_{t+1}=Q_t + H_t (U'_t+U_t) H_t^\top.$
\end{lemma}

The proof of Lemma~\ref{lem:replay_residual_nonlinear} is deferred to Appendix~\ref{app:lem1}.

\begin{remark}\label{remark:linear}
    For the LTI system dynamics with honest sensory network, $\vy_{t+1}=A\vy_t+B\vu_t+\ve_t$ with $\ve_t\sim\N(0,Q)$, the Jacobians are constant with $G_t=A$, $H_t=B$, and $\Tilde{\vu}_t=0$. Lemma~\ref{lem:replay_residual_nonlinear} then reduces to $\mathcal{S}_{t+1}=Q + B(U'_t+U_t)B^\top$.
\end{remark}

\begin{theorem}\label{thm:1}
    With DWM in Eq.~\eqref{eq:wm_scheme}, under a replay attack, the test statistics in Eq.~\eqref{eq:test} follow a generalized $\chi^2$ distribution, \(g_{t}^A \sim \Tilde{\chi}(\mathbf{\omega}_{t}, \mathbf{\kappa}, \mathbf{\lambda}_{t}, s, m)\), where \(s=0\), \(m=0\), \(\mathbf{\omega}_{t} = (\Lambda_{t}(1), \dots, \Lambda_{t}(n))^{\top}\), \(\mathbf{\kappa} = \Vec{\pmb{1}}\), and \(\mathbf{\lambda}_{t} = \mathbf{b}_{t}\). Additionally, \(\mathcal{S}_{t}^{1/2} Q_{t-1}^{-1} \mathcal{S}_{t}^{1/2} = P_{t}^\top \Lambda_{t} P_{t}\) and \(\mathbf{b}_{t} = P_{t} \mathcal{S}_{t}^{-1/2} \mathbf{m}_{t}\).
\end{theorem}

The proof of Theorem~\ref{thm:1} is deferred to Appendix~\ref{app:thm1}. Moreover, by Lemma~\ref{lem:replay_residual_nonlinear}, $\mathbf{m}_{t}=\mathbf{0}$ (hence $\mathbf{b}_{t}=\mathbf{0}$), so $g_{t}^A$ is a central generalized $\chi^2$ (a weighted sum of independent $\chi^2$ variables). Theorem~\ref{thm:1} shows that DWM improves replay attack detection on MTCs. However, choosing the right watermarking intensity is vital for balancing control performance and detection accuracy. These insights will aid in characterizing the feedback from the detector and creating an intelligent DWM framework to address this trade-off.

\begin{figure*}[!ht]
    \centering
    \includegraphics[width=.75\linewidth]{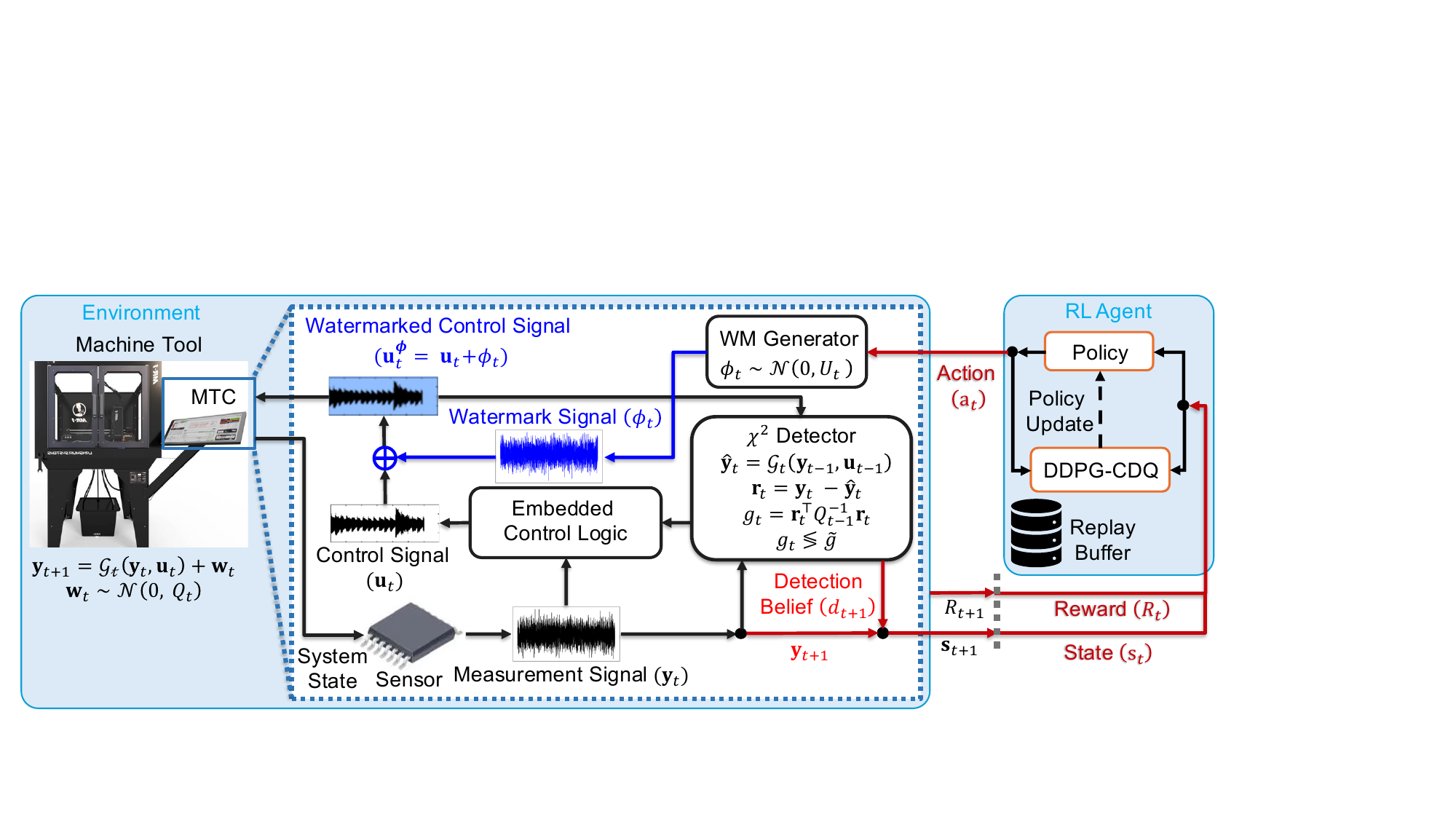}
    \caption{DynaMark framework; blue line shows watermarking watermark-generation and injection, and red lines show agent–environment interaction.}
    \vspace{-3ex}
    \label{fig:fw}
\end{figure*}

\section{DynaMark Methodology}\label{sec:3}
This section outlines the main components of the Markov Decision Process (MDP) model for the DWM problem in MTCs, and the method used to solve it. The DynaMark framework is denoted in Fig.~\ref{fig:fw}. A typical MTC and its sensors form the physical environment. At each sampling interval, the agent observes the system state, including the latest $\vy_t$ and the detector's belief, and chooses $U_t$ as an action. The watermark generator draws a signal from the distribution, integrates it with the control input, and transmits it to the machine tool. The detector assesses the gap between predicted and actual outputs, updates its belief of replay attacks, and provides an updated state and a reward (balancing detection confidence, control performance, and energy consumption). An RL algorithm uses entities in the replay buffer to refine the policy, enabling the watermark to dynamically adapt to changing operational contexts. Crucially, this continuous modulation of the covariance renders the watermark non-stationary, effectively preventing adversaries from learning the static statistical properties required to design replay attacks.

\subsection{Markov Decision Process Formulation}
Formally, an MDP problem~\cite{Puterman528623,sutton2018reinforcement} is defined as a tuple $\mathcal{M}=(\mathcal{S},\mathcal{A},\mathcal{T},\mathcal{R},\rho_{0})$ where $\mathcal{S}$ denotes the \textit{state space}, $\mathcal{A}$ is the \textit{action space}, $\mathcal{T}:\mathcal{S} \times \mathcal{A} \to \mathcal{P}(\mathcal{S})$ represents the \textit{state transition function} that is the probability distribution over states given being in state $s\in\mathcal{S}$ and taking action $a\in\mathcal{A}$; $\mathcal{R}:\mathcal{S}\times\mathcal{A}\times \mathcal{S} \to \R$ is the \textit{reward function} (or expected immediate reward) that is received after transitioning from state $s\in\mathcal{S}$ to state $s^{\prime}\in\mathcal{S}$ by taking action $a\in\mathcal{A}$; and $\rho_{0}:\mathcal{S}\to[0,1]$ is the starting state distribution. Next, we detail each component in the DynaMark context.

\textbf{State Space:} Let $s_{t} \in \mathcal{S}$ be the MDP state at time $t$. We define $s_{t}=(\vy_{t},d_{t})$ where $\vy_{t}$ is the measurement signal, and $d_{t}$ denotes the $\chi^2$ detector's confidence at time $t$. The detection confidence, $d_{t}$, indicates the detector's belief if the system is under attack, using observed $\vy_{t}$. In particular,$d_{t}$ indicates the detector's confidence in an attack when an alarm is triggered. On the other hand, if no alarm sounds, $d_{t}$ illustrates the certainty that there is no attack. The belief is updated sequentially based on the results of the statistical tests. To formalize $d_t$, define the random variables as follows: let $\sigma=1$ represent the occurrence of an actual ongoing replay attack, with a value of 0 otherwise; \(I_{t}=1\) if the detector raises an alarm and 0 otherwise. Subsequently, \(d_t := \Pr(\sigma=1|I_{1:t})\).

\textbf{Action Space:} Using DWM requires careful design of the covariance of the watermarking distribution $\N(0,U_t)$ since the controller performance degrades due to the need for high accuracy in cyberattack detection. Specifically, if $U_t$ produces $\phi_t$ with high intensity, the control performance deteriorates, although this results in a marked improvement in the detection accuracy. Conversely, if $U_t$ yields $\phi_t$ with low intensity, detection accuracy decreases, but the controller's performance remains intact. Another advantage of the RL-based DWM is that, unlike static approaches, the RL agent adaptively modulates \(U_t\) over time, creating a non-stationary profile that significantly increases the complexity of the adversarial learning problem. We define the MDP action space $\mathcal{A}$ as the set of bounded $c\times c$ symmetric PSD matrices, specifically,
\begin{equation}\label{eq:action_space}
    \mathcal{A} = \{U \in \mathbb{S}_+^c : \|U\|_F \le U_{\max} \}.
\end{equation}
Here, $U_{\max}$ is a design parameter reflecting the physical saturation limits of the MTC actuators and a stability budget. In \S~\ref{sec:stability}, we show that, under a locally contractive nominal closed loop, bounding $\E\left[\|\phi\|^2\right]$ via the defined action set in~\eqref{eq:action_space} yields a uniform mean-square bound on the tracking error. Then, observing $s_t\in\mathcal{S}$, the action at time $t$, $a_t\in \mathcal{A}$, is the covariance matrix of the DWM distribution ($U_t$).

\textbf{State Transition Function:} At time $t$, observing $s_t=(\vy_t,d_t)$ and taking action $a_t:=U_t$, a watermark signal $\boldsymbol{\phi}_t\sim\mathcal{N}(0,U_t)$ is added to the control input as in Eq.~\eqref{eq:wm_scheme}. The plant or simulator then generates $\vy_{t+1}$, and the detector updates $d_{t+1}$ from the alarm output. The RL agent thus observes sampled transitions $(s_t,a_t,r_t,s_{t+1})$ without access to an analytic MDP transition kernel.

\begin{theorem}\label{thm:2}
    Let \(\tau\in\R_{>0}\) denotes the replay attack onset random variable with CDF $\mathcal{Q}_\tau(t)=p(\tau \leq t \mid \sigma = 1)$. The detection confidence, \(d_t\), evolves using Bayesian rule,
    \begin{equation}\label{eq:dt}
        d_t = \frac{d_{t-1} \kappa_{1,t}}{d_{t-1}\kappa_{1,t} + (1 - d_{t-1}) \kappa_{0,t}},
    \end{equation}
    where $\kappa_{0,t}= \alpha^{I_{t}} (1 - \alpha)^{1-I_{t}}$ and $\kappa_{1,t} = \alpha^{I_{t}} (1 - \alpha)^{1-I_{t}} \ (1 - \mathcal{Q}_\tau(t)) + (1 - \beta_{t})^{I_{t}} \beta_{t}^{1-I_{t}} \ \mathcal{Q}_\tau(t)$. 
\end{theorem}

The proof of Theorem~\ref{thm:2} appears in Appendix~\ref{app:thm2}. Equation~\eqref{eq:dt} defines the recursive detector-belief update for the next state component $d_t$. Under the nominal model in \S~\ref{sec:2}, this update is implemented by propagating the detector statistic through Eqs.~\eqref{eq:wm_scheme},~\eqref{eq:sys},~\eqref{eq:pred},~\eqref{eq:res},~\eqref{eq:test}, and~\eqref{eq:dt}. This model-based characterization enables detector feedback but is not supplied as an analytic transition model for RL optimization.

\begin{lemma}\label{lemma_beta}
    Let \(\tau\in\R_{>0}\) denotes the replay attack onset random variable with CDF $\mathcal{Q}_\tau(t)=p(\tau \leq t \mid \sigma = 1)$. For the system described by Eq.~\eqref{eq:sys} with Gaussian noise, and its $\chi^2$ detector in Eq.~\eqref{eq:test}, the Type-II error is 
    \begin{equation}\label{eq:beta}
        \beta_t = \mathcal{H}_{g_t^A}(\Tilde{g}) \ \mathcal{Q}_\tau(t) + \mathcal{H}_{g_t}(\Tilde{g}) \ (1 - \mathcal{Q}_\tau(t)),
    \end{equation}
    where $\mathcal{H}_{g_t^A}$ denotes the CDF of the (central) generalized $\chi^2$ distribution, and $\mathcal{H}_{g_t}$ is the CDF of the $\chi^2$ distribution. 
\end{lemma}

The proof of Lemma~\ref{lemma_beta} is in Appendix~\ref{app:lem2}. Equation~\eqref{eq:beta} gives a model-based Type-II error when the assumptions in \S~\ref{sec:2} hold. Procedure~\ref{proc:online_belief_update} summarizes this computation: given $I_t$, it uses the current and replayed watermark covariances to form $\mathcal{S}_t$, estimate $\mathcal{H}_{g_t^A}(\widetilde{g})$, compute $\beta_t$, and update $d_t$. The RL policy observes only the resulting belief $d_t$, not the closed-form model used to compute it.

\begin{algorithm}[t]
\caption{Online belief update under nominal model in \S~\ref{sec:2}}
\label{proc:online_belief_update}
\begin{algorithmic}[1]
\Require Detector: $I_t\in\{0,1\}$, $\widetilde{g}$, $d_{t-1}$, $\alpha$; onset prior $\mathcal{Q}_\tau$; System: $Q_{t-1}$, $H_t$, $U_{t-1}$; history $\mathcal{U}_{\{t-1\}}$, replay delay $\Delta$, number of Monte-Carlo samples $M$
\Ensure Updated belief $d_{t}$

\State Set $U'_{t-1} \leftarrow U_{t-\Delta-1}$ from $\mathcal{U}_{\{t-1\}}$
\State Form $\mathcal{S}_t \leftarrow Q_{t-1} + H_{t-1}(U_{t-1}+U'_{t-1})H_{t-1}^\top$
\For{$m=1,\ldots,M$}\Comment{{\small estimate $\mathcal{H}_{g_t^A}(\Tilde{g})$ by Monte-Carlo}}
    \State Draw $\vr^{(m)}\sim \mathcal{N}(0,\mathcal{S}_t)$.
    \State Compute $g^{A,(m)}_t \leftarrow
        \left(\vr^{(m)}\right)^\top Q_t^{-1} \vr^{(m)}$
\EndFor
\State Set $\mathcal{H}_{g_t^A}(\Tilde{g})
    \leftarrow
    \frac{1}{M}\sum_{m=1}^{M}
    \mathbf{1}\!\left\{g^{A,(m)}_t < \widetilde{g}\right\}$

\State Compute onset probability $\mathcal{Q}_{\tau}(t)$

\State Compute $\beta_t$ using Eq.~\eqref{eq:beta}
\State Compute $\kappa_{0,t}, \kappa_{1,t}$ in Theorem~\ref{thm:2}

\State Update $d_t$ using Eq.~\eqref{eq:dt}

\State \Return $d_t$.
\end{algorithmic}
\end{algorithm}

\begin{remark}\label{remark:beta_w}
The Gaussian closed-form belief characterization using Lemma~\ref{lemma_beta} is local and relies on the operating-regime linearization in Lemma~\ref{lem:replay_residual_nonlinear} and Gaussian residual assumptions. When these assumptions are not appropriate, $\beta_t$ can be estimated empirically by Monte-Carlo evaluation of the replay statistic using simulator rollouts or block bootstrap from empirical residuals, while preserving the same belief-update interface in Eq.~\eqref{eq:dt} to the RL agent; see Appendix~\ref{app:beta_calibration}.
\end{remark}

\textbf{Reward Function:} Recall the controller performance degrades due to the need for high accuracy in a replay attack detection that requires a careful choice of the DWM covariance. The goal of adaptively deciding on $U_t$ is to establish a trade-off between detection accuracy and control performance while reacting to the changes in the environment. Let $r_t(s,a)$ denote the random reward of being in state $s$ and taking action $a$ at time $t$ that is defined as
\begin{equation}\label{eq:reward}
    r_t(s,a) = -\omega_{1} 
        \| \phi_{t} \|_{1}
        - \omega_{2} \| \vy_{t+1}^{\star} - \vy_{t+1} \|_{2} + \omega_{3} |0.5 - d_{t+1}|.
\end{equation}

This reward is a heuristic shaping objective to balance energy overhead, control performance, and detection confidence; convergence or optimality are not implied by the reward design alone, and are instead addressed via numerical validation later. The term $\| \phi_{t} \|_{1}$ is used as an actuation-overhead regularizer that encourages sparse watermark excitations (occasional stronger excitation rather than persistent small excitation). This choice reduces wear or overhead while preserving the ability to increase excitation when detection confidence is low. The term $\| \vy_{t+1}^{\star} - \vy_{t+1} \|_{2}$ penalizes deviations from the un-watermarked trajectory, maintaining control performance. This deviation term directly shapes the learning objective to reduce the error metric that appears in the stability bound analyzed in~\S~\ref{sec:stability}. The term \(|0.5 - d_{t+1}|\) increases detection confidence by pushing $d_{t+1}$ toward either 0 or 1, avoiding uncertain detection outcomes. A confidence value close to 0.5 indicates high uncertainty, which weakens detection effectiveness.

\begin{remark}
The term \(\vy_{t+1}^{\star}\) denotes the reference no-watermark measurement at time \(t+1\), obtained under the same operating condition and baseline controller (with \(\phi_t=0\)). This reference can be generated from nominal no-watermark data, a calibrated simulator, or a parallel shadow model.
\end{remark}

While the reward encourages stable operation by penalizing trajectory deviation and watermark magnitude, stability is formally addressed in~\S~\ref{sec:stability}. There, the stability theorem shows that if the nominal closed loop is locally contractive and the DWM covariance is bounded, then the expected tracking error remains uniformly bounded. Thus, the reward acts as a performance-shaping mechanism, while boundedness follows from the closed-loop property and the action constraint.

\subsection{Closed-Loop System Stability under DynaMark}\label{sec:stability}
The watermark generated by DynaMark perturbs the commanded control signal by Eq.~\eqref{eq:wm_scheme}, where $\phi_t\sim\mathcal{N}(0,U_t)$ and $U_t\in\mathcal{A}$ is selected by the RL policy. This subsection establishes a local mean-square boundedness result for the deviation between the watermarked trajectory and the nominal (un-watermarked) trajectory under nominal model in~\S\ref{sec:2}.

Let the tracking error be $\xi_t := \vy_t-\vy_t^\star$. The following assumptions capture MTCs' standard operating conditions around a nominal regime and formalize the notion that watermarking stays within actuator limits.

\begin{enumerate}
\item \textit{Local smoothness of plant dynamics.} The closed-loop dynamics $\mathcal{G}_t$ is continuously differentiable for all $t$ in a neighborhood of the nominal operating regime. Consequently, the Jacobian of $\mathcal{G}_t$ with respect to the control input is uniformly bounded along the nominal trajectory: there exists $\bar H>0$ such that $\sup_{t\ge 0}\|H_t^\star\|^2\le \bar H$, where $H_t^\star:=\nabla_u \mathcal{G}_t(\vy_t^\star,\vu_t^\star)$. Physically, $\bar H$ quantifies the worst-case sensitivity of the next measurement to small perturbations in the commanded input within the intended operating region.

\item \textit{History-dependent controller with bounded local gains.} The MTC control logic $\vu_t=f_c(\vh_t;\eta)$ uses a finite measurement history $\vh_t=[\vy_t,\vy_{t-1},\ldots,\vy_{t-w}]^\top$ and is differentiable on a compact neighborhood of the nominal histories. Hence, its history Jacobians are uniformly bounded: there exist constants $\{\bar K_i\}_{i=0}^w$ such that $\sup_{t\ge 0}\|K_{t,i}\|^2\le \bar K_i$ for all $i$, where $K_{t,i}:=\nabla_{\vy_{t-i}} f_c(\vh_t^\star;\eta)$. This assumption captures that, in normal operation, the controller does not exhibit unbounded amplification of measurement perturbations across the finite memory window.

\item \textit{Finite-variance unmodeled disturbances.} The aggregate disturbance term (including differences between nominal and actual noise as well as unmodeled effects) has bounded second moment: $\E\|\varrho_t\|^2\le \sigma_\varrho^2$ for all $t$. This assumes process noise, sensor noise, and modeling mismatch stay within finite bounds during normal operation.

\item \textit{Local contractivity of the nominal closed loop.} Define $A_t := G_t^\star + H_t^\star K_{t,0}$, where $G_t^\star:=\nabla_y\mathcal{G}_t(\vy_t^\star,\vu_t^\star)$ and $K_{t,0}$ is the instantaneous history gain. Assume the nominal closed loop is locally contractive in the sense that $\bar A:=\sup_{t\ge0}\|A_t\|<1$. This condition formalizes that, around the operating trajectory, the baseline feedback loop attenuates small deviations rather than amplifying them.
\end{enumerate}

\begin{theorem}\label{thm:stability}
Consider the closed-loop system in Eq.~\eqref{eq:sys} with the history-dependent controller and DWM in Eq.~\eqref{eq:wm_scheme} operated under DynaMark with the action space in~\eqref{eq:action_space}. Under Assumptions~($i$)--($iv$), suppose $(C_1):\Bar{A} < \sqrt{1 - \left(\frac{3 (w+2)^{w+2}\Bar{H}\Bar{K}}{w^{w-2}}\right)^{\frac{1}{w+2}}}$ and $(C_2):\Bar{H}\Bar{K} < \frac{w^{w-2}}{3(w+2)^{w+2}}$ where $\Bar{K}=\max_{i=1,\dots,w} \Bar{K}_{i}$. Then, there exist design parameters $\epsilon>0$ and $\rho\in(0,1)$ such that
\[
\nu(\epsilon,\rho) := (1+\epsilon)\Bar{A}^2 + \rho + 3(1+\epsilon^{-1})w^2\rho^{-w}\Bar{H}\Bar{K} \in (0,1),
\]
and, with $V_t := \sum_{i=0}^{w}\rho^i\| \vy_{t-i} - \vy^\star_{t-i}\|^2$, we have $\E[V_{t+1}] \le \nu(\epsilon,\rho)\E[V_t] + \vartheta(\epsilon,U_{\max})$, where $\vartheta(\epsilon,U_{\max}) := 3(1+\epsilon^{-1}) \left(\Bar{H}U_{\max}\sqrt{c} +\sigma_\varrho^2\right)$. Consequently, for all $t\ge 0$, $\E\|\vy_{t} - \vy^\star_{t}\|^2 \le \nu(\epsilon,\rho)^t V_0 + \left[\vartheta(\epsilon,U_{\max})/\left(1-\nu(\epsilon,\rho)\right)\right]$, and, in particular,
\begin{equation}\label{eq:stability_proof}
\limsup_{t\to\infty}\E\|\vy_{t} - \vy^\star_{t}\|^2 \le \frac{\vartheta(\epsilon,U_{\max})}{1-\nu(\epsilon,\rho)}.
\end{equation}
\end{theorem}

The proof of Theorem~\ref{thm:stability} is deferred to Appendix~\ref{app:thm3}. 

\begin{remark}\label{rem:feasibility}
Theorem~\ref{thm:stability} shows that, in the nominal regime where the baseline MTC closed loop is locally contractive, adaptive DWM with bounded covariance preserves mean-square bounded tracking performance. The necessary and sufficient conditions in Theorem~\ref{thm:stability} have a transparent interpretation: the condition $(C_2)$ limits the effective loop gain accumulated over the $w$-step history, while the condition $(C_1)$ enforces a local contraction margin; together they ensure that suitable design parameters $(\epsilon,\rho)$ exist such that $\nu(\epsilon,\rho)<1$ and the recursion holds. Moreover, the steady-state bound~\eqref{eq:stability_proof} separates (i) a disturbance-driven floor through $\sigma_\varrho^2$ and (ii) a watermark-induced contribution that scales with watermark energy (e.g., $\E\|\phi_t\|^2=\tr(U_t)$). Consequently, $U_{\max}$ acts as a stability and performance budget: increasing $U_{\max}$ enlarges the admissible set of non-stationary watermark covariances and can improve replay detectability, but it increases the bound in~\eqref{eq:stability_proof} proportionally through the watermark-induced term. This clarifies the role of the reward in~\eqref{eq:reward}: it shapes the detection-overhead-tracking trade-off within the stability envelope enforced by the action constraint $U_t\in\mathcal{A}$.
\end{remark}

\begin{remark}
Theorem~\ref{thm:stability} is not intended to establish global stabilization of an unstable system. Instead, it establishes a local boundedness guarantee by demonstrating that, provided the nominal closed-loop system is already locally contractive around the intended operating regime, the bounded DWM signals selected by DynaMark preserve mean-square boundedness. If the nominal closed-loop is unstable, the system should first be stabilized or retuned before applying DynaMark for replay-attack detection. Thus, the theorem should be interpreted as a safety-envelope result for watermark adaptation, not as a replacement for nominal controller design.
\end{remark}

\subsection{Policy Optimization Method}\label{sec:policy_opt}
Let $\pi_\theta:\mathcal{S}\rightarrow\mathcal{A}$ denote a parameterized (deterministic) DWM policy with parameters $\theta$, where the action $a_t=\pi_\theta(s_t)$ specifies the covariance $U_t\in\mathcal{A}$ of the watermarking distribution. For a given policy $\pi_\theta$, define the discounted return from time $t$ as $G_t := \sum_{k=t}^{\infty}\gamma^{k-t} r_k(s,a)$, $\gamma\in(0,1)$, and, the performance objective as the expected return under the trajectory distribution induced by $\pi_\theta$ and the environment dynamics, $J(\theta) := \E_{\pi_\theta}\left[G_0 \big| a=\pi_\theta(s)\in\mathcal{A},s_0\sim\rho_0\right]$. The optimal policy is any $\pi_{\theta^\star}$ satisfying $\theta^\star\in\arg\max_\theta J(\theta)$.

Since the action is continuous and the environment is stochastic due to plant disturbances and measurement noise, we adopt a deterministic actor--critic approach that learns both (i) an actor $\pi_\theta$ and (ii) two critic networks $Q_{\psi^1},Q_{\psi^2}:\mathcal{S}\times\mathcal{A}\rightarrow\mathbb{R}$ approximating the action-value function $Q^{\pi_\theta}(s,a)
:= \E_{\pi_\theta}\left[G_t \big| s_t=s,\ a_t=a\right].$ We use a DDPG-based policy optimization method~\cite{Lillicrap2015ContinuousLearningb} with a clipped double-Q critic target inspired by TD3~\cite{fujimoto2018addressing}. We refer to this implementation as DDPG-CDQ. The method keeps the deterministic structure of DDPG, while using two critics and a conservative minimum target to reduce critic overestimation. The actor is optimized using the deterministic policy gradient $\nabla_\theta J(\theta)
=
\mathbb{E}_{s\sim d^{\pi_\theta}}
\left[
\nabla_\theta \pi_\theta(s)
\nabla_a Q_{\psi^1}(s,a)\big|_{a=\pi_\theta(s)}
\right],$ where $d^{\pi_\theta}$ is the discounted state-visitation distribution induced by $\pi_\theta$.

The choice of DDPG-CDQ is an implementation choice rather than an algorithmic contribution; other continuous-control RL algorithms, such as PPO, SAC, or full TD3, could be used to solve the same DynaMark MDP problem. We use a DDPG-based deterministic actor because the action is a continuous watermark covariance, and the actor can naturally output a factorized representation of this covariance.

Due to the definition of action space in~\eqref{eq:action_space}, we reparameterize $U_t$ through a learned matrix factor to enforce PSD by construction. Specifically, the actor outputs a lower-triangular matrix $L_t=L_\theta(s_t)$ and we set $U_t = \Pi_{\mathcal{A}}\left(L_t L_t^\top\right)$, where $\Pi_{\mathcal{A}}$ denotes a projection onto $\mathcal{A}$, ensuring both $U_t\in\mathbb{S}_+^c$ and $\|U_t\|_F\le U_{\max}$. This parameterization yields a smooth mapping from policy parameters to feasible watermark covariances while respecting the actuator-safe watermark budget.

Given a transition $(s_t,a_t,r_t,s_{t+1})$ sampled from the replay buffer, the DDPG-CDQ target is $y_t
:= r_t + \gamma
\min_{i\in\{1,2\}}
Q_{\psi^i_{\mathrm{tgt}}}
\left(s_{t+1},\pi_{\theta_{\mathrm{tgt}}}(s_{t+1})\right).$
The critic parameters are updated by minimizing the Bellman error $\mathcal{L}(\psi^i)
=
\mathbb{E}
\left[
\left(
Q_{\psi^i}(s_t,a_t)-y_t
\right)^2
\right]$, for $i\in\{1,2\}$.
The actor is updated by ascending an estimate of $J(\theta)$ through the critic gradient above. Target networks $(\theta_{\mathrm{tgt}},\psi^1_{\mathrm{tgt}},\psi^2_{\mathrm{tgt}})$ are updated via Polyak averaging, which mitigates oscillations induced by bootstrapping and function approximation. We emphasize that DDPG-CDQ is not a full TD3 implementation. It adopts the clipped double-Q target, but does not use target-policy smoothing or delayed actor updates. As in DDPG, the exploration is introduced by an Ornstein-Uhlenbeck (OU) process~\cite{PhysRev.36.823} prior to projection onto $\mathcal{A}$, producing a temporally correlated exploration process while preserving feasibility, i.e., $U_t
= \Pi_{\mathcal{A}}\left( (L_t + \Delta_t)(L_t+\Delta_t)^\top \right)$, where $\Delta_t$ is a OU process.

\section{Evaluation Results}\label{sec:4}
We evaluate DynaMark across three representative settings to assess detection effectiveness and the security–performance trade-off under linear, nonlinear, and hardware-in-the-loop conditions. Implementation details, training hyperparameters, and code to reproduce the experiments are available at: [\hyperlink{https://github.com/navidaftabi/DynaMark}{https://github.com/navidaftabi/DynaMark}]. Additional DDPG training details are provided in Appendix~\ref{app:ddpg}.

\subsection{Baseline Watermarking Policies}\label{sec:baselines}
We compare DynaMark against three classes of watermarking baselines to isolate the value of learning an adaptive covariance policy from detector feedback.

\begin{itemize}[leftmargin=10pt]
    \item \textit{Static-covariance watermarking.}
    This baseline uses a fixed watermark covariance throughout the experiment, independent of the measurement state or detector belief. For case studies~1 and~2, the low- and high-intensity constants are selected from empirical lower and upper quantiles of the DynaMark adaptive covariance; for case~2, we also include the empirical mean covariance as an intermediate static operating point.

    \item \textit{Non-RL belief-adaptive watermarking.}
    This baseline uses the same detector belief as DynaMark but replaces RL with the deterministic rule $U_t = U_{\min} + (U_{\max}-U_{\min})d_t$. It tests whether directly increasing watermark intensity with attack belief suffices, without optimizing a long-term reward over energy, control performance, and detection confidence.

    \item \textit{LTI-based offline and online watermarking.}
    For the physical stepper-motor case, we also compare against the offline covariance design and online watermark update method in~\cite{Liu9061046}. These baselines rely on an LTI surrogate of the motor dynamics and provide a comparison with model-based watermarking designs under mode-switching, time-varying hardware behavior.
\end{itemize}

\subsection{Evaluation Metrics}
We use the following metrics to evaluate nominal-operation cost and replay-detection effectiveness. When computing a norm over vector quantities with components in different units, each component is first normalized by its reference unit so the resulting metric is dimensionless.

\begin{itemize}[leftmargin=10pt]
    \item \textit{Watermark energy.}
    The watermarking overhead, $\|\phi_t\|_1$, measures the instantaneous magnitude of the injected watermark signal; lower values indicate less actuation effort and lower watermarking cost.

    \item \textit{Control-performance degradation.}
    The nominal performance loss, $\|\vy_{t}^{\mathrm{\star}}-\vy_{t}\|_2$, measures the instantaneous trajectory deviation; lower values indicate smaller impact on nominal control behavior.

    \item \textit{Cumulative control-performance degradation (CPD).}
    To summarize nominal performance, we report
    $\mathrm{CPD}=\frac{1}{N}\sum_{i=1}^{N}\sum_t \|\mathbf{y}_{t}^{\mathrm{\star},(i)}-\mathbf{y}_{t}^{(i)}\|_2$.
    This metric summarizes the total nominal performance cost over the full evaluation horizon.

    \item \textit{False-alarm control ($\mathrm{ARL}_0$).}
    We set $\tilde{g}=\mathcal{H}_{g_t}^{-1}(1-\alpha)$, giving nominal false-alarm probability $\alpha$ and the average run-length $\mathrm{ARL}_0\approx 1/\alpha$. When the empirical statistic does not follow the theoretical distribution, we set $\tilde{g}$ using the empirical $(1-\alpha)$-quantile of nominal $g_t$.

    \item \textit{Detection delay ($\mathrm{ARL}_1$).}
    Under attack, detection performance is measured by
    $\mathrm{ARL}_1=\frac{1}{N}\sum_{i=1}^{N}(T_{d,i}-\tau)$,
    where $T_{d,i}=\min\{t\geq \tau:I_t^{(i)}=1\}$ is the first alarm time in replication $i$. Smaller $\mathrm{ARL}_1$ indicates faster detection.

    \item \textit{Cumulative post-onset detection uncertainty (CDU).}
    To summarize post-onset belief evolution, we report
    $\mathrm{CDU}=\frac{1}{N}\sum_{i=1}^{N}\sum_{t\geq\tau}(1-d_t^{(i)})$. Lower values indicate that the belief moves toward the attack state more quickly after replay begins.

    \item \textit{Post-onset alarm frequency (AF).}
    We report the fraction of post-onset time steps that trigger alarms,
    $\mathrm{AF}=\frac{1}{N}\sum_{i=1}^{N}\frac{1}{T-\tau+1}\sum_{t=\tau}^{T}\mathbb{I}\{I_t^{(i)}=1\}$.
    Higher values indicate more persistent replay detection after attack onset.
\end{itemize}

\subsection{Case Study 1: DE of the Siemens Sinumerik 828D MTC}\label{sec:numerical}
\subsubsection{Experiment Setup}
\begin{figure}[!ht]
    \centering
    \includegraphics[width=0.7\linewidth]{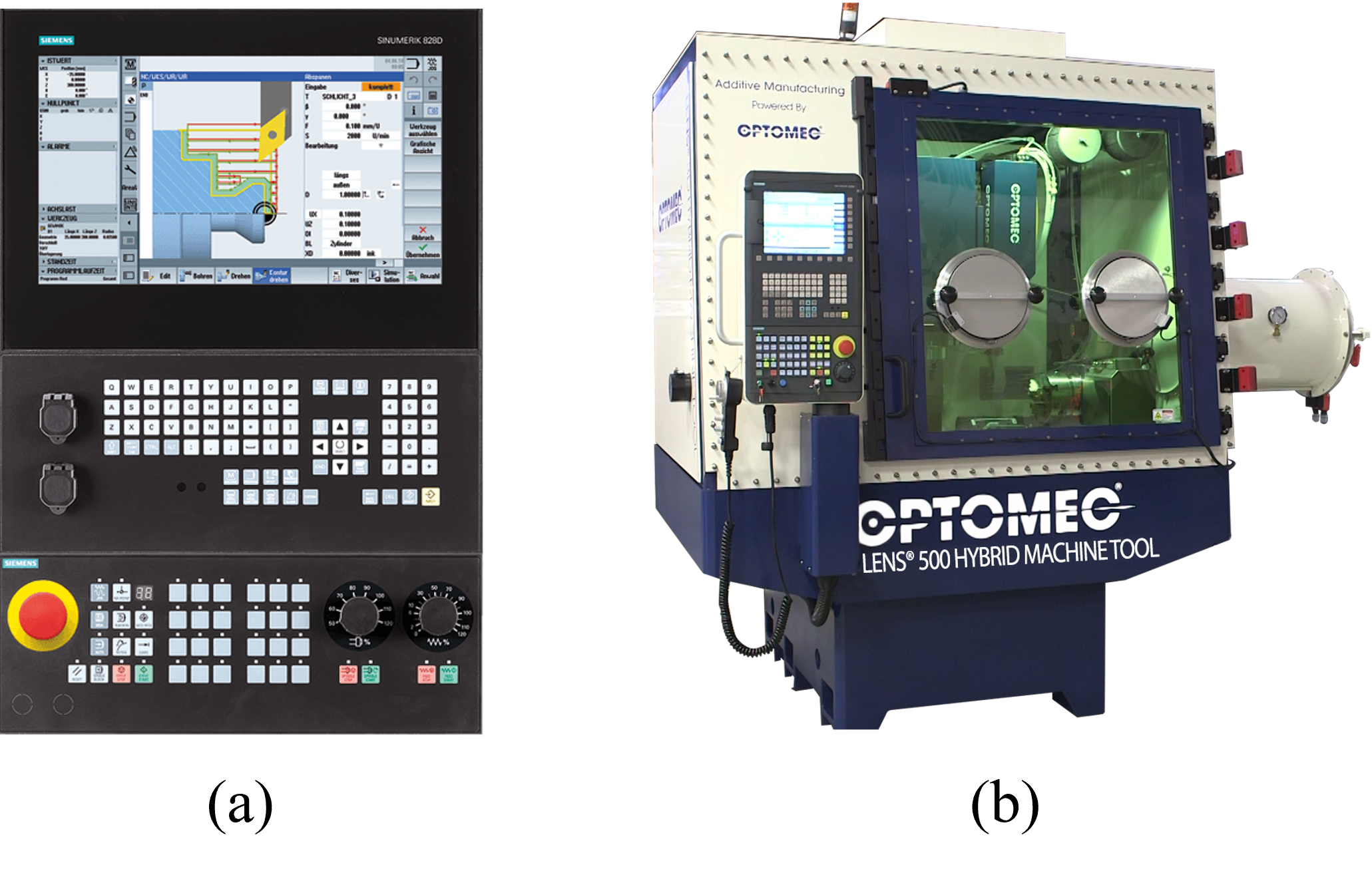}
    \caption{(a) Siemens Sinumerik 828D controller, (b) Optomec LENS\textregistered MTS 500 hybrid machine tool.}
    \label{fig:3}
\end{figure}

The Siemens Sinumerik 828D MTC (see Fig.~\ref{fig:3}) connects operators and machinery, handling tool changes and processes~\cite{TIWARI2023695}. This DE replicates the MTC's 2-axis motion control with high precision by tuning parameterized transfer functions based on real controller data. In our experiments, the motion was analyzed along the $y-$axis. The DE uses a linear dynamic model on the measurement space (see Remark~\ref{remark:linear}) with $A=1.0$, $B=0.010$, and $Q=1.3741\times10^{-13}$. The baseline controller is proportional, $\vu_t = K_p(\bar \vy - \vy_t)$, with $K_p=1.0$ and $\bar \vy=0.012$. Each rollout has horizon $T=1200$. The replay-attack prior is $\sigma\sim\mathrm{Ber}(0.05)$, with onset $\tau\mid\sigma=1\sim\mathrm{Geom}(1/T)$, and the attacker uses full-history replay. The reward weights are set as $\omega_1=\omega_2=0.35$ and $\omega_3=0.3$. Given the data, Theorem~\ref{thm:stability} applies to this case study with $\Bar{A}=0.99$, $\Bar{H}=10^{-4}$, and $\Bar{K}=0$. By selecting $\epsilon=0.01$ and any $\rho<0.00995$, we obtain $\nu(\epsilon,\rho)<1$, following the verification procedure detailed in Appendix~\ref{app:thm3_verification}.

We evaluate DynaMark over 40 replications with $\alpha=0.005$, corresponding to the target nominal false-alarm level used for this case study. The attack scenario involves the attacker collecting full-system measurements, launching a replay attack \(\tau=600\) by replacing true sensor data with recorded data $\vy_t^{'}$, and manipulating the controller action to the machine as $\vu_t^A = -\vu_t$ by the end of the horizon. The control logic $\vu'_t$ uses $\vy'_t$, and the detector monitors $(\vy'_t,\vu'_t)$, while the machine tool receives $\vu_t^A$. This attack cannot be detected without DWM.

\subsubsection{Performance Evaluation}
\begin{figure*}[!ht]
    \centering
    \subfloat[]{\includegraphics[width=0.25\linewidth]{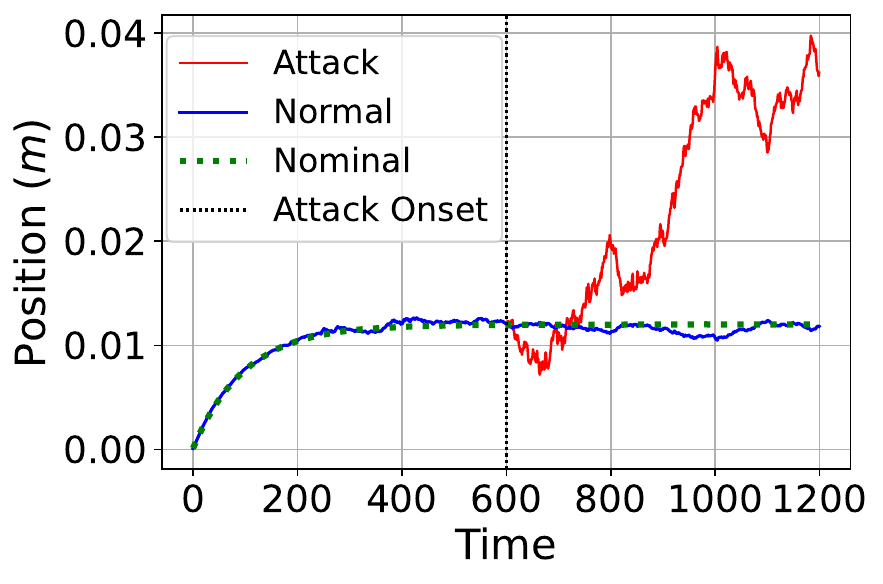}\label{fig:4a}}
    \subfloat[]{\includegraphics[width=0.25\linewidth]{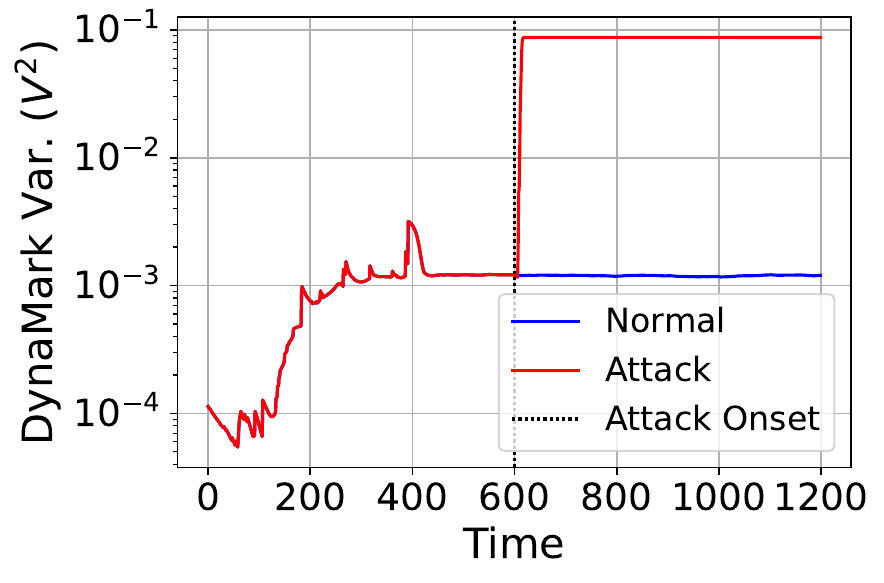}\label{fig:4b}}
    \subfloat[]{\includegraphics[width=0.25\linewidth]{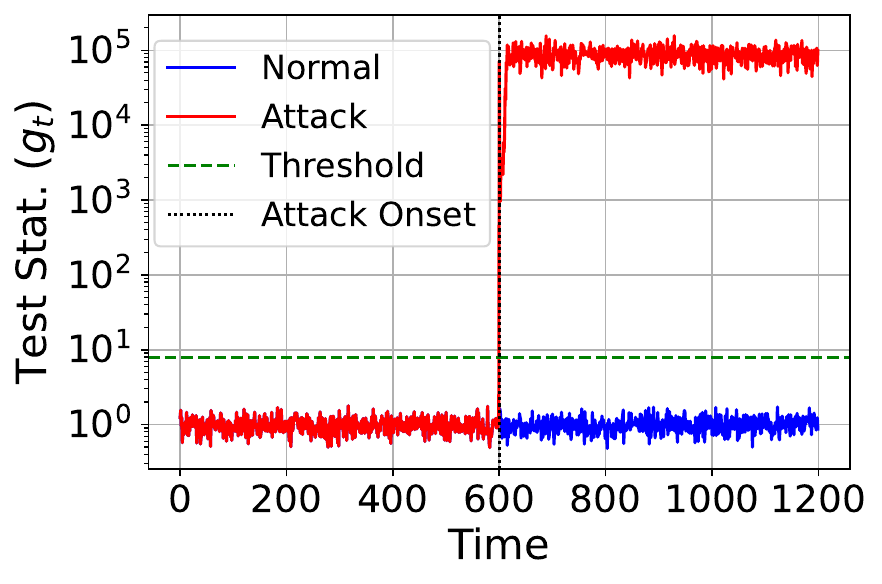}\label{fig:4c}}
    \subfloat[]{\includegraphics[width=0.25\linewidth]{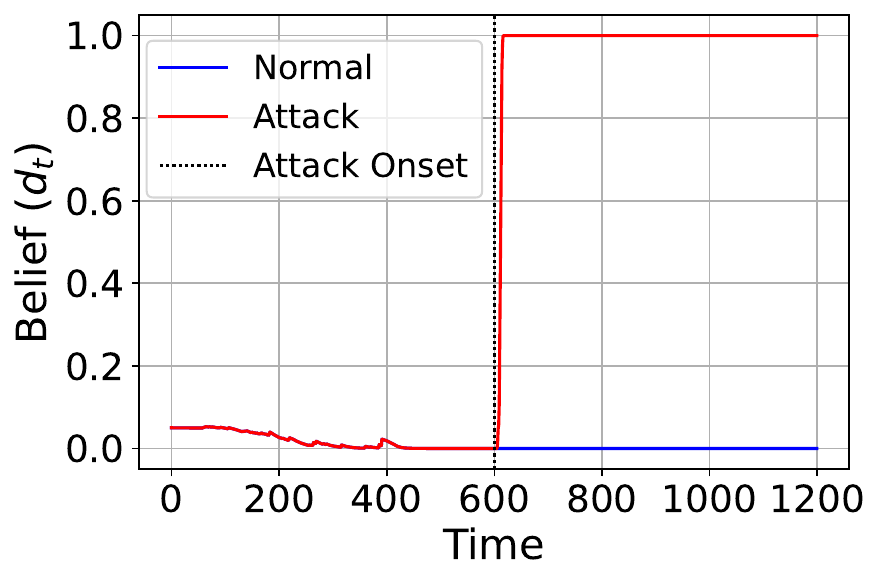}\label{fig:4d}}
    \caption{Average performance of DynaMark on the DE under nominal operation and replay attack ($\tau=600$): (a) position, (b) DynaMark's learned policy, (c) test statistic, and (d) detector belief.}
    \label{fig:4}
\end{figure*}

Figure~\ref{fig:4} summarizes DynaMark’s average behavior on the DE under nominal operation and replay attack. Under nominal operation, DynaMark policy remains low, the detector statistic stays below the threshold, and the trajectory closely follows the no-watermark baseline. After attack onset, DynaMark increases the variance, producing a residual shift that drives the detector statistic above the threshold and raises the attack belief to at least $0.95$ within 15 time steps. These results confirm DynaMark's capacity to balance detection accuracy and control performance in MTCs. 

\begin{figure*}[!ht]
    \centering
    \subfloat[]{\includegraphics[width=0.25\linewidth]{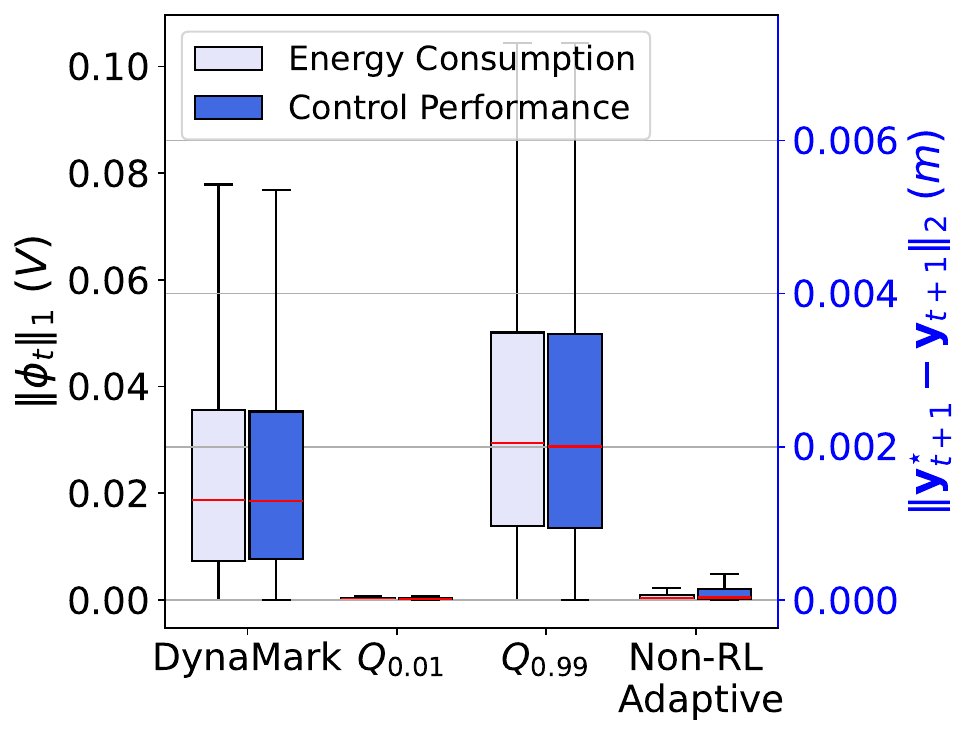}%
    \label{fig:5a}}
    \subfloat[]{\includegraphics[width=0.215\linewidth]{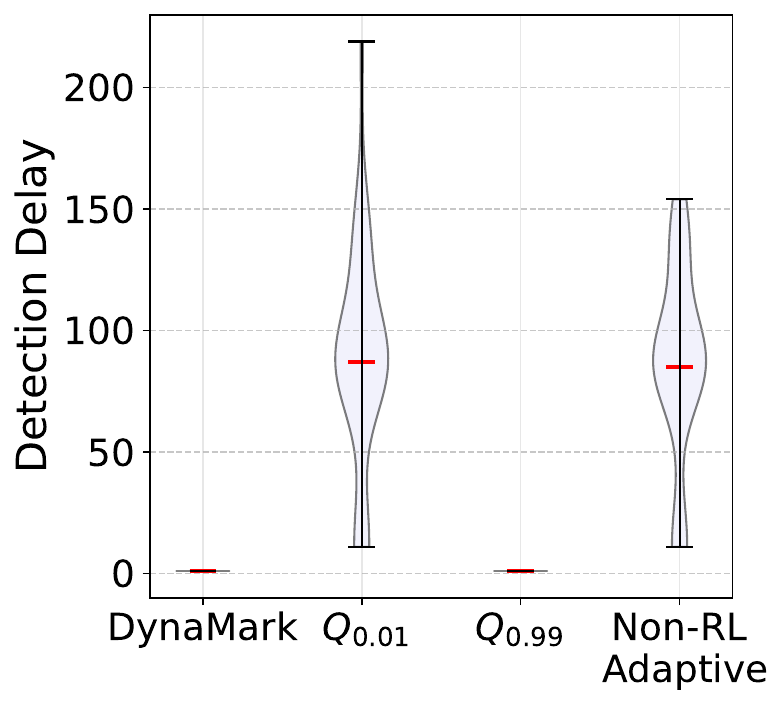}%
    \label{fig:5b}}
    \subfloat[]{\includegraphics[width=0.26\linewidth]{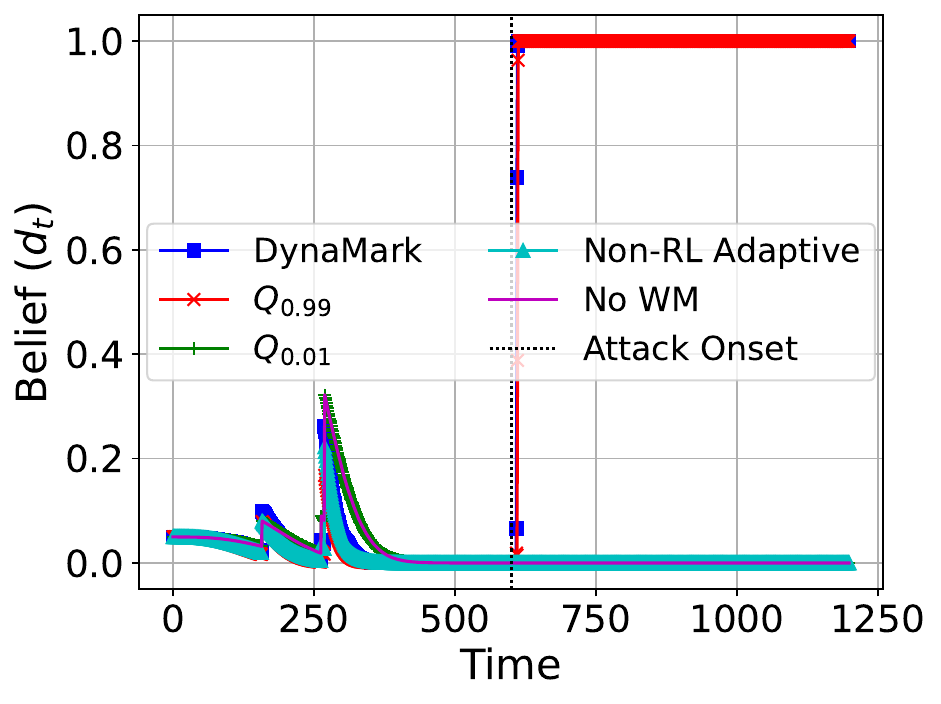}%
    \label{fig:5c}}
    \subfloat[]{\includegraphics[width=0.26\linewidth]{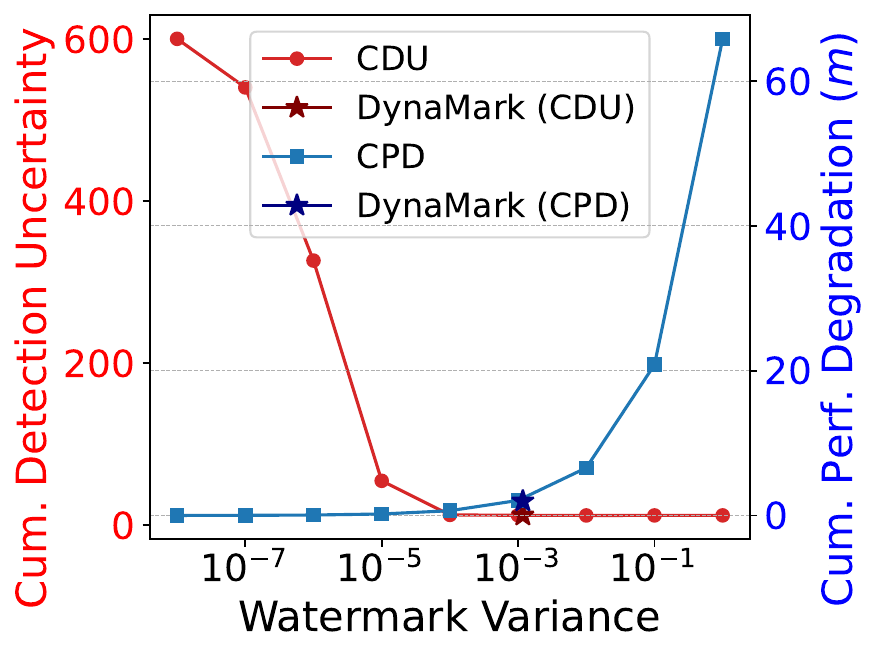}%
    \label{fig:5d}}
    \caption{Comparison of DynaMark with static-variance and non-RL adaptive baselines on DE: (a) nominal watermark energy and control-performance degradation, (b) detection delay under replay, (c) detector belief for one representative replay trial, and (d) security--performance frontier based on $\mathrm{CDU}$ and $\mathrm{CPD}$ across static-variance values.}
    \label{fig:5}
\end{figure*}

Figure~\ref{fig:5} compares DynaMark with the static-variance and non-RL adaptive baseline watermarking policies defined in \S~\ref{sec:baselines} under nominal operation and replay attack ($U_{\min}=U_{Q_{0.01}}=10^{-7}$ and $U_{\max}= U_{Q_{0.99}}=1.9{\times}10^{-3}$).
\begin{enumerate}
\item Under nominal operation, DynaMark keeps watermark energy and control-performance degradation well below the $Q_{0.99}$ baseline while maintaining substantially stronger excitation than the $Q_{0.01}$ case. The non-RL adaptive rule also has low nominal cost because the detector belief remains near zero during normal operation.

\item Under replay, DynaMark achieves $\mathrm{ARL}_1=1$ in all replications and drives the detector belief to one. The $Q_{0.99}$ baseline also detects quickly but with higher nominal cost, while the $Q_{0.01}$ and non-RL adaptive baselines are less reliable or slower because their watermark intensity remains too weak before sufficient belief accumulation.
\end{enumerate}

To map the security--performance frontier, we sweep a range of static-variance values, $ 10^{-8},10^{-7},\dots,10^{-1},1.0$, and compute $\mathrm{CDU}$ under attack and $\mathrm{CPD}$ under nominal operation. Figure~\ref{fig:5d} shows the expected knee: $\mathrm{CDU}$ decreases rapidly as watermark intensity increases, with limited performance loss up to the knee, after which additional detection gains level off while $\mathrm{CPD}$ increases sharply. DynaMark aligns near this knee, indicating that it balances detection and performance without relying on a preset static-variance.

\subsubsection{Ablation and Sensitivity Analysis}\label{sec:ablation}
We conduct ablation and sensitivity studies on the DE case to assess the contribution of each reward component and the robustness of DynaMark to key design parameters. For the reward ablation, we train separate policies by selectively enabling or disabling the energy, performance, and belief terms in the reward. For the watermark-budget sensitivity, we train separate policies with different values of $U_{\max}$ while keeping the full reward weights fixed. In contrast, the attack-prior and attack-onset-prior studies are evaluation-only: we use the trained DynaMark policy and vary only $q_{\mathrm{prior}}$ or $p_{\mathrm{geom}}$ in the online belief update, while keeping the actual replay attack onset fixed at $\tau=600$.

\begin{figure}[!ht]
    \centering
    \includegraphics[width=0.5\linewidth]{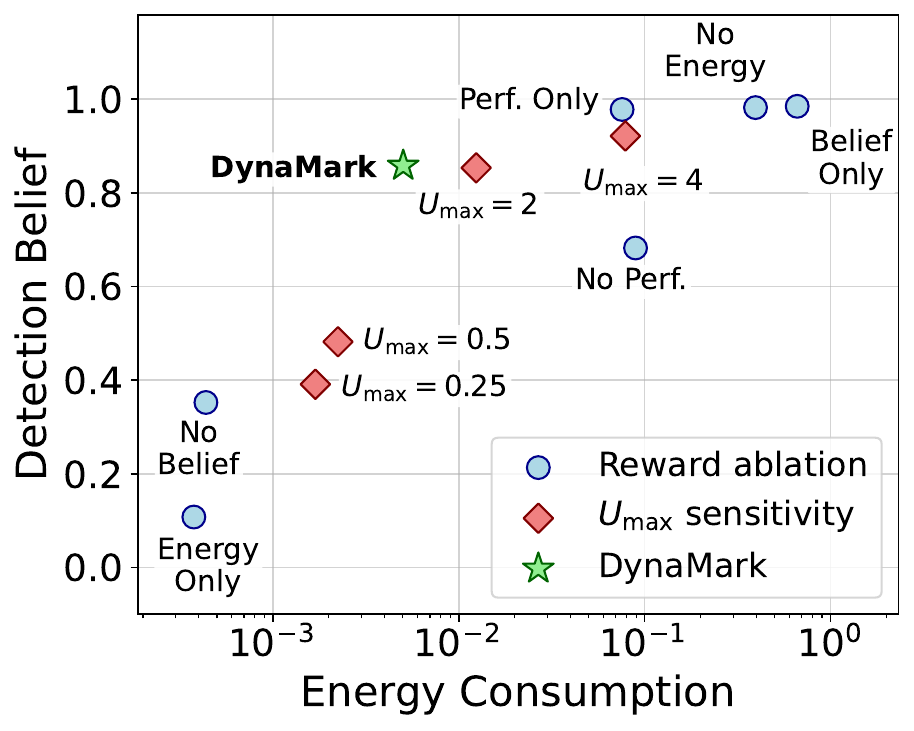}%
    \includegraphics[width=0.5\linewidth]{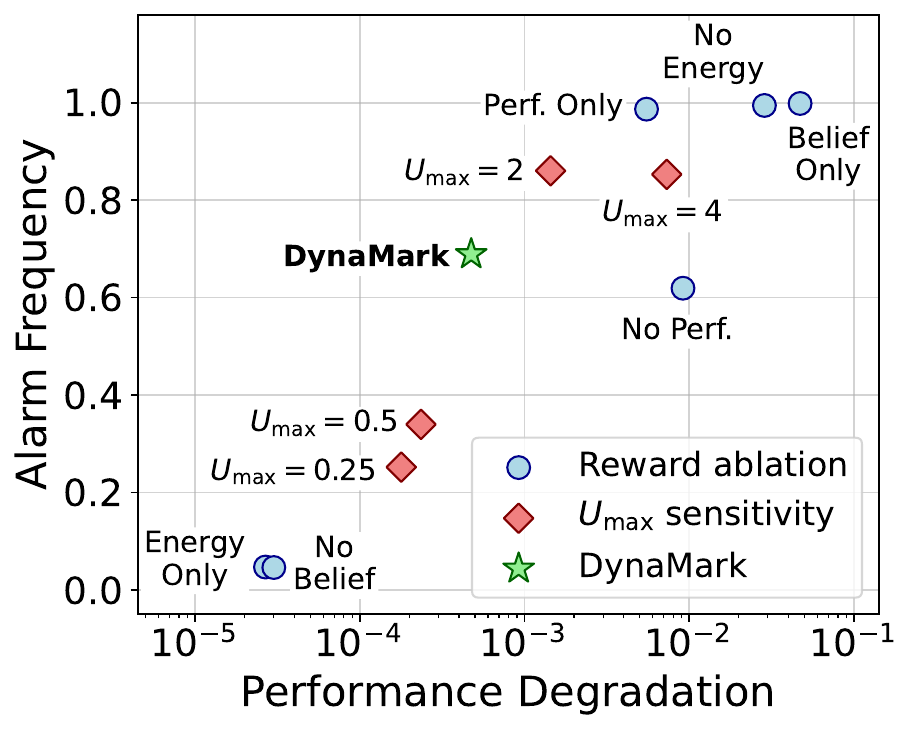}
    \caption{Ablation and sensitivity trade-offs for DynaMark. Each point represents aggregated policy evaluation results over seeds and replications}
    \label{fig:ablation}
\end{figure}

Fig.~\ref{fig:ablation} visualizes the aggregated trade-offs between nominal cost and detection effectiveness. The DynaMark policy provides a balanced operating point: it achieves substantially stronger detection than energy-only and no-belief variants, while avoiding the large energy and performance costs of belief-only and no-energy variants. Table~\ref{tab:ablation} reports the corresponding aggregated metrics, averaged over evaluation replications and summarized as mean $\pm$ standard error across seeds. The results show that increasing $U_{\max}$ improves detection persistence but increases nominal cost, while the evaluation-only prior studies indicate that DynaMark remains effective under moderate prior misspecification, with more conservative priors producing slower belief growth and larger onset-prior values reducing post-onset detection confidence.

\begin{table*}[!ht]
    \caption{Aggregated ablation and sensitivity results. Values are reported as mean $\pm$ standard error across seeds and replications.}
    \label{tab:ablation}
    \resizebox{\linewidth}{!}{
    \begin{tabular}{lllllllllll}
    \toprule
    Study & Variant & $(\omega_1,\omega_2,\omega_3)$ & $U_{\max}$ & $q_{\mathrm{prior}}$ & $p_{\mathrm{geom}}$ & Energy & \makecell{Performance \\ Degradation} & ARL$_1$ & Post-Onset AF & Detection Belief \\
    \midrule
    Baseline & DynaMark & $(0.35, 0.35, 0.3)$ & 1.0 & 0.05 & $1/T$ & 5.03e-3 $\pm$ 1.59e-3 & 4.75e-4 $\pm$ 1.09e-4 & 0.972 $\pm$ 0.506 & 0.689 $\pm$ 0.064 & 0.858 $\pm$ 0.062 \\
    \midrule
    \multirow{6}{*}{Reward} & Energy Only & $(0.35, 0, 0)$ & 1.0 & 0.05 & $1/T$ & \textbf{3.76e-4 $\pm$ 1.43e-4} & \textbf{2.68e-5 $\pm$ 1.04e-5} & 34 $\pm$ 11.3 & 0.047 $\pm$ 0.0341 & 0.108 $\pm$ 0.0966 \\
     & Performance Only & $(0, 0.35, 0)$ & 1.0 & 0.05 & $1/T$ & 7.56e-2 $\pm$ 3.15e-2 & 5.50e-3 $\pm$ 2.23e-3 & 0.125 $\pm$ 0.0447 & 0.987 $\pm$ 0.00385 & 0.978 $\pm$ 0.00344 \\
     & Belief Only & $(0, 0, 0.3)$ & 1.0 & 0.05 & $1/T$ & 6.62e-1 $\pm$ 3.97e-2 & 4.71e-2 $\pm$ 2.36e-3 & \textbf{0 $\pm$ 0} & \textbf{0.998 $\pm$ 9.26e-05} & \textbf{0.985 $\pm$ 7.25e-05} \\
     & No Belief & $(0.35, 0.35, 0)$ & 1.0 & 0.05 & $1/T$ & 4.36e-4 $\pm$ 9.66e-5 & 3.01e-5 $\pm$ 6.71e-6 & 9.29 $\pm$ 4.09 & 0.046 $\pm$ 0.0154 & 0.352 $\pm$ 0.143 \\
     & No Performance & $(0.35, 0, 0.3)$ & 1.0 & 0.05 & $1/T$ & 8.93e-2 $\pm$ 4.80e-2 & 9.16e-3 $\pm$ 4.44e-3 & 4.96 $\pm$ 4.2 & 0.619 $\pm$ 0.148 & 0.682 $\pm$ 0.146 \\
     & No Energy & $(0, 0.35, 0.3)$ & 1.0 & 0.05 & $1/T$ & 3.95e-1 $\pm$ 9.24e-2 & 2.86e-2 $\pm$ 6.15e-3 & 0.01 $\pm$ 0.00764 & 0.994 $\pm$ 0.00275 & 0.982 $\pm$ 0.00245 \\
    \midrule
    \multirow{4}{*}{Budget} & $U_{\max}=0.25$ & $(0.35, 0.35, 0.30)$ & 0.25 & 0.05 & $1/T$ & 1.69e-3 $\pm$ 5.12e-4 & 1.79e-4 $\pm$ 4.80e-5 & 16.5 $\pm$ 9.89 & 0.252 $\pm$ 0.113 & 0.391 $\pm$ 0.142 \\
     & $U_{\max}=0.5$ & $(0.35, 0.35, 0.30)$ & 0.5 & 0.05 & $1/T$ & 2.24e-3 $\pm$ 6.61e-4 & 2.35e-4 $\pm$ 7.65e-5 & 17.9 $\pm$ 10.8 & 0.34 $\pm$ 0.13 & 0.482 $\pm$ 0.16 \\
     & $U_{\max}=2$ & $(0.35, 0.35, 0.30)$ & 2.0 & 0.05 & $1/T$ & 1.24e-2 $\pm$ 5.28e-3 & 1.44e-3 $\pm$ 4.91e-4 & 0.979 $\pm$ 0.501 & 0.86 $\pm$ 0.0956 & 0.854 $\pm$ 0.0954 \\
     & $U_{\max}=4$ & $(0.35, 0.35, 0.30)$ & 4.0 & 0.05 & $1/T$ & 7.88e-2 $\pm$ 4.51e-2 & 7.30e-3 $\pm$ 4.09e-3 & 0.987 $\pm$ 0.412 & 0.853 $\pm$ 0.07 & 0.922 $\pm$ 0.0204 \\
    \midrule
    \multirow{2}{*}{\makecell{Attack \\ Prior}} & $q_{\mathrm{prior}}=0.01$ & $(0.35, 0.35, 0.3)$ & 1.0 & 0.01 & $1/T$ & 4.71e-3 $\pm$ 2.07e-3 & 3.35e-4 $\pm$ 1.39e-4 & 1.84 $\pm$ 0.996 & 0.643 $\pm$ 0.0766 & 0.81 $\pm$ 0.0866 \\
    & $q_{\mathrm{prior}}=0.1$ & $(0.35, 0.35, 0.3)$ & 1.0 & 0.1 & $1/T$ & 9.3e-3 $\pm$ 1.67e-3 & 1.09e-3 $\pm$ 2.49e-4 & 0.495 $\pm$ 0.319 & 0.67 $\pm$ 0.0677 & 0.84 $\pm$ 0.0714 \\
    \midrule
    \multirow{2}{*}{\makecell{Attack \\ Onset}} & $p_{\mathrm{geom}}=0.5/T$ & $(0.35, 0.35, 0.3)$ & 1.0 & 0.05 & $0.5/T$ & 5.12e-3 $\pm$ 1.42e-3 & 4.86e-4 $\pm$ 1.03e-4 & 0.782 $\pm$ 0.241 & 0.794 $\pm$ 0.0701 & 0.973 $\pm$ 7.76e-3\\
    & $p_{\mathrm{geom}}=2/T$ & $(0.35, 0.35, 0.3)$ & 1.0 & 0.05 & $2/T$ & 5.12e-3 $\pm$ 1.74e-3 & 4.79e-4 $\pm$ 1.19e-4 & 1.73 $\pm$ 0.719 & 0.43 $\pm$ 0.12 & 0.57 $\pm$ 0.155 \\
    \bottomrule
    \end{tabular}}
\end{table*}

\subsection{Case Study 2: Nonlinear Mass-Spring-Damper (MSD)}\label{sec:msd}

\subsubsection{Experiment Setup}
To assess DynaMark beyond linear dynamics and Gaussian uncertainty, we consider a single-axis nonlinear mass-spring-damper (MSD) system, a canonical reduced-order surrogate for compliant MTC axes and transmission dynamics where structural flexibility and damping shape tracking performance and vibration responses, e.g., axis drive compliance, guideway friction, and coupling elasticity~\cite{ALTINTAS2011779}. Such second-order models are widely used in MTC and monitoring because they capture the dominant low-frequency mechanics that interact with feedback loops and are directly exposed to cyber--physical manipulation of sensor feedback and commanded inputs. We adopt the nonlinear dynamics $m\ddot p + b(\dot p) + k(p) = F(t) + u(t)$ with damping ($b(v)=b_1v+b_2v^3$) and spring ($k(p)=k_1p+k_2p^3$) forces, where $m$ is the mass, $p$ is displacement from equilibrium position, $v=\dot p$, $F(t)$ is a measurable disturbance, and $u(t)$ is the commanded input (see~\cite{MatlabMSD_online} for more details). With sampling time $T_s=\Delta t$ and honest sensing, we instantiate Eq.~\eqref{eq:sys} with $\vy_t=[p_t,v_t]^\top$, $\vu_t=[0, u_t+F_t]^\top$, and $\mathcal{G}_t(\vy_t,\vu_t)=\left[p_t + T_s v_t;\ v_t - \frac{T_s}{m}\left(b(v_t) + k(p_t) -u_t-F_t\right)\right]$. To test robustness beyond the Gaussian setting used in Lemma~\ref{lemma_beta}, we model the uncertainty term $\ve_t$ using a zero-mean heavy-tailed Student-$t$ distribution with 5 degrees of freedom scaled to match covariance $Q$. We set $m=1.0$ kg, $k_1=0.5\;N/m$, $k_2=1.0\; N/m^3$, $b_1=1.0\; N/(m/s)$, $b_2=0.1\; N/(m/s)^3$, $T_s=0.01\; s$, and $Q=10^{-7}I$. The exogenous disturbance is constant ($F_t=F_0=2.0 \; N$). Each rollout has horizon $T=4000$. The replay attack's prior and onset distribution, and its mechanism, follow the MTC threat model in \S~\ref{sec:numerical} with $\tau=1000$. The nominal command is a chirp excitation to induce time-varying operating conditions, i.e., $u_t^\star=0.1\sin\left(t \omega_t\right) \; N$, $\omega_t=0.1(\frac{1+F_0}{0.1})^{\frac{t}{T\times T_s}}$. The reward weights are set as $\omega_1=0.1,\omega_2=0.35$, and $\omega_3=0.55$.

This case study is intended as an empirical stress test beyond the sufficient closed-loop condition in Theorem~\ref{thm:stability} and beyond the Gaussian residual setting used in the analytical detector belief characterization. Since the nominal command is programmed rather than measurement-feedback, we do not invoke Theorem~\ref{thm:stability} here. Instead, stability is enforced through the bounded watermark action constraint and verified by bounded trajectories over all training and evaluation rollouts. Because the residuals are non-Gaussian in this experiment, the miss-detection probability $\beta_t$ is not computed using the Gaussian closed form in Lemma~\ref{lemma_beta}; instead, it is estimated offline by Monte Carlo calibration over finite grids of candidate watermark variances and replay-onset times, and then used during training via a lookup table; see Appendix~\ref{app:beta_calibration} for details.

\begin{figure*}[!ht]
    \centering
    \subfloat[]{\includegraphics[width=0.25\linewidth]{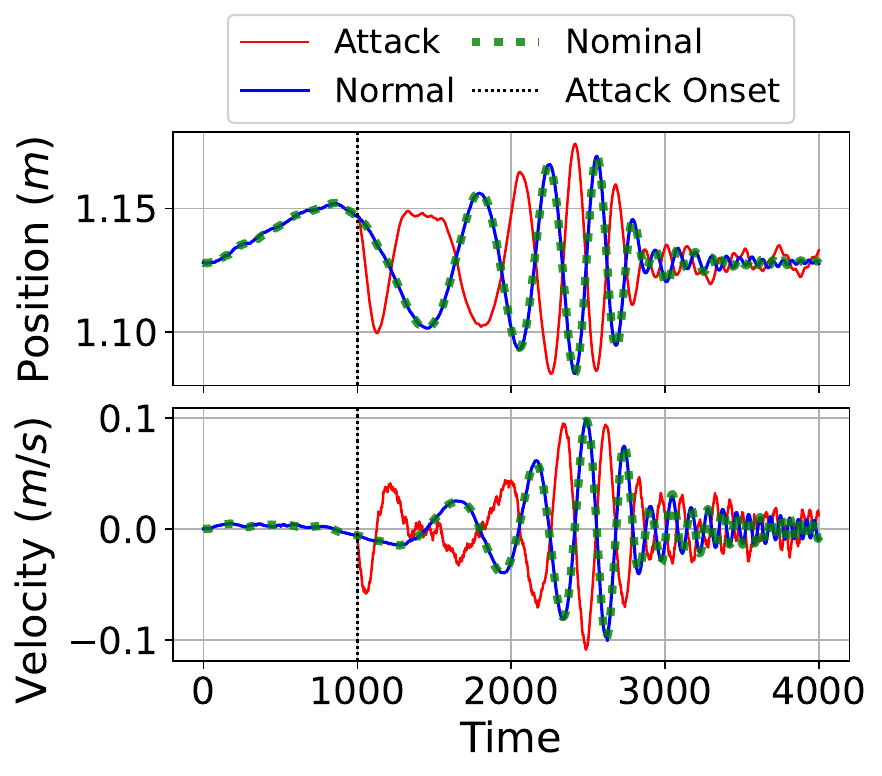}%
    \label{fig:6a}}
    \subfloat[]{\includegraphics[width=0.245\linewidth]{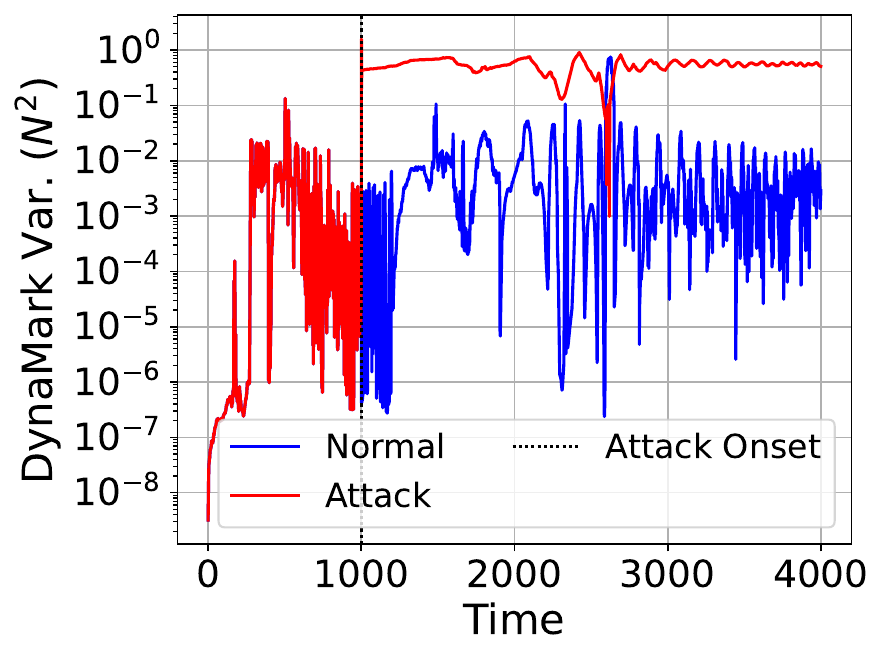}%
    \label{fig:6b}}
    \subfloat[]{\includegraphics[width=0.245\linewidth]{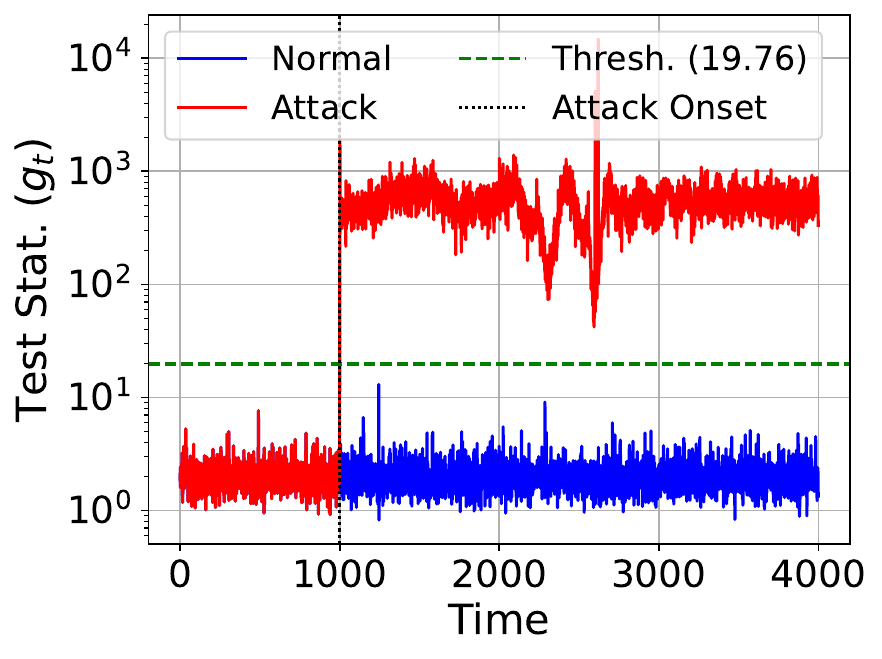}%
    \label{fig:6c}}
    \subfloat[]{\includegraphics[width=0.245\linewidth]{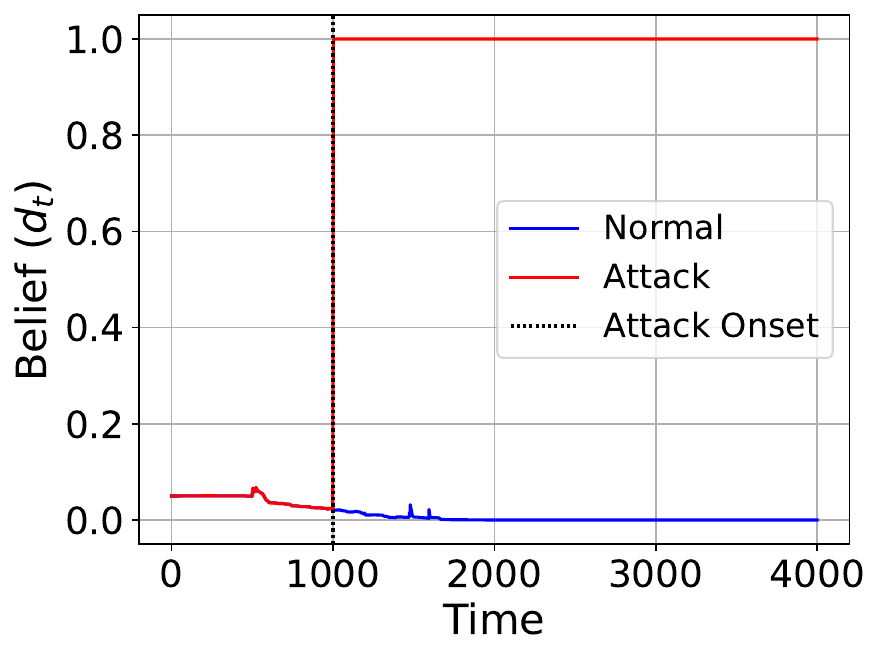}%
    \label{fig:6d}}
    \caption{Average performance of DynaMark on the MSD under normal operation and replay attack ($\tau=1000$): (a) trajectories under nominal operation, normal operation with DynaMark, and under attack; (b) DynaMark-selected watermark covariance (action) over time; (c) test statistic with the empirically calibrated alarm threshold; (d) attack belief inferred by the detector.}
    \label{fig:6}
\end{figure*}

\begin{figure}[!ht]
    \centering
    \includegraphics[width=0.525\linewidth]{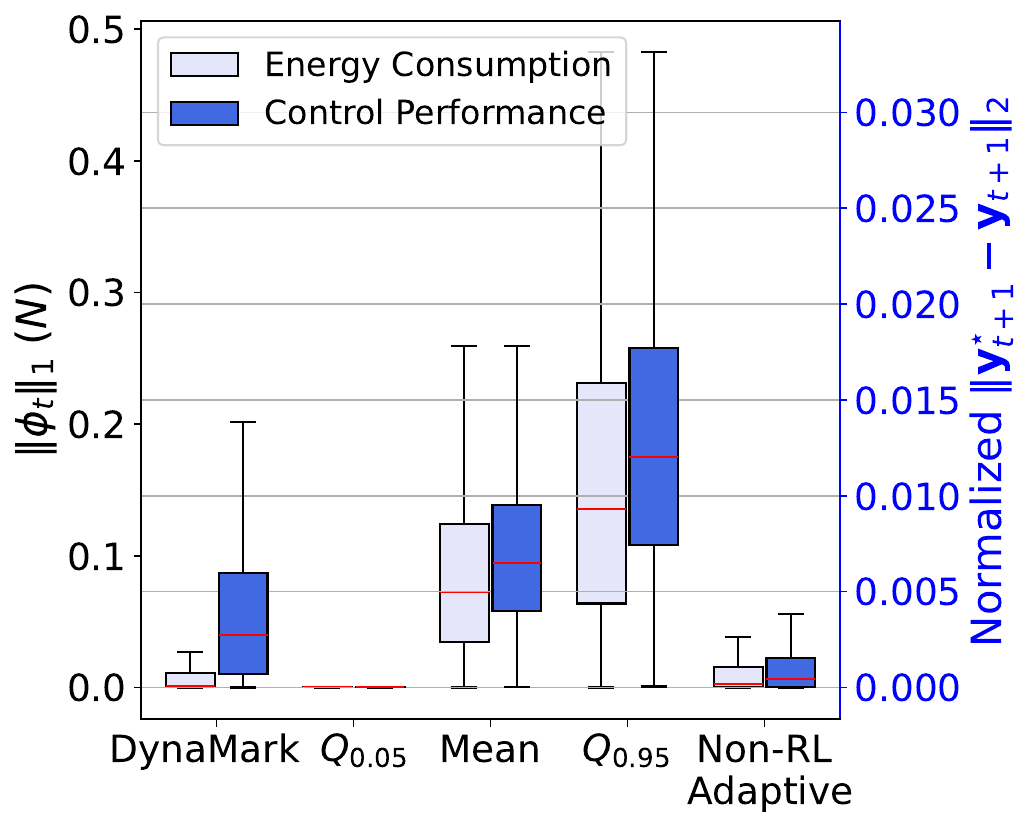}%
    \includegraphics[width=0.475\linewidth]{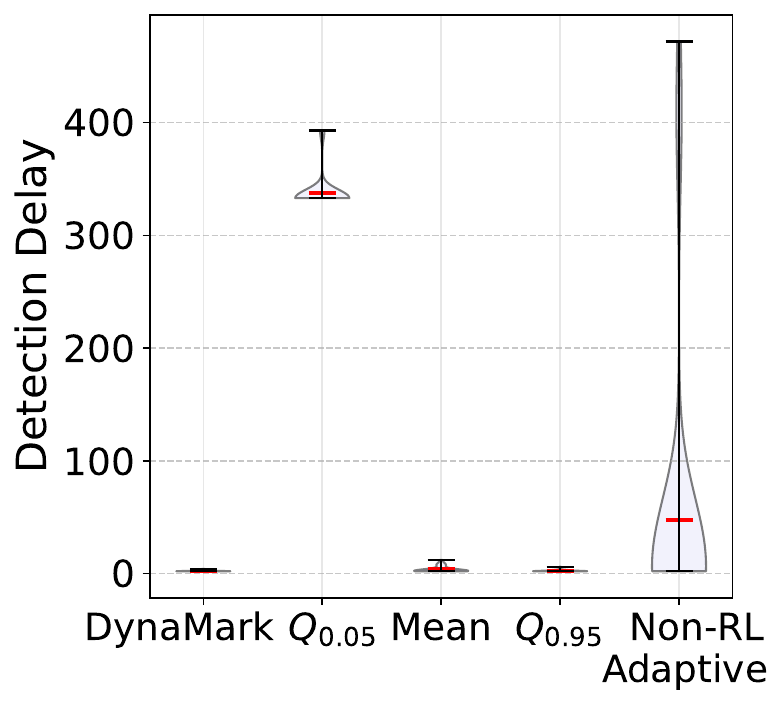}
    \caption{Comparison of DynaMark with static-variance and non-RL adaptive baselines on the nonlinear MSD: (a) nominal watermark energy and control-performance degradation, and (b) detection delay under replay attack.}
    \label{fig:7}
\end{figure}

\subsubsection{Performance Evaluation}
Figure~\ref{fig:6} shows DynaMark’s behavior on the nonlinear MSD under nominal operation and a replay attack starting at $\tau=1000$. In nominal operation, the learned policy keeps the DWM variance low, the trajectory remains close to the nominal response, and the detector statistic stays below the calibrated alarm threshold, keeping the posterior attack belief near zero. After replay begins, DynaMark increases the variance, producing a clear residual shift; consequently, the test statistic rises above the threshold and the attack belief quickly saturates to one.

To contextualize the trade-off, Fig.~\ref{fig:7} benchmarks against static-variance and non-RL adaptive baselines ($U_{\min}=U_{Q_{0.05}}=5.7e-7$, $U_{\text{mean}}=9e-4$, $U_{\max}=U_{Q_{0.95}}=7e-3$). As expected, the $Q_{0.95}$ baseline improves detection performance by strengthening excitation, but it incurs higher watermark energy and control-performance degradation; conversely, the $Q_{0.05}$ baseline has low nominal cost but delayed and unreliable detection. The mean baseline provides an intermediate static trade-off, reducing $\mathrm{ARL}_1$ relative to $Q_{0.05}$ but with higher nominal cost than DynaMark. The non-RL adaptive baseline keeps nominal cost low because the belief remains near zero in normal operation, but under attack its detection delay is more variable, indicating that directly scaling $U_t$ with $d_t$ is less reliable than learning a systematic DWM policy.

\subsection{Case Study 3: Physical Testbed Implementation}\label{sec:case}
This section evaluates DynaMark on a physical smart stepper-motor testbed representative of MTC-controlled motion systems used in drilling, milling, laser cutting, and 3D printing. Although the motor plant is simple, the implementation preserves practical controller artifacts such as discrete command updates, communication delays, buffering, and timing jitter, all of which affect replay detection and watermark injection. The firmware and motion-control stack is also closely related to those used in 3D-printer controllers, supporting the relevance of this testbed to manufacturing control pipelines.

\begin{figure}[!ht]
  \centering
  \includegraphics[width=.9\linewidth]{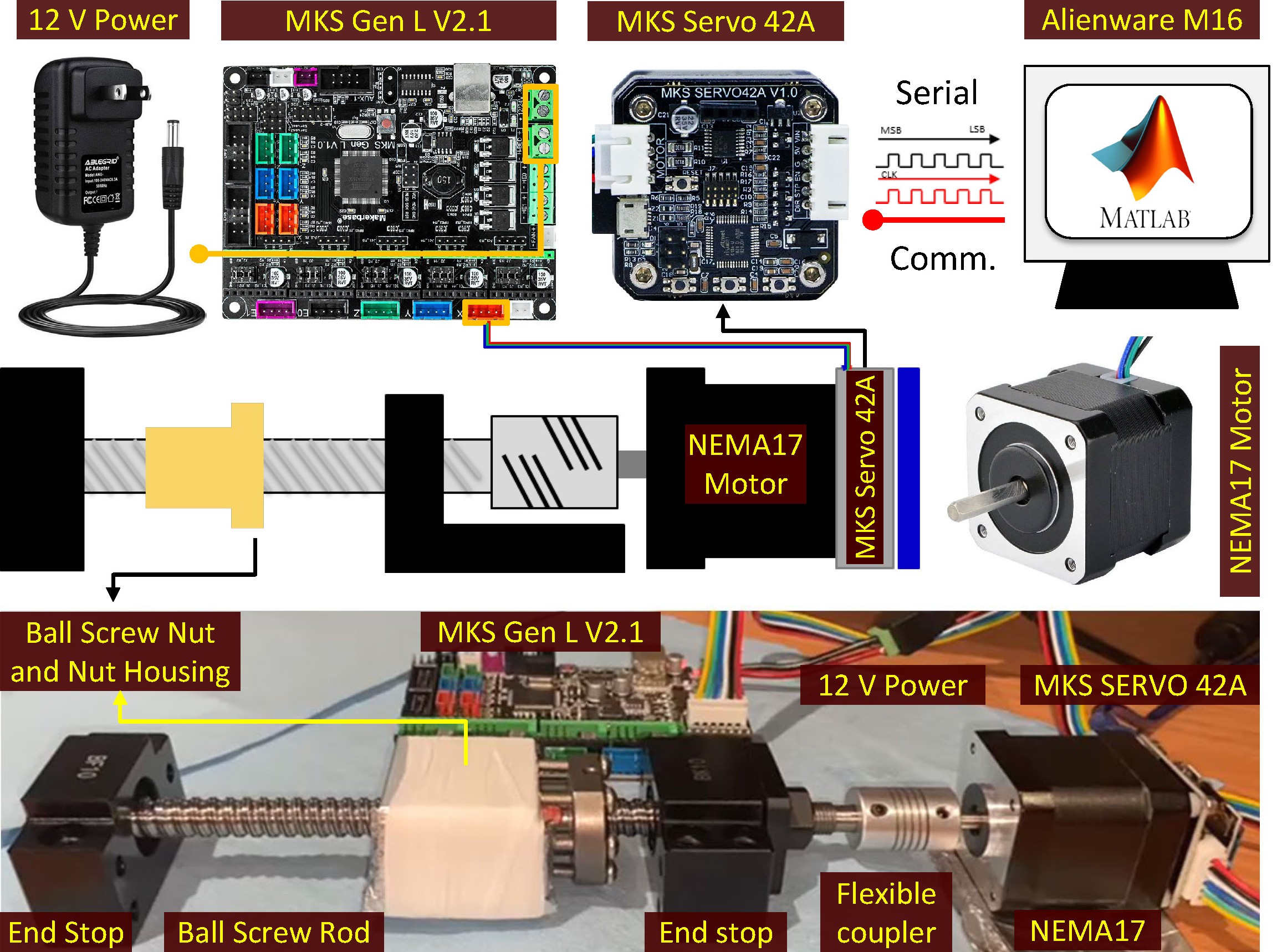}
  \caption{Smart stepper-motor physical implementation.}
  \label{fig:stepper}
\end{figure}

\subsubsection{Experiment Setup}\label{sec:4.2.1}

The experimental platform integrates a $\rm NEMA17$ closed-loop stepper-motor with magnetic encoder, driven by an $\rm MKS$ $\rm Gen$ $\rm L$ $\rm V2.1$ control board and an $\rm MKS$ $\rm Servo$ $\rm 42A$ smart driver as illustrated in Fig~\ref{fig:stepper}. The $\rm MKS$ $\rm Servo$ $\rm 42A$'s firmware allows for servo-like control logic, which was then modified to accept $U_{t}$ updates via serial commands during motion using a command ``watermark $U_{t}$'', while ensuring synchronization with the encoder’s closed-loop feedback. This is accomplished via serial communication with a PC running $\rm MATLAB$ at $115200$ baudrate with $\rm CR+LF$ termination. Additional firmware modifications enabled scripted cyberattack scenarios triggered via their respective commands. At each control tick, the driver emits a timestamped state vector $\langle \boldsymbol{\tau}_t, \vy_t, \vu_t, \boldsymbol{\phi}_t, \mathrm{attack}_{flag} \rangle$, where $\boldsymbol{\tau}_t$ are time steps, $\vy_t$ are position measurements (encoder counts converted to mm), $\vu_t$ is the control signal, $\boldsymbol{\phi}_t$ is the watermark signal, and $\mathrm{attack}_{flag}$ is a 1/0 variable indicating an active attack. 

\begin{figure*}[!ht]
    \centering
    \subfloat[]{\includegraphics[width=.2\linewidth]{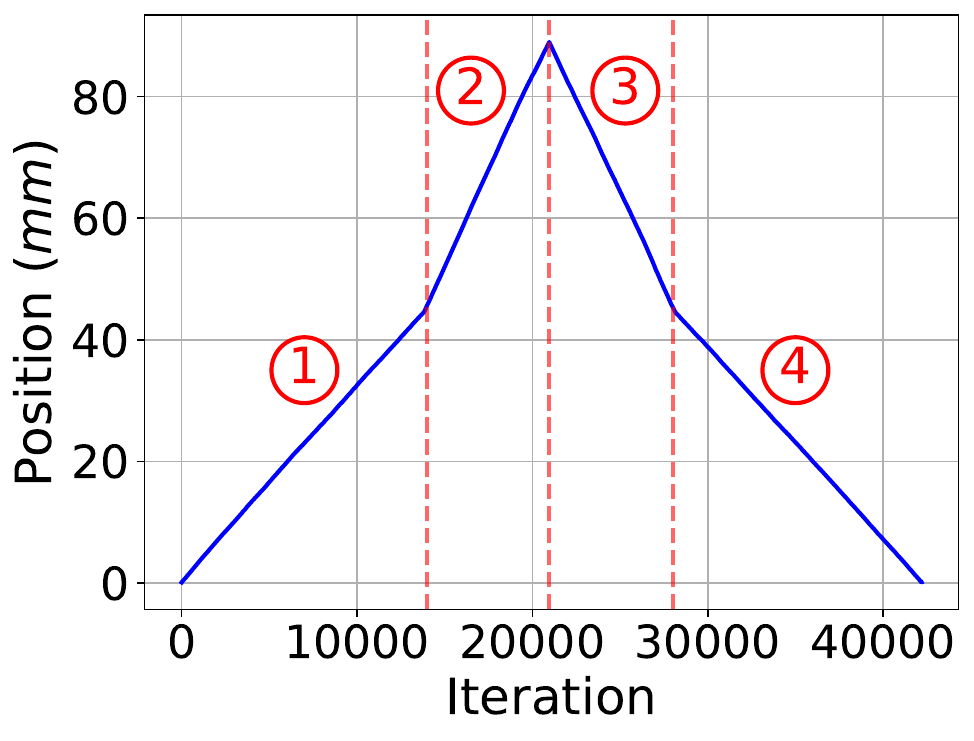}%
    \label{fig:motorparatraj}}
    \subfloat[]{\includegraphics[width=.2\linewidth]{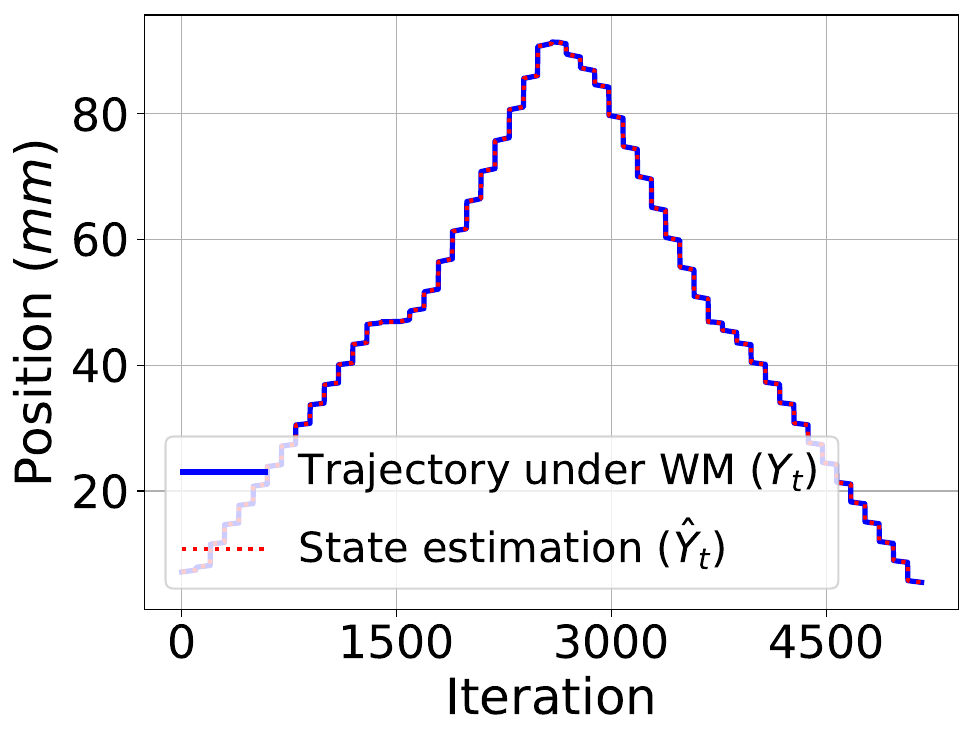}%
    \label{fig:motor1}}
    \subfloat[]{\includegraphics[width=.2\linewidth]{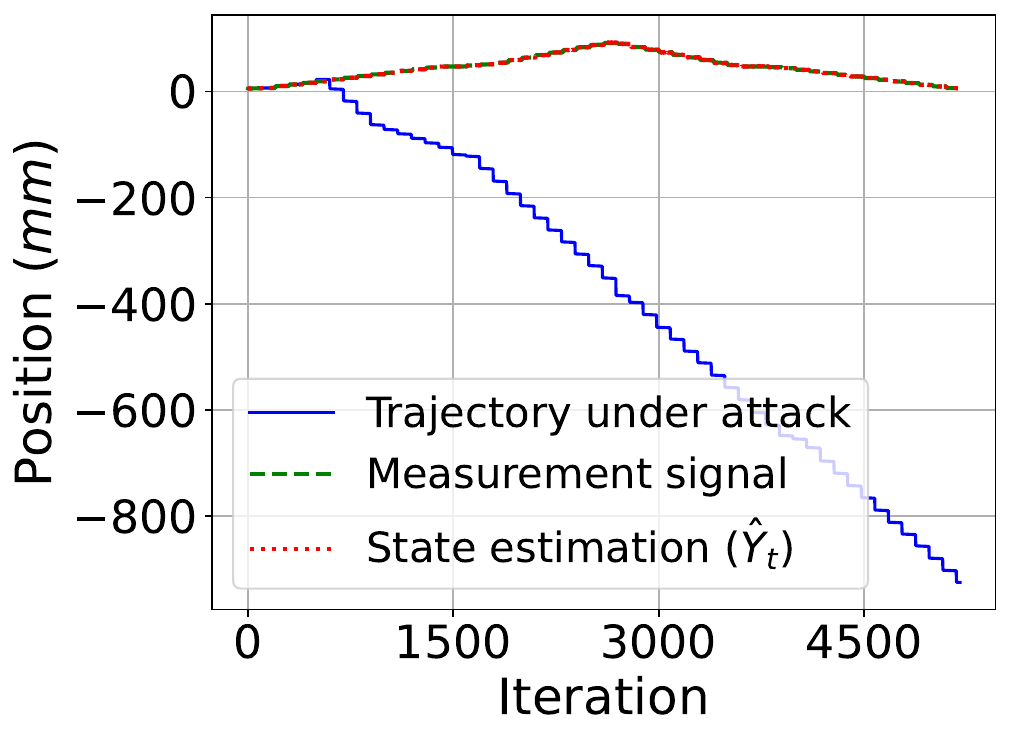}%
    \label{fig:motor2a}
    }
    \subfloat[]{\includegraphics[width=.2\linewidth]{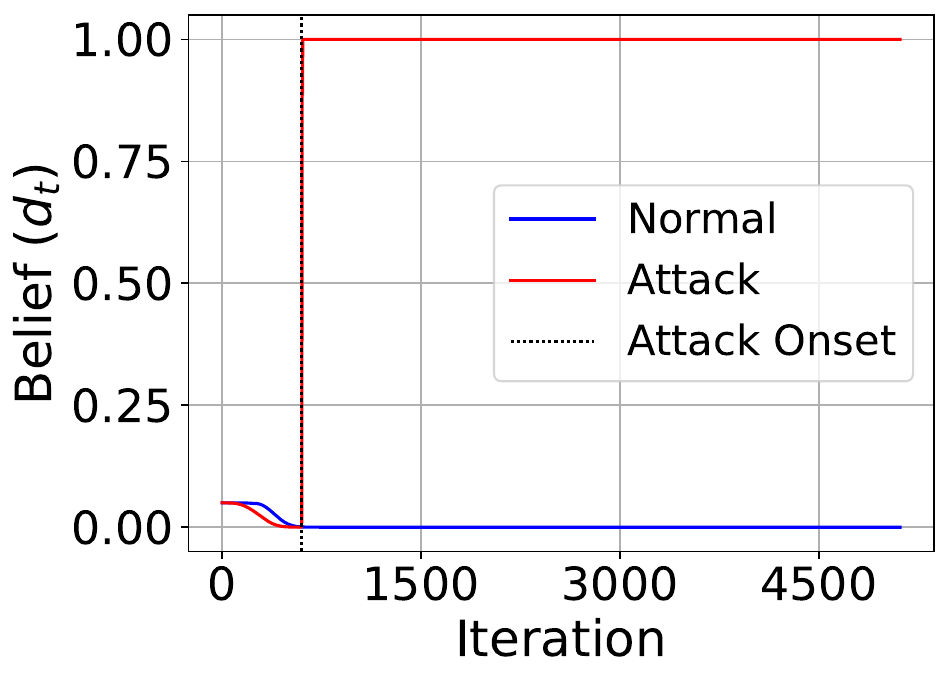}%
        \label{fig:motor2b}}
    \subfloat[]{\includegraphics[width=.2\linewidth]{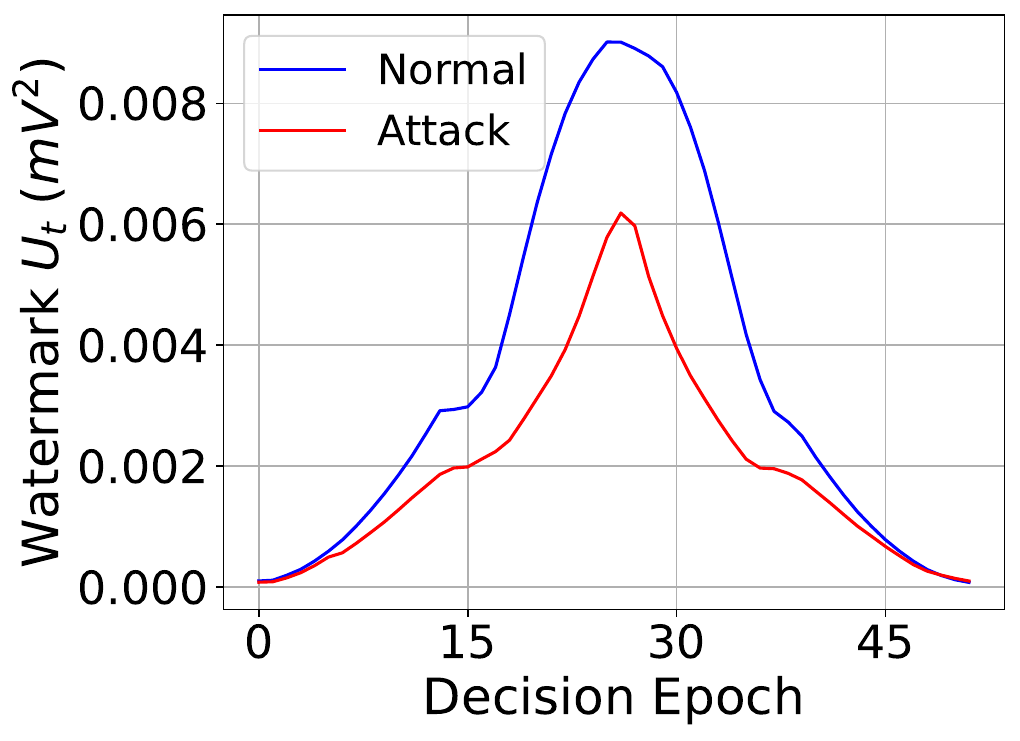}%
        \label{fig:motor2c}}
    \caption{The stepper-motor position under normal conditions (a) continuous, no watermark, with four numbered segments based on four motor commands, and (b) discretized and under DynaMark's DWM. The stepper‑motor response to a replay attack with onset at decision epoch 7, processed index $(\approx600)$. (c) Motor position under DynaMark (blue), the replayed measurement stream presented to the detector (green), and the state estimate $\widehat{\vy}_t$ (red). (d) Evolution of the detector belief $d_t$ under normal operation (blue) and during the attack (red). (e) Watermark covariance~$U_t$ chosen by the RL policy for the two conditions.}
\end{figure*}

DynaMark deployment on the motor involves three steps: system identification, DE construction for RL training, and attack monitoring on the physical testbed. For identification, we model the motor as a piecewise linear system over operating points defined by speed and direction. At each operating point, the parameters $A$, $B$, $Q$, and desired set-point $\overline{\vy}$ are estimated by fitting an {\fontfamily{qcr}\selectfont ARX(1,1)} model to watermark-free data $\{\vy_t,\vu_t\}$ collected at the firmware's native acquisition rate of $\rm 1 \ kHz$. The resulting local models are combined to represent the motor's nonlinear, mode-dependent behavior. Table~\ref{tab:motorparatable} reports the identified parameters for the four command segments, and Fig.~\ref{fig:motorparatraj} shows the corresponding continuous motion profile with segment boundaries.

\begin{table}[ht]
\centering
\caption{{\fontfamily{qcr}\selectfont ARX(1,1)} parameters identified from watermark-free continuous data.}
\label{tab:motorparatable}
\begin{tabular}{@{}lcccc@{}}   % <- five column spec: l c c c c
\toprule
Motor commands &
\(A\) &
\(B \;(\mathrm{mm/mV})\) &
\(Q \;(\mathrm{mm^{2}})\) &
\(\overline{\vy} \;[\mathrm{mm}]\) \\ \midrule
M1) 4000$^{\circ}$, 200 RPM & \(1\) & \(0.0075\) & \(5.57\times10^{-6}\) & 46.94 \\
M2) 8000$^{\circ}$, 300 RPM  & \(1\) & \(0.0108\) & \(9.81\times10^{-6}\) & 91.38 \\
M3) 4000$^{\circ}$, 300 RPM  & \(1\) & \(0.0107\) & \(9.38\times10^{-6}\) & 46.92 \\
M4) 0$^{\circ}$, 200 RPM & \(1\) & \(0.0076\) & \(5.57\times10^{-6}\) &  2.48 \\ 
\bottomrule
\end{tabular}
\end{table}

For the physical testbed, Theorem~\ref{thm:stability} can be verified using the identified local {\fontfamily{qcr}\selectfont ARX(1,1)} models and the implemented firmware controller following steps in Appendix~\ref{app:thm3_verification}. Although the firmware has a PID structure, the runtime integer-scaled gains are $K_p=921$, $K_i=0$, and $K_d=10$ with scaling factor $1024$; hence, the implemented controller is a finite-history PD controller with $w=1$ and local measurement gains $K_{t,0}=-(921+10)/1024=-0.9092$ and $K_{t,1}=10/1024=0.0098$. Using Table~\ref{tab:motorparatable}, we obtain $\Bar{H}=\max_j B_j^2=1.1664\times 10^{-4},$ $\Bar{K}=K_{t,1}^2=9.54\times 10^{-5},$ and $\Bar{A}=\max_j |A_j+B_jK_{t,0}|=0.9932$. These values satisfy conditions $(C_1)$ and $(C_2)$. Choosing $\epsilon=0.005$ gives $\epsilon<\Bar{A}^{-2}-1$, $c_\epsilon=6.71\times10^{-6}$, and $\rho_\epsilon^\star=0.00259$. With $\rho=\rho_\epsilon^\star$, we obtain $g_\epsilon(\rho_\epsilon^\star)<\delta_\epsilon$ and $\nu(\epsilon,\rho)=0.9965<1$.

During online decision-making or attack monitoring, trajectory updates are issued in discrete batches rather than as a fully continuous stream. This discretization leads to stepwise profiles in the recorded motion, even when the commanded trajectory is smooth. Figure~\ref{fig:motor1} illustrates the stepwise behavior from batchwise collection and processing. Nevertheless, the stepper-motor maintains accurate tracking across the entire motion profile under these normal operating conditions. To mitigate the RL agent training cost on the physical testbed, we construct the DE of the motor using these same stepwise trajectories and trained the agent on this DE with the reward weights as $\omega_1=\omega_2=0.35$ and $\omega_3=0.3$. This ensures the DE accurately reflects the operational characteristics observed during online decision-making. The last task is performing attack monitoring on the actual testbed. A replay attack scenario is considered wherein the adversary replays the measurements while flipping the control signal. For this physical implementation, the nominal alarm threshold is empirically calibrated from residual data with $\alpha=0.005$. 

The online computational cost is dominated by the belief update. For residual dimension $d$ and $K$ Monte-Carlo samples used in replay-side $\beta_t$ estimation, each detector update costs $O(Kd^2+d^3)$, while the trained actor is evaluated once per outer decision epoch with fixed cost $C_{\pi}$. Thus, for $M$ detector samples and $T$ decision epochs, the total online cost is $O(M(Kd^2+d^3)+TC_{\pi})$. In the stepper-motor implementation, $d=1$ and $K=2000$ are fixed, so the detector update has constant per-sample cost and total runtime scales linearly with the number of acquired samples. Memory is $O(M)$ when logs are stored and can be reduced to $O(1)$ with a sliding-window implementation.

\subsubsection{Performance Evaluation}
\label{sec:4.2.2}

Figures~\ref{fig:motor1}--\ref{fig:motor2c} show that the trained RL policy successfully transfers to the physical testbed. Under the replay attack starting at decision epoch $7$ (iteration $\sim600$), the physical trajectory in Fig.~\ref{fig:motor2a} diverges from the spoofed measurements, while the state estimate $\widehat{\vy}_t$ tracks the replayed input. The discrepancy between the spoofed measurement and state estimates triggers the detector belief $d_t$ in Fig.~\ref{fig:motor2b}, which reaches $1$ within five samples. Figure~\ref{fig:motor2c} shows DynaMark adapting $U_t$: under nominal motion, it peaks near $\rm 0.009 \ mV^2$ during acceleration and tapers on descent; after belief updating to 1, DynaMark adjusts $U_t$ to maintain a strong residual gap. These results confirm that DynaMark is deployable on the physical implementation while successfully distinguishing replayed measurements from the actual motor response in real time.

\subsubsection{Comparative Analysis}

To instantiate the LTI-based baselines for the stepper-motor DE, we form a single LTI surrogate despite the mode-switching behavior shown in Fig.~\ref{fig:motorparatraj}. Specifically, we estimate an {\fontfamily{qcr}\selectfont ARX(1,1)} model from watermark-free data over the full motion cycle, obtaining $A=1.0$, $B=0.00372$, and $Q=3.1911{\times}10^{-5}$. Using this surrogate, we compute offline static watermark variances by solving the covariance-design optimization in~\cite{Liu9061046} under loss budgets $\delta\in\{100\%,50\%,20\%,10\%,5\%\}$ relative to the no-watermark LQG cost $J_0$ with $X=1.0$, and we compute the online baseline by implementing the adaptive method in~\cite{Liu9061046} and reporting its converged variance. Table~\ref{tab:opt_u} lists the resulting $U_\star$. Each baseline is then evaluated on the DE under the same nominal and replay settings as in Section~\ref{sec:4.2.1}.

\begin{table}[!ht]
\centering
\caption{Offline and online (converged) watermark variances $U_\star$ for the stepper-motor LTI surrogate under varied loss budgets.}
\label{tab:opt_u}
\resizebox{\linewidth}{!}{
\begin{tblr}{
  cells = {c},
  cell{1}{1} = {c=2}{},
  cell{2}{1} = {r=2}{},
  vline{2} = {-}{},
  vline{2-3} = {2}{},
  vline{3} = {-}{},
  hline{1-2,4} = {-}{},
}
$\delta$        &         & $100\%J_0$ & $50\%J_0$ & $20\%J_0$ & $10\%J_0$ & $5\%J_0$ \\
$U_\star(mV^2)$ & Offline & 0.8655     & 0.4327    & 0.1731    & 0.0865    & 0.0432   \\
                & Online  & 0.03005    & 0.00203   & $5.4e-5$  & $1.2e-5$  & $2.1e-6$ 
\end{tblr}
}
% \vspace{-3ex}
\end{table}

\begin{figure}[!ht]
    \centering
    \subfloat[]{\includegraphics[width=.333\linewidth]{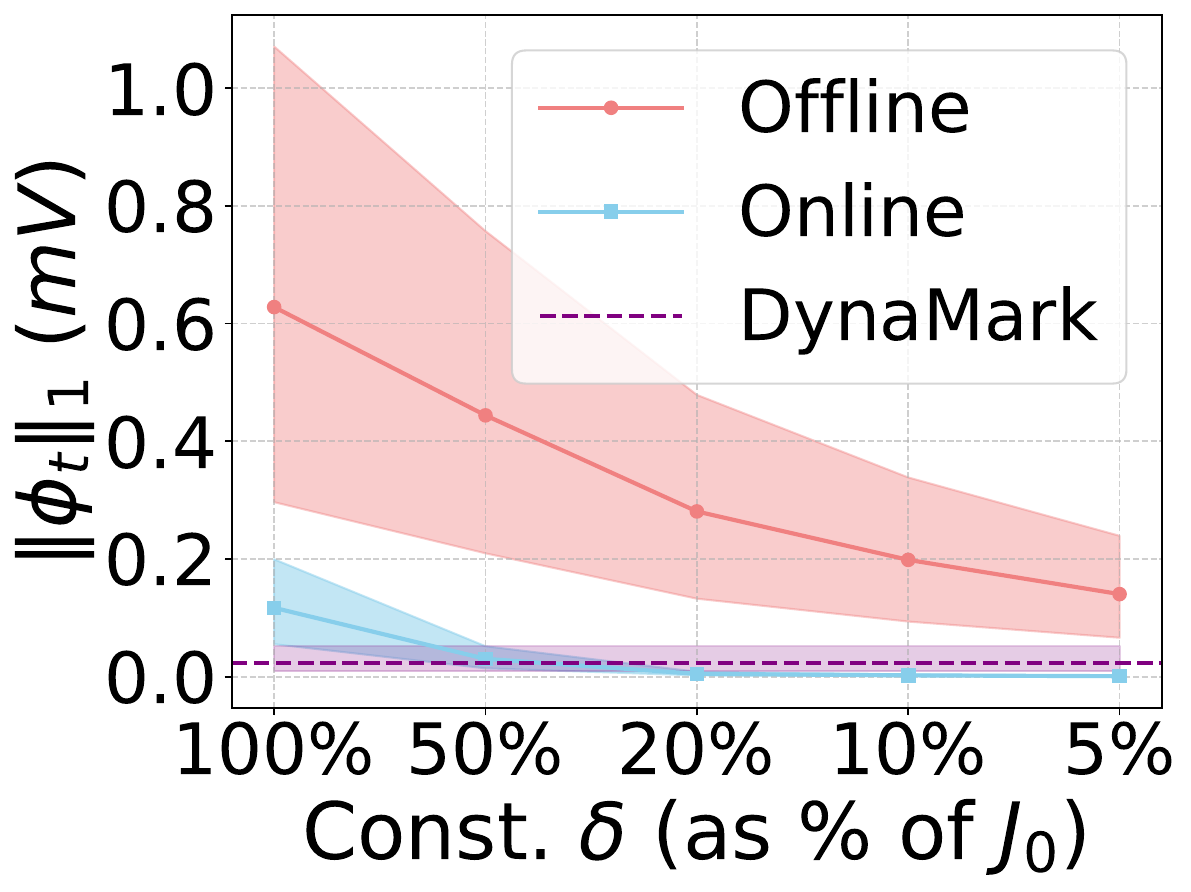}%
        \label{fig:compa}}
    \hfill
    \subfloat[]{\includegraphics[width=.333\linewidth]{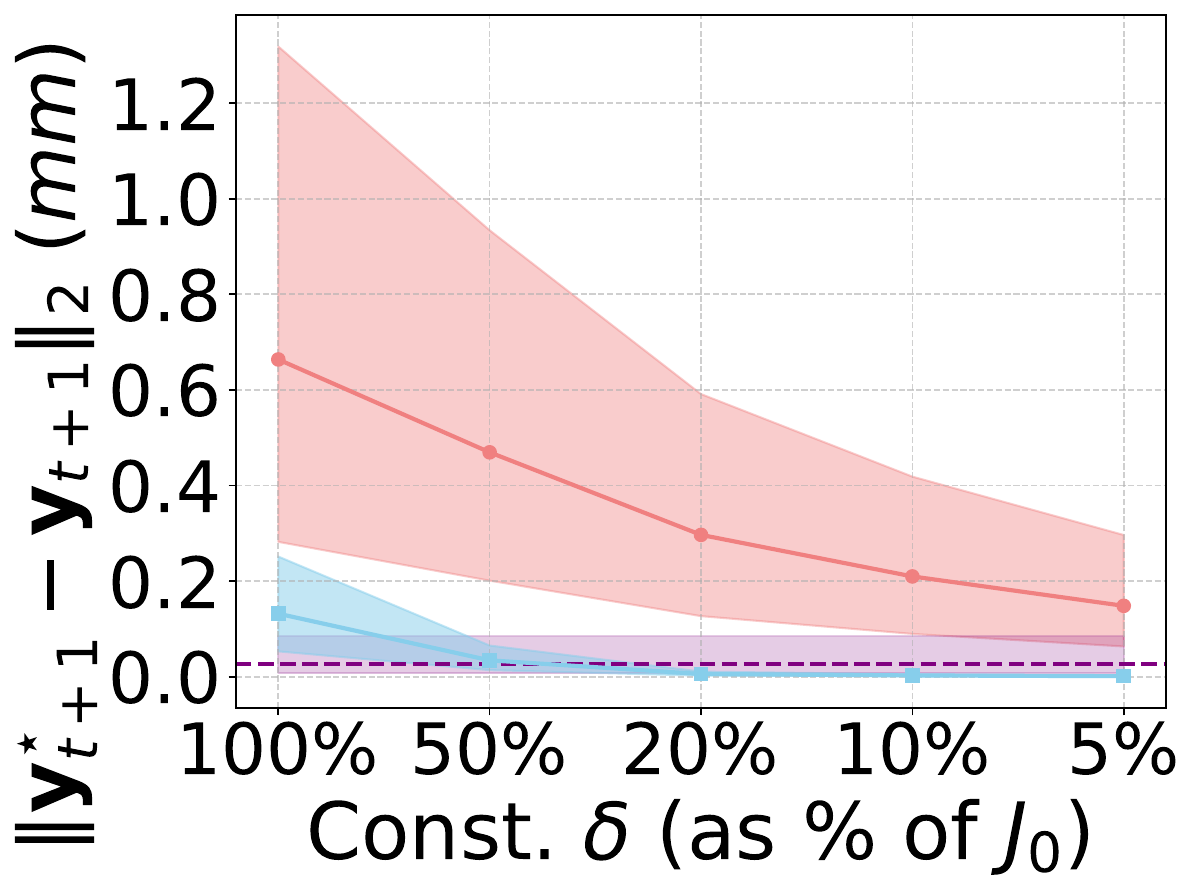}%
        \label{fig:compb}}
    \hfill
    \subfloat[]{\includegraphics[width=.333\linewidth]{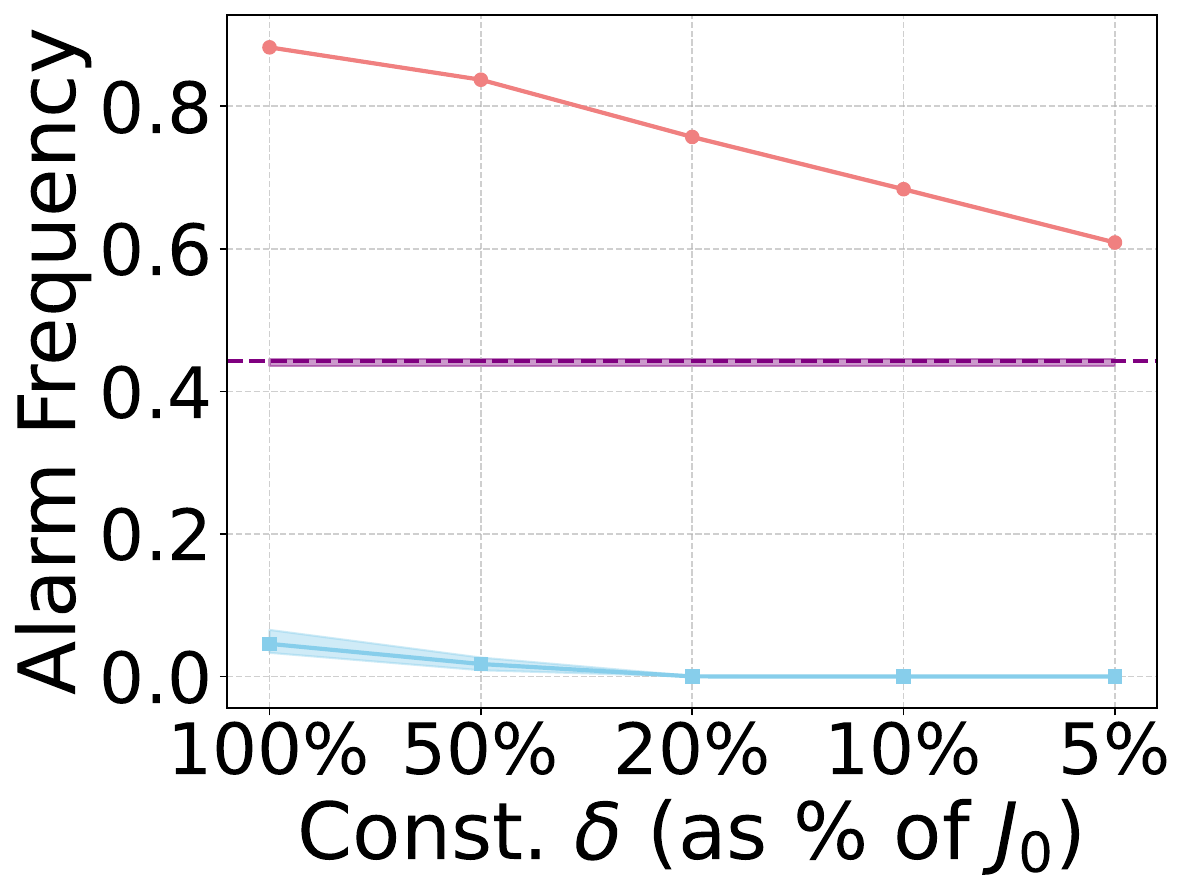}%
        \label{fig:compc}}
    \caption{Stepper-motor DT comparison of LTI-based baselines and DynaMark. Solid lines show the mode and shaded bands the inter-quartile range across trials varying the LQG-loss budget $\delta$ as a percentage of $J_0$.}
    \label{fig:comp}
\end{figure}

\begin{enumerate}
\item For the offline LTI-designed static watermark, tightening $\delta$ reduces watermark energy and control-performance degradation, but the wide inter-quartile ranges indicate sensitivity to the time-varying motor regime. The online LTI baseline operates with smaller nominal cost across all budgets, but DynaMark maintains consistently low watermark energy and nominal performance degradation without relying on a static covariance tuned to an LTI surrogate.

\item The post-onset $\mathrm{AF}$ decreases as $\delta$ tightens for the offline baseline, showing the expected trade-off between nominal cost and detection performance. The online LTI baseline yields near-zero $\mathrm{AF}$ across budgets, suggesting that its converged watermark intensities are too weak for reliable replay detection on the time-varying motor trajectory. DynaMark achieves substantially higher $\mathrm{AF}$ with no dependence on $\delta$, indicating stronger attack sensitivity while remaining performance-aware in nominal operation.
\end{enumerate}

\section{Conclusion}\label{sec:5}
This paper developed \emph{DynaMark}, an RL-based DWM framework to detect replay attacks in industrial MTCs. By casting adaptive watermark design as an MDP, DynaMark learns an adaptive covariance policy from measured data and detector feedback without supplying an analytic plant transition model during policy optimization. The detector and belief-update modules are model-augmented: under Gaussian residuals, theoretical results characterize how input watermarking preserves the nominal $\chi^{2}$ detector structure and yields a tractable generalized $\chi^{2}$ statistic under replay attack, enabling a Bayesian belief update used as a compact detection-confidence state. In addition, under locally contractive nominal closed-loop operation and bounded watermark covariance, we established a mean-square boundedness guarantee for the tracking error, providing a principled safety envelope for learned watermark policies.

Experiments on the Siemens Sinumerik 828D DE, the nonlinear MSD benchmark, and a physical stepper-motor testbed show that DynaMark learns an adaptive covariance policy $U_t$ from measurements and detector beliefs, improving the detection--performance trade-off without relying on a static watermark intensity. In the DE and MSD studies, DynaMark outperforms both static-variance baselines and a non-RL belief-adaptive baseline, showing that reward-optimized covariance adaptation is more reliable under replay than direct belief-based scaling. The MSD case further demonstrates applicability under time-varying and non-Gaussian conditions by replacing the Gaussian closed-form belief computation with offline Monte-Carlo calibration, without invoking the closed-loop boundedness theorem. On the physical stepper-motor, DynaMark transfers from DE to hardware and maintains post-onset alarm activity under mode-switching dynamics, whereas LTI-surrogate baselines can become either too costly or too weak for reliable detection. Future work will integrate DynaMark with online watermark recovery to synthesize authenticated, energy-efficient recovery inputs after detection.

\section*{Acknowledgments}
This work was supported by the U.S. National Science Foundation under Grant 2622141.

\bibliography{references}
\bibliographystyle{IEEEtran}

\begin{IEEEbiography}[{\includegraphics[width=1in,height=1.25in,clip,keepaspectratio]{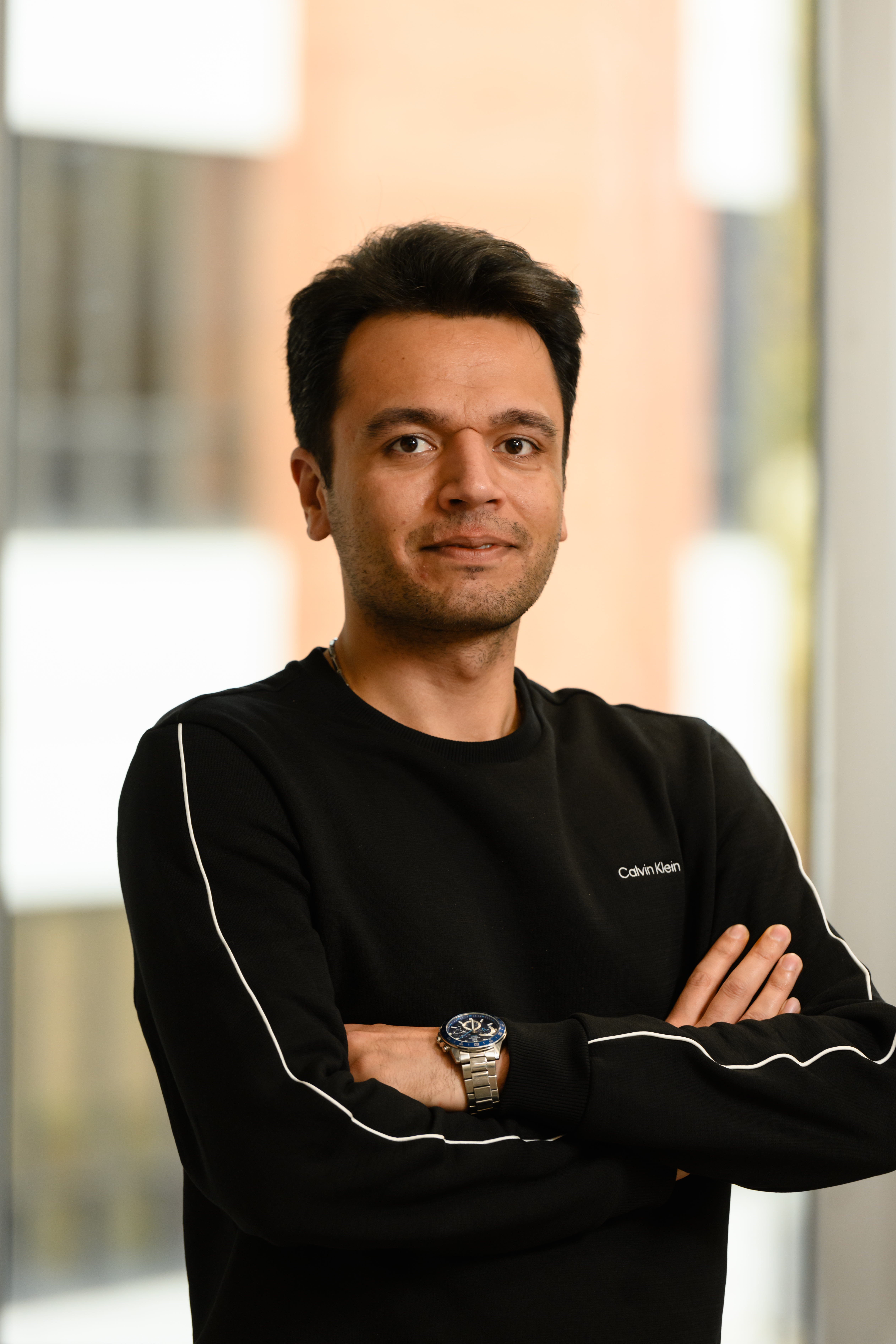}}]{Navid Aftabi}
received his B.Sc. degree in Computer Engineering from University of Tabriz, Tabriz, Iran. He earned an M.Sc. in Industrial Engineering from Sharif University of Technology, Tehran, Iran. He is currently a Ph.D. student in the Department of Industrial \& Systems Engineering at University Washington. His research leverages AI and develops data-driven and model-based methods for cyberattack and anomaly detection, diagnosis, mitigation, and resiliency in complex cyber-physical systems. He is a member of Society of Manufacturing Engineers (SME), Institute for Industrial and Systems Engineers (IISE), and Institute for Operations Research and the Management Sciences (INFORMS).
\end{IEEEbiography}

\begin{IEEEbiography}[{\includegraphics[width=1in,height=1.25in,clip,keepaspectratio]{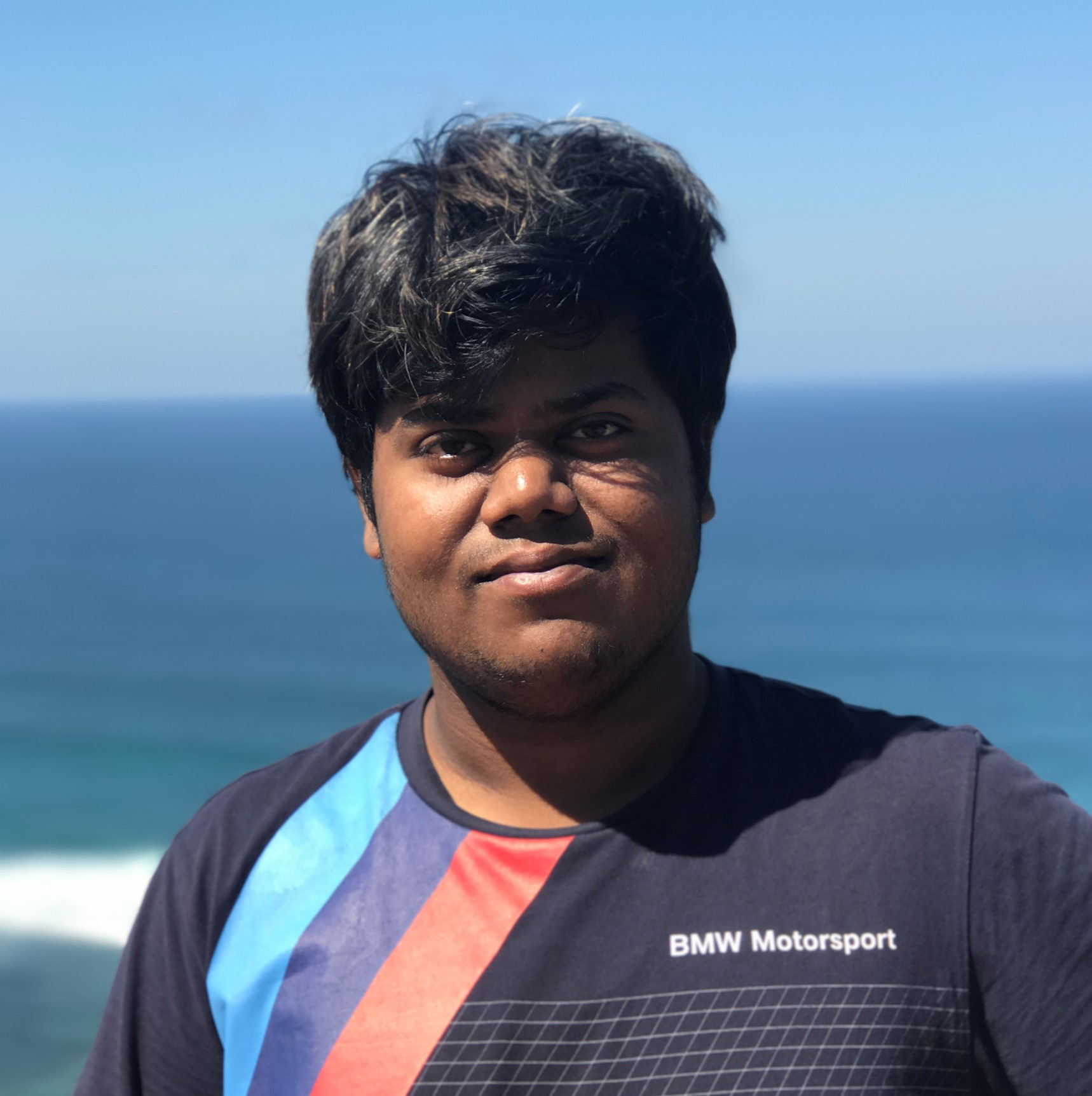}}]{Abhishek Hanchate} (Student Member, IEEE) received his B.Tech. degree in Industrial Engineering from College of Engineering Pune (COEP) Technological University, India, in 2017, wherein he was also an exchange student at Nanyang Technological University (NTU), Singapore. He earned his M.S. degree in Industrial Engineering from Texas A\&M University, TX, USA in 2020. He is currently a Ph.D. student at Texas A\&M University, TX, USA, where he is pursuing a doctorate in Industrial Engineering and an M.S. in Electrical Engineering, specializing in data science and machine learning for smart and secure manufacturing. His research integrates cybersecurity for manufacturing networks and machine tool controllers with dynamic watermarking, digital twins, federated learning, and multimodal data fusion to enable resilient, privacy-preserving, and adaptive industrial systems. He has applied these methods to real-time anomaly detection, process monitoring, and industrial Internet of Things deployments, as well as to computer vision-based industrial inspection, active learning, Bayesian optimization, and recommendation systems. He is a member of Institute of Electrical and Electronics Engineers (IEEE), Society of Manufacturing Engineers (SME), Institute for Industrial and Systems Engineers (IISE), and Institute for Operations Research and the Management Sciences (INFORMS). 
\end{IEEEbiography}

\begin{IEEEbiography}[{\includegraphics[width=1in,height=1.25in,clip,keepaspectratio]{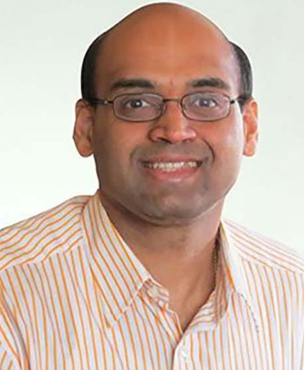}}]{Satish T. S. Bukkapatnam} received the bachelor’s degree from S. V. University, Tirupati, India, and the master’s and Ph.D. degrees from Pennsylvania State University, State College, PA, USA. He has served as an AT\&T Professor with Oklahoma State University and as an Assistant Professor with the University of Southern California. He is currently the Director of Texas A\&M Engineering Experimentation Station (TEES) the Institute for Manufacturing Systems. He also holds an affiliate faculty appointment at Ecole Nationale Superior Arts et Metier (ENSAM), France. He also serves as a Rockwell International Professor with the Department of Industrial and the Systems Engineering Department, Texas A\&M University, College Station, TX, USA. His research addresses the harnessing of high-resolution nonlinear dynamic information, especially from wireless MEMS sensors, to improve the monitoring and prognostics, mainly of ultraprecision and nanomanufacturing processes and machines, and cardiorespiratory processes. His research has led to 151 peer-reviewed publications (87 published/accepted in journals and 64 in conference proceedings), five pending patents,14 completed Ph.D. dissertations, \$5 million in grants as PI/Co-PI from the National Science Foundation, the U.S. Department of Defense, and the private sector, and 17 best-paper/poster recognitions. He is a Fellow of the Institute for Industrial and Systems Engineers (IISE) and the Society of Manufacturing Engineers (SME). He has been recognized with Oklahoma State University regents distinguished research, Halliburton outstanding college of engineering faculty, IISE Boeing technical innovation, IISE Eldin outstanding young industrial engineer, and SME Dougherty outstanding young manufacturing engineer awards. He also serves as an Editor of the IISE Transactions, Design and Manufacturing Focused Issue.
\end{IEEEbiography}

\begin{IEEEbiography}[{\includegraphics[width=1in,height=1.25in,clip,keepaspectratio]{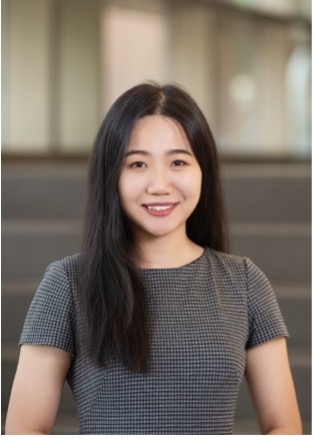}}]{Dan Li} is an Assistant Professor in the Department of Industrial and Systems Engineering at the University of Washington. She received her Ph.D. in Industrial Engineering and her M.S. in Statistics from the Georgia Institute of Technology, and her B.S. in Automotive Engineering from Tsinghua University in Beijing, China. Her research interests lie in developing new data-driven algorithms tailored to enhance the cyber-physical resilience and security of critical infrastructures. Dan is the recipient of the NSF CAREER Award and the IISE Transactions Best Application Paper Award. She has also been recognized in multiple Best Track Paper and Best Student Paper Awards in Energy Systems, DAIS, and QCRE divisions at the IISE Annual Meetings, as well as the INFORMS QSR community.
\end{IEEEbiography}

\clearpage
\setcounter{page}{1}
\appendix
\subsection{Proof of Lemma~\ref{lem:replay_residual_nonlinear}}\label{app:lem1}
By the replay dynamics in Eq.~\eqref{eq:replay_dyn}, the detector forms $\widehat{\vy}'_{t+1}=\mathcal{G}_t(\vy'_t,\vu^\phi_t)$. Therefore, $\vr_{t+1}^A
= \ve'_t + \left(\mathcal{G}_t(\vy'_t,\vu^{\phi'}_t)-\mathcal{G}_t(\vy'_t,\vu^\phi_t)\right)$. Using Eq.~\eqref{eq:lin_sys} in the input argument around $(\widehat{\vy}_t,\Bar{\vu}_t)$ yields
$\mathcal{G}_t(\vy'_t,\vu^{\phi'}_t)-\mathcal{G}_t(\vy'_t,\vu^\phi_t)\approx H_t(\vu^{\phi'}_t-\vu^\phi_t)$.
Substituting Eq.~\eqref{eq:wm_scheme} gives
$\vr_{t+1}^A\approx \ve'_t + H_t(\phi'_t-\phi_t)$.
By distribution of $\ve'_t$, $\phi_t$, and $\phi'_t$, and their independence, the affine Gaussian form implies $\vr_{t+1}^A$ is (approximately) Gaussian with zero mean and covariance
$Q_t + H_t(U'_t+U_t)H_t^\top$.\qed

\subsection{Proof of Theorem~\ref{thm:1}}\label{app:thm1}
Under a replay attack, Lemma~\ref{lem:replay_residual_nonlinear} shows $\vr_{t}^A\sim\N(0, \mathcal{S}_{t})$. We prove the theorem for a generic case where $\vr_{t}^A\sim\N(\mathbf{m}_{t}, \mathcal{S}_{t})$. Using standard normal random variables, we can rewrite $\vr_{t}=\mathbf{m}_{t} + \mathcal{S}_{t}^{1/2} \mathbf{Z}$ where $\mathbf{Z}\sim\N(0,I)$. Then, the test statistics under attack follow
\begin{align*}
    g_{t}^A &= \left(\mathbf{Z} + \mathcal{S}_{t}^{-1/2} \mathbf{m}_{t}\right)^\top  \mathcal{S}_{t}^{1/2} Q_{t-1}^{-1} \mathcal{S}_{t}^{1/2} \left(\mathbf{Z} + \mathcal{S}_{t}^{-1/2} \mathbf{m}_{t}\right) .
\end{align*}
Applying the spectral theorem, $\mathcal{S}_{t}^{1/2} Q_{t-1}^{-1} \mathcal{S}_{t}^{1/2} = P_{t}^\top  \Lambda_{t}P_{t}$ where $P_{t}$ is an orthogonal matrix, i.e., $P_{t}^\top  P_{t}= P_{t}P_{t}^\top =I$; and, $\Lambda_{t}$ is a diagonal matrix with positive diagonal elements. Let $\mathbf{Y}_{t} = P_{t}\mathbf{Z}$ and $\mathbf{b}_{t}=P_{t}\mathcal{S}_{t}^{-1/2} \mathbf{m}_{t}$. Since $\mathbf{Z}$ is a multivariate standard normal distribution, $\mathbf{Y}_{t}$ also follows the same distribution. Then,
\begin{align*}
    g_{t}^A =& (\mathbf{Y}_{t} + \mathbf{b}_{t})^\top  \Lambda_{t} (\mathbf{Y}_{t} + \mathbf{b}_{t}) \\
    =& \sum_{i=1}^{n} \Lambda_{t}(i) \left(\mathbf{Y}_{t}(i) + \mathbf{b}_{t}(i)\right)^{2},
\end{align*}
that implies, test statistics under replay follow the generalized $\chi^{2}$ distribution (weighted sum of noncentral chi-square variables) $g_{t}^A\sim\Tilde{\chi}(\omega_{t},\kappa, \lambda_{t}, s, m)$
where $s=0$, $m=0$, $\kappa =\Vec{\pmb{1}}$, $\omega_{t} =\left(\Lambda_{t}(1),\dots,\Lambda_{t}(n)\right)^\top$, and $\lambda_{t}= \mathbf{b}_{t} $.  Moreover, by Lemma~\ref{lem:replay_residual_nonlinear}, $\mathbf{m}_{t}=\mathbf{0}$ (hence $\mathbf{b}_{t}=\mathbf{0}$), so $g_{t}^A$ is a central generalized $\chi^2$ (a weighted sum of independent $\chi^2$ variables).\qed

\subsection{Proof of Theorem~\ref{thm:2}}\label{app:thm2}

By the definition of the random variables $\sigma$ and $I_t$, the detector's belief at time $t$ is defined as  $d_t = \Pr(\sigma=1|I_{1:t})$. Observing $\vy_{t}$ and $I_t$, and applying the Bayesian rule,
\begin{align*}
    p(\sigma | I_{1:t}) &= p(\sigma|I_{1:t-1}) \frac{p(I_{t}|I_{1:t-1},\sigma)}{p(I_{t}|I_{1:t-1})}.
\end{align*}
Conditioning on $\sigma$, the conditional evidence is computed as
\begin{align*}
    p(I_{t}|I_{1:t-1}) =& \sum_{k\in\{0,1\}} p(I_{t},\sigma=k|I_{1:t-1}) \\
    =& \sum_{k\in\{0,1\}} p(I_{t}|I_{1:t-1},\sigma=k) \Pr(\sigma=k|I_{1:t-1}).
\end{align*}
Under our assumptions, $\ve_t \perp \ve_s$ for $t\ne s$. When $\sigma=0$, we have $\vr_{t+1}=\ve_t$ that implies $\vr_t \perp \vr_s$ for $t\ne s$ given $\sigma=0$; this implies $g_t\perp g_s$, and, accordingly, $I_t\perp I_s$ for $t\ne s$ given $\sigma=0$. Therefore, conditional independence under the no-attack hypothesis, the conditional PDF simplifies to $p(I_{t}|I_{1:t-1},\sigma=0) = p(I_{t}|\sigma=0).$ Therefore, the  probability of raising an alarm depend only on the Type-I error, i.e., $I_{t}|\sigma=0\sim \text{Bernoulli}(\alpha)$.

Define the random variable $\tau\in\R_{>0}$ as the attack onset. Condition on whether the attack has started by time $t$:
\begin{align}
    p(I_{t} &\mid I_{1:t-1},\sigma=1) = \notag\\ 
     &p(I_{t}\mid I_{1:t-1},\sigma=1, \tau > t) \ p(\tau > t \mid I_{1:t-1},\sigma=1) \notag\\
    &+ p(I_{t}\mid I_{1:t-1},\sigma=1, \tau \leq t) \ p(\tau \leq t \mid I_{1:t-1},\sigma=1) \label{eq:prob1}
\end{align}
For $\tau > t$, since attack has not started yet, the residuals follow the nominal ones. Therefore, by $\vr_t \perp \vr_s$, and, accordingly, $I_t\perp I_s$ for $t\ne s$ given $\sigma=1$ and $\tau > t$, it implies $p(I_{t}\mid I_{1:t-1},\sigma=1, \tau > t) = p(I_{t}|\sigma=0)$.
When $\tau \leq t$, the attack has started and residuals follow the Gaussian distribution in Lemma~\ref{lem:replay_residual_nonlinear}. Therefore, by Theorem~\ref{thm:1}, definition of $\vr_t^A$, $\ve_t \perp \ve_s$ and $\phi_t \perp \phi_s$ for $t \ne s$, and $\ve_t \perp \phi_t$ for all $t$, $p(I_{t}\mid I_{1:t-1},\sigma=1, \tau \leq t) = p(I_t \mid \sigma=1)$. Therefore, depending on $\tau$, probability of raising an alarm depends on both Type-I and Type-II errors, i.e.,
\[
I_{t}|\sigma=1,\tau=k \sim 
\begin{cases}
    \text{Ber}(\alpha) & \text{if }  t < k\\
    \text{Ber}(1 - \beta_{t}) & \text{if } t \geq k 
\end{cases}.
\]
Substituting these results in Eq.~\eqref{eq:prob1} yields
\begin{align}
    p(I_{t} &\mid I_{1:t-1},\sigma=1) = \notag\\ 
     &\alpha^{I_{t}} (1 - \alpha)^{1-I_{t}} \ p(\tau > t \mid I_{1:t-1},\sigma=1) \notag\\
    &+ (1 - \beta_{t})^{I_{t}} \beta_{t}^{1-I_{t}} \ p(\tau \leq t \mid I_{1:t-1},\sigma=1),
\end{align}
that yields the conditional evidence $p(I_{t}|I_{1:t-1})$ and $d_t$. Before the attack onset, the alarm sequence $I_{1:t}$ contains only false alarms; after the attack onset, it no longer affects the onset time. Thus, the attack onset time is conditionally independent of the alarm sequence given $\sigma = 1$, i.e., $p(\tau \leq t \mid I_{1:t-1}, \sigma = 1) = p(\tau \leq t \mid \sigma = 1).$

Assuming a prior distribution on the random variable $\sigma$, which by its definition is a Bernoulli distribution with some parameter $q$, and the random variable $\tau|\sigma=1$, the detector's belief updates at each time using the relation in Eq.~\eqref{eq:dt}.\qed

\subsection{Proof of Lemma~\ref{lemma_beta}}\label{app:lem2}
Note that $\alpha=\Pr(I_t=1|\sigma=0)$ and $\beta_t=\Pr(I_t=0|\sigma=1)$. It is assumed that the Type-I error is time-invariant and can be controlled by carefully selecting the test statistic threshold, $\Tilde{g}$. However, an operator could choose to vary $\Tilde{g}$ over time, which would make the Type-I error time-dependent. The Type-II error, on the other hand, can vary depending on the occurrence of a replay attack. Let the random variable $\tau\in\R_{>0}$ be the attack onset. Given $\sigma=1$, if $\tau \leq t$, then by Theorem~\ref{thm:1}, the probability of not raising an alarm depends on $\Tilde{\chi}(\omega_{t},\kappa, \lambda_{t}, s, m)$. If $\tau > t$, the probability of not raising an alarm is based on the $\chi^2$ distribution. By law of total probability, this relationship is given by
\begin{align*}
    \beta_t =& \Pr(I_t=0|\sigma=1) \\
    =&\Pr(I_t=0, \tau\leq t|\sigma=1) + \Pr(I_t=0, \tau > t|\sigma=1) \\
    =& \Pr(g_{t} < \Tilde{g} | \sigma = 1, \tau\leq t) \Pr(\tau\leq t | \sigma = 1) \\
    &+ \Pr(g_{t} < \Tilde{g}) \Pr(\tau > t | \sigma = 1) \\
    =& \mathcal{H}_{g_t^A}(\Tilde{g}) \Pr(\tau\leq t | \sigma = 1) + \mathcal{H}_{g_t}(\Tilde{g}) \Pr(\tau > t | \sigma = 1),
\end{align*}
where $\mathcal{H}_{g_t^A}$ is the CDF of the (central) generalized $\chi^2$ distribution, and $\mathcal{H}_{g_t}$ is the CDF of the $\chi^2$ distribution.\qed

\subsection{Estimating \texorpdfstring{$\beta_t$} \; under non-Gaussian noise}
\label{app:beta_calibration}
Although a Gaussian distribution was assumed, it remains feasible to obtain an estimate of $\beta_t$. One possible approach is to conduct numerical simulations, provided that a sufficiently accurate digital twin of the system is available. Procedure~\ref{proc:beta_calibration} estimates the lookup table $\{\hat{\beta}_t(U)\}$ offline for a system under non-Gaussian noise. The detector threshold $\Tilde{g}$ is first calibrated from nominal rollouts to achieve the target false-alarm level $\alpha$. Then, for each candidate watermark variance $U$ and replay onset $\tau$, replay-attack simulations are performed to estimate the probability of no alarm at each time step. Averaging over the chosen onset-time distribution yields $\hat{\beta}_t(U)$, which replaces $\beta_t$ in the online belief update.

During online learning, if the selected action $U_t$ coincides with a grid point in $\mathcal{U}$, then $\hat{\beta}_t(U_t)$ is read directly from the lookup table. Otherwise, if $U^{(k)} \leq U_t \leq U^{(k+1)}$, we use piecewise linear interpolation:
\begin{align*}
    \hat{\beta}_t(U_t)
    =&
    \frac{U^{(k+1)}-U_t}{U^{(k+1)}-U^{(k)}} \, \hat{\beta}_t\!\left(U^{(k)}\right) \\
    &+
    \frac{U_t-U^{(k)}}{U^{(k+1)}-U^{(k)}} \, \hat{\beta}_t\!\left(U^{(k+1)}\right).
\end{align*}
The interpolated value $\hat{\beta}_t(U_t)$ is then substituted into $\kappa_{1,t}$ in Theorem~\ref{thm:2}.

If the watermark action is scalar, or the covariance matrix is parameterized by a scalar design variable, then $\hat{\beta}_t(U_t)$ can be computed by piecewise linear interpolation over the calibrated action grid. For matrix-valued actions, the same idea applies after replacing $U_t$ with its low-dimensional parameterization; otherwise, multidimensional interpolation would be required.

\begin{algorithm}[t]
\caption{Offline calibration of $\hat{\beta}_t(U)$ under non-Gaussian noise}
\label{proc:beta_calibration}
\begin{algorithmic}[1]
\Require horizon $T$, action grid $\mathcal U=\{U^{(1)},\dots,U^{(K)}\}$, onset grid $\mathcal T=\{\tau_1,\dots,\tau_L\}$, number of nominal rollouts $M_0$, number of attack rollouts $M$, target false-alarm level $\alpha$, onset weights $\{w_\tau\}_{\tau\in\mathcal T}$, matrix $Q$
\Ensure threshold $\tilde g$ and lookup table $\{\hat{\beta}_t(U^{(k)})\}_{t=1,\dots,T}^{k=1,\dots,K}$

\State \textbf{Nominal calibration:}
\For{$m=1$ to $M_0$}
    \State Simulate one nominal rollout with the chosen non-Gaussian noise model for $T$ time-step
    \State Compute $\vr_t$ and $g_t^{(m)} \gets \vr_t^\top Q^{-1} \vr_t$
\EndFor
\State Set $\tilde g$ as the empirical $(1-\alpha)$-quantile of the pooled nominal statistics $\{g_t^{(m)}\}$

\State \textbf{Attack-side calibration:}
\ForAll{$U \in \mathcal U$}
    \State Simulate one nominal rollout under watermark $\mathcal{N}(0,U)$ and store the trajectory for replay
    \ForAll{$\tau \in \mathcal T$}
        \For{$m=1$ to $M$}
            \State Simulate replay attack with onset $\tau$ under the same watermark variance $U$
            \State Compute $\vr_t^{A,(m)}$ 
            \State $g_t^{A,(m)} \gets (\vr_t^{A,(m)})^\top Q^{-1} \vr_t^{A,(m)}$
            \State Set $I_t^{(m)} \gets \mathbf{1}\{g_t^{A,(m)} > \tilde g\}$
        \EndFor
        \For{$t=1$ to $T$}
            \State Estimate $\hat{\beta}_t(U,\tau) \gets \frac{1}{M}\sum_{m=1}^M \mathbf{1}\{I_t^{(m)}=0\}$
        \EndFor
    \EndFor
    \For{$t=1$ to $T$} \Comment{average over onset times}
        \State $\hat{\beta}_t(U) \gets \sum\nolimits_{\tau \in \mathcal T} w_\tau \hat{\beta}_t(U,\tau)$
    \EndFor
\EndFor

\State \Return $\tilde g$ and $\hat{\beta}_t(U)$ for all $t=1,\dots,T$ and $U \in \mathcal U$
\end{algorithmic}
\end{algorithm}

\subsection{Proof of Theorem~\ref{thm:stability}}\label{app:thm3}
Under the assumptions in \S~\ref{sec:stability}, let $\vy^\star_t$ be the nominal trajectory and $\vu^\star_t = f_c(\vh^\star_t;\eta)$ the ideal control input that keeps the system on this path, i.e., $\vy^\star_{t+1} = \mathcal{G}_t(\vy^\star_t, \vu^\star_t) + \ve^\star_t$. Define the tracking error $\xi_t=\vy_t - \vy^\star_t$, and, the control deviation $\Delta\vu_t = \vu_t - \vu^\star_t$. In a neighborhood of $(\vy^\star_t,\vu^\star_t)$, a first-order Taylor expansion gives
\begin{equation}\label{eq:taylor_g}
    \mathcal{G}_t(\vy,\vu)\approx \mathcal{G}_t(\vy^\star_t,\vu^\star_t) + G_t^\star(\vy-\vy^\star_t) + H_t^\star(\vu-\vu^\star_t),
\end{equation}
where higher-order terms are negligible in the operating region of interest, $G_t^\star:=\nabla_y \mathcal{G}_t(\vy^\star_t,\vu^\star_t)$ and $H_t^\star:=\nabla_u \mathcal{G}_t(\vy^\star_t,\vu^\star_t)$. By differentiability of $\mathcal{G}_t$ on a compact neighborhood of $(\vy^\star_t,\vu^\star_t)$ for all $t$, there exists $\Bar{H}>0$ such that $\sup_{t\ge0} \|H_t^\star\|^2\leq\Bar{H}$, for all $(\vy,\vu)$ in this neighborhood. By differentiability of the controller in a neighborhood of $\vh_t^\star$, we obtain the linear approximation
$f_c(\vh;\eta) \approx f_c(\vh_t^\star;\eta) + K_t(\vh - \vh_t^\star)$, where $K_t = [K_{t,0},\dots,K_{t,w}] \in \R^{c\times n(w+1)}$ and $K_{t,i} = \nabla_{\vy_{t-i}} f_c(\vh_t^\star;\eta)$ for $i=0,\dots,w$. This yields $\vh - \vh_t^\star = \left[\xi_t, \xi_{t-1}, \dots, \xi_{t-w}\right]^\top$ and $\Delta\vu_t \approx \sum_{i=0}^{w} K_{t,i} \xi_{t-i}$. Since the nominal histories lie in a compact set, then $\sup_{t>0}\|K_{t,i} \|^2 \leq \Bar{K}_i$ for all $i=0,\dots,w$.

Then, using Eq.~\eqref{eq:taylor_g}, locally around $(\vy^\star_t,\vu^\star_t)$, we have $\xi_{t+1} \approx A_t \xi_t + d_t,$ where $A_t = G_t^\star + H_t^\star K_{t,0}$, $d_t = H_t^\star \sum_{i=1}^{w} K_{t,i} \xi_{t-i} + H_t^\star \phi_t + \varrho_t$, and $\varrho_t=\ve_t - \ve^\star_t + \text{unmodeled noises}$. For any $\epsilon>0$, Young’s inequality gives:
\begin{equation}\label{eq:ineq1}
    \|\xi_{t+1}\|^2 \leq (1+\epsilon) \|A_t \xi_t\|^2 + \left(1+\epsilon^{-1}\right) \|d_t\|^2.
\end{equation}

By Cauchy–Schwarz (CS) and boundedness of Jacobians, $\|d_t\|^2 \leq 3 \left(w\Bar{H} \sum_{i=1}^{w} \Bar{K}_i \|\xi_{t-i}\|^2 + \Bar{H} \|\phi_t\|^2 + \|\varrho_t\|^2\right)$ that depends on $\{\xi_\tau\}_{\tau=t-w}^{t}$. Define
\begin{equation}\label{eq:lyp}
    V_t = \sum\nolimits_{i=0}^{w} \rho^i \ \|\xi_{t-i}\|^2 , \qquad \rho\in(0,1).
\end{equation}
By Eq.~\eqref{eq:lyp}, $\|\xi_{t-i}\|^2 \leq \rho^{-i} V_t$. Hence, $\sum_{i=1}^{w} \Bar{K}_i\|\xi_{t-i}\|^2 \leq V_t \sum_{i=1}^{w} \Bar{K}_i \rho^{-i}.$ Moreover, with $\Bar{K}=\max_{i=1,\dots,w}\Bar{K}_i$ and $\rho\in(0,1)$, $\sum_{i=1}^{w} \Bar{K}_i \rho^{-i} \leq w\Bar{K} \rho^{-w}$. Combining together,
\begin{equation}\label{eq:ineq3}
    \|d_t\|^2 \leq 3 \left(w^2 \Bar{H} \Bar{K} \rho^{-w} V_t + \Bar{H} \|\phi_t\|^2 + \|\varrho_t\|^2\right).
\end{equation}

By CS, $\|A_t \xi_t\|^2 \leq \|A_t \|^2 \ \|\xi_t\|^2$, and by Eq.~\eqref{eq:lyp}, $\|\xi_t\|^2 \leq V_t$. Then, $\|A_t \xi_t\|^2 \leq \|A_t \|^2 V_t$. Together with~\eqref{eq:ineq1} and~\eqref{eq:ineq3}
\begin{align}\label{eq:ineq4}
    \|\xi_{t+1}\|^2 \leq& \left((1+\epsilon) \|A_t \|^2 + 3\left(1+\epsilon^{-1}\right)w^2 \rho^{-w} \Bar{H} \Bar{K} \right)V_t \notag \\
    &+ 3\left(1+\epsilon^{-1}\right) \left(\Bar{H} \|\phi_t\|^2 + \|\varrho_t\|^2\right).
\end{align}

By Eq.~\eqref{eq:lyp},
\begin{align}
    V_{t+1} =\|\xi_{t+1}\|^2  + \rho \sum_{i=0}^{w-1} \rho^{i} \ \|\xi_{t-i}\|^2\leq \|\xi_{t+1}\|^2  + \rho V_t.\label{eq:ineqv}
\end{align}
Adding $\rho V_t$ to both sides of~\eqref{eq:ineq4}, using~\eqref{eq:ineqv}, and taking the expectation yield
\begin{align*}
    \E[V_{t+1}] \leq& \; \left((1+\epsilon) \|A_t \|^2 + \rho + 3\left(1+\epsilon^{-1}\right)w^2 \rho^{-w} \Bar{H} \Bar{K} \right) \E[V_t] \\
    &+ 3\left(1+\epsilon^{-1}\right) \left(\Bar{H} \E\left[\|\phi_t\|^2\right] + \E\left[\|\varrho_t\|^2\right]\right).
\end{align*}
As $\phi_t\sim\N(0,U_t)$ and $U_t\in\mathcal{A}$, by CS and ~\eqref{eq:action_space}, we have that $\E\left[\|\phi_t\|^2\right] = \tr(U_t) = \left\langle U_t,I\right\rangle \leq \|U_t\|_F \|I\|_F \leq U_{\max}\sqrt{c}$. Also, assume $\E\left[\|\varrho_t\|^2\right] \leq \sigma^2_\varrho$. Altogether,
\begin{align*}
    \E[V_{t+1}] \leq& \left((1+\epsilon) \|A_t \|^2 + \rho + 3\left(1+\epsilon^{-1}\right)w^2 \rho^{-w} \Bar{H} \Bar{K} \right) \E[V_t] \\
    &+ 3\left(1+\epsilon^{-1}\right) \left(\Bar{H} U_{\max}\sqrt{c} + \sigma^2_\varrho\right).
\end{align*}
Let $\Bar{A}=\sup_{t\ge0} \|A_t \|$. Then, 
\begin{equation}\label{eq:ineq6}
    \E[V_{t+1}] \le \nu(\epsilon,\rho)\E[V_t] + \vartheta(\epsilon, U_{\max}),
\end{equation}
where
\begin{align*}
    \nu(\epsilon,\rho) &= (1+\epsilon) \Bar{A}^2 + \rho + 3\left(1+\epsilon^{-1}\right)w^2 \rho^{-w} \Bar{H} \Bar{K}, \\
    \vartheta(\epsilon, U_{\max}) &= 3\left(1+\epsilon^{-1}\right) \left(\Bar{H} U_{\max}\sqrt{c} + \sigma^2_\varrho\right).
\end{align*}
To ensure~\eqref{eq:ineq6} is a contraction, it must $\nu(\epsilon,\rho)<1$.

Assume the nominal closed loop is locally contractive, i.e., $\Bar{A}<1$. Then, there exists $\epsilon>0$ such that $(1+\epsilon)\Bar{A}^2<1$. Define the residual margin $\delta_\epsilon = 1-(1+\epsilon)\Bar{A}^2>0$ and let
\begin{equation}\label{eq:c_epsilon}
    c_\epsilon=3(1+\epsilon^{-1})w^2 \Bar{H}\Bar{K}.
\end{equation}
A sufficient condition for $\nu(\epsilon,\rho)<1$ is the existence of $\rho\in(0,1)$ with $g_\epsilon(\rho)=\rho + c_\epsilon \rho^{-w} < \delta_\epsilon$. $g_\epsilon$ is strictly convex on $(0,1)$ and has a unique minimizer $\rho_\epsilon^\star$ (if $\rho_\epsilon^\star\in(0,1)$). Thus, it suffices that some $\rho_\epsilon^\star\in\arg\min_{\rho\in(0,1)} g_\epsilon(\rho)$ satisfies $g_\epsilon(\rho_\epsilon^\star)<\delta_\epsilon$. Differentiating $g_\epsilon$ and setting the derivative to zero gives 
\begin{equation}\label{eq:rho_g_star}
    \rho_\epsilon^\star = (w c_\epsilon)^{\frac{1}{w+1}}, \qquad g_\epsilon(\rho_\epsilon^\star) = (1+w^{-1})\rho_\epsilon^\star.
\end{equation}
Note that, by definition of $\delta_\epsilon$, $g_\epsilon(\rho_\epsilon^\star)<\delta_\epsilon$ implies $\rho_\epsilon^\star\in(0,1)$. Therefore, we have the following conditions for $\nu(\epsilon,\rho)<1$:
\begin{itemize}[leftmargin=10pt]
    \item \textit{Existence of $\epsilon>0$:} A necessary and sufficient condition is $(1+\epsilon)\Bar{A}^2 < 1$ for some $\epsilon>0$, which holds iff $\Bar{A}<1$ (e.g., any $\epsilon\in(0,\Bar{A}^{-2}-1)$ is admissible). Without this, $\nu(\epsilon,\rho)<1$ is not guaranteed.
    \item \textit{Existence of $\rho\in(0,1)$ for a given $\epsilon>0$:} It suffices that $g_\epsilon(\rho_\epsilon^\star)<\delta_\epsilon$. 
\end{itemize} 
For the latter condition, let $\epsilon_\alpha=\alpha(\Bar{A}^{-2} - 1)$ be admissible, with $\alpha \in (0,1)$ and $\Bar{A} < 1$. Substituting $\epsilon_\alpha$ into Eq.~\eqref{eq:c_epsilon} yields the corresponding coefficient $c_{\epsilon_\alpha}$. In turn, substituting $c_{\epsilon_\alpha}$ into the expression for $\rho_{\epsilon}^\star$ in Eq.~\eqref{eq:rho_g_star} leads to
\[
\rho_{\epsilon_\alpha}^\star
= \left(3 w^3 \Bar{H} \Bar{K} \,\frac{\alpha + (1-\alpha)\Bar{A}^2}{\alpha(1 - \Bar{A}^2)}\right)^{\frac{1}{w+1}}.
\]
By $\alpha + (1-\alpha)\Bar{A}^2 < 1$, it follows that
\[
\rho_{\epsilon_\alpha}^\star \;<\; \left(\frac{3w^3 \Bar{H} \Bar{K}}{\alpha\bigl(1 - \Bar{A}^2\bigr)}\right)^{\frac{1}{w+1}}.
\] 
Therefore, by the expression for $g_\epsilon(\rho_\epsilon^\star)$ in Eq.~\eqref{eq:rho_g_star} and with some $\alpha\in(0,1)$, if 
\begin{equation}\label{eq:ineq7}
    \left(1+w^{-1}\right)\left(\frac{3w^3 \Bar{H} \Bar{K}}{\alpha(1 - \Bar{A}^2)}\right)^{\frac{1}{w+1}} < \delta_{\epsilon_\alpha},
\end{equation}
then defined $\epsilon$ satisfies $g_{\epsilon}(\rho_{\epsilon_{\alpha}}^\star) < \delta_{\epsilon_\alpha}$ that implies $\rho$ exists, and $\rho=\rho^\star_\epsilon$. Substituting $\epsilon_\alpha$ in the expression for $\delta_{\epsilon_\alpha}$ yields $\delta_{\epsilon_\alpha}=(1-\alpha)(1-\Bar{A}^2)$. In turn, substituting $\delta_{\epsilon_\alpha}$ in~\eqref{eq:ineq7}, the inequality~\eqref{eq:ineq7} holds if 
\begin{subequations}\label{eq:ineq9}
\begin{align}
    &\Bar{A} < \sqrt{1 - \left(\frac{3w^3 \left(1+w^{-1}\right)^{w+1} \Bar{H} \Bar{K}}{\alpha(1-\alpha)^{w+1}}\right)^{\frac{1}{w+2}}}, \label{eq:ineq9a}  \\
    &3w^3 \left(1+w^{-1}\right)^{w+1} \Bar{H} \Bar{K} < \alpha(1-\alpha)^{w+1}.
\end{align}
\end{subequations}
Define $f_\epsilon(\alpha)=\alpha(1-\alpha)^{w+1}$. The largest $\alpha$ for which~\eqref{eq:ineq9} holds is the maximizer of $f_\epsilon$. $f_\epsilon$ is concave on $(0,1)$, and differentiating and setting the derivative to zero gives $\alpha^\star=\frac{1}{w+2}$, with $f_\epsilon(\alpha^\star)=\frac{1}{w+2}\left(\frac{w+1}{w+2}\right)^{w+1}$. Substituting $f_\epsilon(\alpha^\star)$ in~\eqref{eq:ineq9}, the necessary and sufficient conditions for the existence of $\epsilon>0$ and $\rho\in(0,1)$ are (since~\eqref{eq:ineq9a} is stricter than $\bar{A}<1$):
\begin{subequations}\label{eq:ineq11}
\begin{align}
    \Bar{A} &< \sqrt{1 - \left(\frac{3 (w+2)^{w+2} \Bar{H} \Bar{K}}{w^{w-2}}\right)^{\frac{1}{w+2}}},  \\
    \Bar{H} \Bar{K} &< \frac{w^{w-2}}{3(w+2)^{w+2}}.
\end{align}
\end{subequations}

Thus, under the conditions in~\eqref{eq:ineq11} with $\nu(\epsilon,\rho)<1$,~\eqref{eq:ineq6} yields a linear recursion: $\E[V_{t}] \le \nu(\epsilon,\rho)^t V_0 + \left[\vartheta(\epsilon,U_{\max})/(1-\nu(\epsilon,\rho))\right]$, and, by $\|\xi_{t}\|^2 \leq V_t$, it yields $\E\|\xi_{t}\|^2 \le \nu(\epsilon,\rho)^t V_0 + \left[\vartheta(\epsilon,U_{\max})/(1-\nu(\epsilon,\rho))\right]$. Therefore,
\[
\limsup_{t\to\infty}\E\|\xi_{t}\|^2 \leq \frac{\vartheta(\epsilon,U_{\max})}{1-\nu(\epsilon,\rho)}. \tag*{\qed}
\]

\subsubsection{Theorem~\ref{thm:stability} Verification}\label{app:thm3_verification}

Given the controller history window $w$, verify Theorem~\ref{thm:stability} as follows:
\begin{enumerate}
    \item Compute $\Bar{H}$, $\Bar{K}$, and $\Bar{A}$ from the local closed-loop model.

    \item Verify $\E\|\varrho_t\|^2\le \sigma_\varrho^2<\infty$.

    \item Check conditions $(C_1)$ and $(C_2)$. If either condition fails, the existence of admissible $(\epsilon,\rho)$ is not guaranteed by Theorem~\ref{thm:stability}.

    \item Choose $\epsilon\in(0,\Bar{A}^{-2}-1)$ and compute $\delta_\epsilon$ and $c_\epsilon$.

    \item If $c_\epsilon>0$, compute $\rho_\epsilon^\star$ and set $\rho=\rho_\epsilon^\star$. Then verify $g_\epsilon(\rho_\epsilon^\star)<\delta_\epsilon$ and $\nu(\epsilon,\rho)<1$.

    \item If $c_\epsilon=0$, choose any $\rho\in(0,\delta_\epsilon)$ that gives $\nu(\epsilon,\rho)<1$.

    \item Specify $U_{\max}$, compute $\vartheta(\epsilon,U_{\max})$, and report the resulting mean-square bound in~\eqref{eq:stability_proof}.
\end{enumerate}

\subsection{Implementation Specifications}\label{app:ddpg}

\begin{table*}[!t]
\centering
\caption{Implementation details for training DDPG-CDQ.}
\label{tab:ddpg_specs}
\resizebox{\linewidth}{!}{
\begin{threeparttable}
\begin{tabular}{llll}
\toprule
\textbf{Setting} & \textbf{Case study 1 (Section~\ref{sec:numerical})} & \textbf{Case study 2 (Section\ref{sec:msd}}) & \textbf{Physical testbed (Section~\ref{sec:case})} \\ \midrule
\textbf{DDPG} & & & \\
Actor/critic architecture & 3 FC\tnote{*} \ layers ($32$ neurons each) & 3 FC layers ($64$ neurons each) & 3 FC layers ($64$ neurons each) \\
Activation \& normalization & Leaky-ReLU + layer norm & Leaky-ReLU + layer norm & Leaky-ReLU + layer norm \\
Number of critics & 2 (clipped double-Q) & 2 (clipped double-Q) & 2 (clipped double-Q) \\
Optimizer \& learning rate & RMSprop, $1\times10^{-3}$ & RMSprop, $1\times10^{-3}$ & RMSprop, $1\times10^{-3}$ \\
Gradient clipping & $\Vert \nabla\Vert_2 \le 1$ & $\Vert \nabla\Vert_2 \le 1$ & $\Vert \nabla\Vert_2 \le 1$ \\
Target-network update & $\tau=5\times10^{-3}$ (soft) & $\tau=5\times10^{-3}$ (soft) & $\tau=5\times10^{-3}$ (soft) \\
Replay buffer size & $10^{6}$ transitions & $10^{6}$ transitions & $10^{8}$ transitions \\
Mini-batch size & $512$ & $512$ & $512$ \\
Exploration noise & OU($\mu{=}0$, $\sigma_0{=}0.995$, $\theta{=}0.15$) & OU($\mu{=}0$, $\sigma_0{=}1.95$, $\theta{=}0.15$) & OU($\mu{=}0$, $\sigma_0{=}0.995$, $\theta{=}0.15$) \\
Noise decay & $\sigma \leftarrow 0.995\,\sigma$ & $\sigma \leftarrow 0.995\,\sigma$ & $\sigma \leftarrow 0.995\,\sigma$ \\
Target update frequency & $1$ & $1$ & $100$ \\
$U_{\max}$ & $1$ & $2$ & $1$ \\
\textbf{Training} & & & \\
Training episodes & $200$ & $50$ & $200$ \\
Max. steps per episode & $10^{3}$ env. steps & $4\times10^{3}$ env. steps & $4.1\times10^{4}$ fast-time steps \\
Eval. policy & every 10 ep. for 10 ep. & every 5 ep. for 5 ep. & every 5 ep. for 5 ep. \\ 
\textbf{Environment} & & & \\
Reward weights & $\omega_1=\omega_2=0.35,\omega_3=0.30$ & $\omega_1=0.1,\omega_2=0.35,\omega_3=0.55$ & $\omega_1=\omega_2=0.35,\omega_3=0.30$ \\
Replay attack prior & $\sigma\sim\mathrm{Ber}(0.05), \tau|\sigma\sim\mathrm{Geom}(1/T)$ & $\sigma\sim\mathrm{Ber}(0.05), \tau|\sigma\sim\mathrm{Geom}(1/T)$ & $\sigma\sim\mathrm{Ber}(0.05), \tau|\sigma\sim\mathrm{Geom}(1/T)$\\
Num. Samples for $\Tilde{\chi}$ & $2000$ & $2000$ & $2000$ \\ \bottomrule
\end{tabular}
\begin{tablenotes}
\item[*] Fully-connected
\end{tablenotes}
\end{threeparttable}
}
\end{table*}

\begin{figure*}[!t]
    \centering
    \subfloat[]{\includegraphics[width=0.3\linewidth]{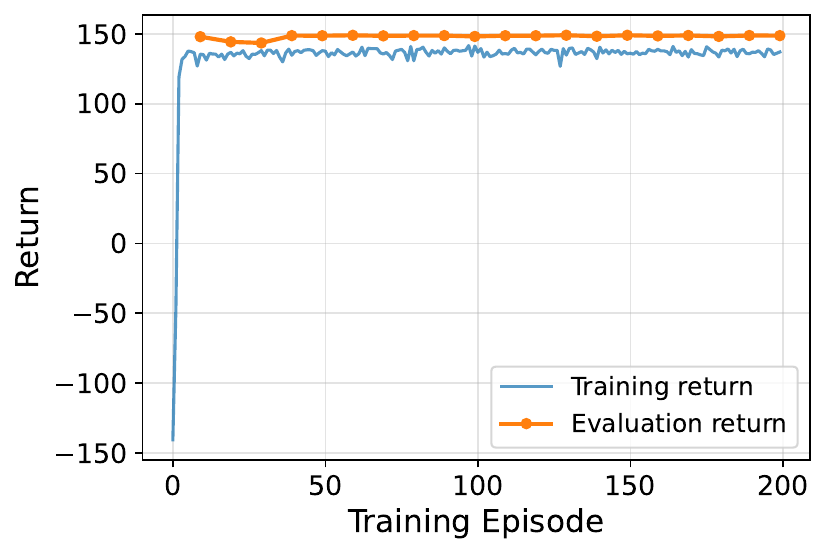}%
    \label{fig:lr-a}}
    \subfloat[]{\includegraphics[width=0.3\linewidth]{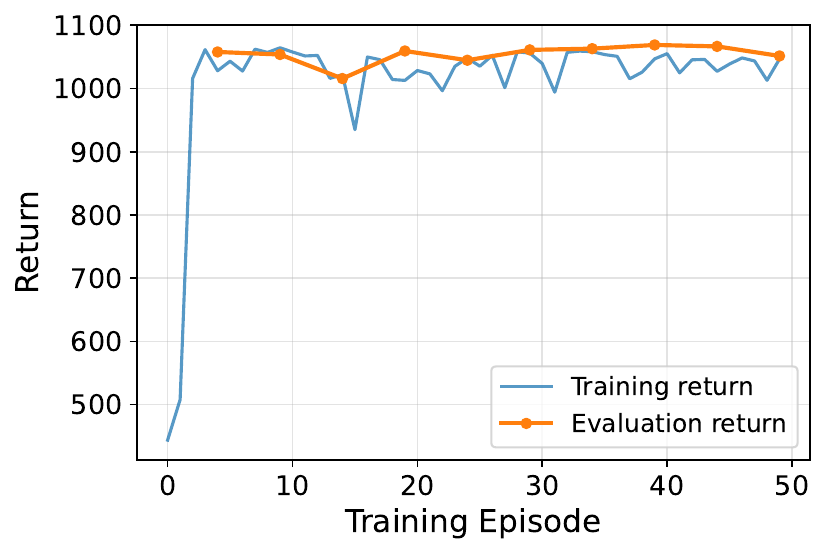}%
    \label{fig:lr-b}}
    \subfloat[]{\includegraphics[width=0.3\linewidth]{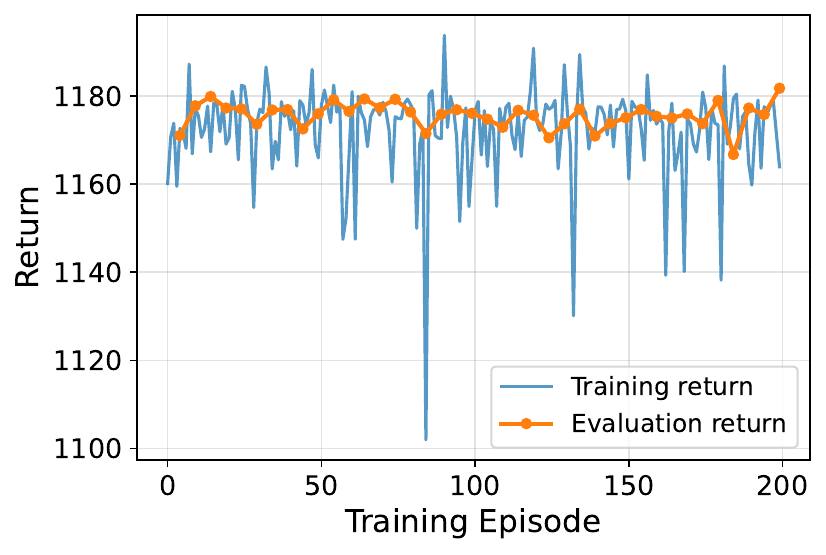}%
    \label{fig:lr-c}}
    \caption{Representative DDPG-CDQ training diagnostics for (a) case study~1, (b) case study~2, and (c) case study~3.}
    \label{fig:lr}
\end{figure*}

To deploy the DynaMark framework, we implemented a DDPG-CDQ agent using a PyTorch-based actor-critic setup. Table~\ref{tab:ddpg_specs} lists the DDPG-CDQ hyperparameters for our experiments. Implementation details, training hyperparameters, and code to reproduce the experiments are available at: [\hyperlink{https://github.com/navidaftabi/DynaMark}{https://github.com/navidaftabi/DynaMark}]. Figure~\ref{fig:lr} reports representative training and evaluation returns for the DDPG-CDQ used to instantiate DynaMark in the three case studies. These curves are provided as empirical training diagnostics, not as a theoretical convergence guarantee. other algorithms such as PPO, SAC, or full TD3 could be used to solve the same framework. Training stability was supported by clipped double-Q targets, target networks with Polyak averaging, gradient clipping, layer normalization, replay-buffer sampling, and decaying OU exploration. Final policy behavior under reward, watermark-budget, and prior variations is evaluated in the ablation and sensitivity study, where the reported metrics are aggregated over seeds and replications. Reward weights $(\omega_1, \omega_2, \omega_3)$ encode the practitioner's preference among actuation overhead, nominal performance, and detection confidence. A practical selection procedure is to first set $\omega_2$ to satisfy a process-specific tracking/quality tolerance under nominal operation (e.g., ensuring $\|\vy_t^{\star} - \vy_t\|_2$ remains within an acceptable band), then increase $\omega_3$ until replay attacks consistently drive the belief away from $0.5$ with acceptable delay. Finally, tune $\omega_1$ to limit watermark energy/wear while preserving the achieved detection behavior. In practice, we found that keeping $\omega_1+\omega_2+\omega_3=1$ and performing a coarse grid search over a small set of candidates yields stable policies, with the final choice guided by the desired operating point on the security–performance frontier.

\end{document}